\documentclass[aps,prd,11pt,notitlepage,longbibliography,nofootinbib,tightenlines,preprintnumbers]{revtex4-1}

\usepackage{booktabs}
\usepackage{amsmath,amssymb,amsfonts}
\usepackage{graphicx}
\usepackage{color}
\usepackage{tikz}
\usetikzlibrary{calc}
 \usetikzlibrary{decorations.text}
 \usetikzlibrary{decorations.pathreplacing}
 \usetikzlibrary{shapes}

\usepackage{dsfont}
\usepackage[pdftex]{hyperref}
\hypersetup{colorlinks=true, linkcolor=darkred, citecolor=blue, linktoc=page}
\definecolor{darkred}{rgb}{0.8,0.1,0.1}

\makeatletter\def\l@subsubsection#1#2{}%
\makeatother

\numberwithin{equation}{section}
\renewcommand\theequation{\arabic{section}.\arabic{equation}} 

\usepackage{subfigure}

\def\cA{{\cal A}}
\def\cB{{\cal B}}
\def\cC{{\cal C}}

\def\cF{{\cal F}}

\def\cG{{\cal G}}
\def\cL{{\cal L}}

\def\cN{{\cal N}}

\def\cZ{{\cal Z}}

\def\id{{\mathds{1}}}

\def\CC{\ensuremath{\mathds C}}
\def\RR{\ensuremath{\mathds R}}
\def\ZZ{\ensuremath{\mathds Z}}

\DeclareMathOperator{\vol}{vol}
\DeclareMathOperator{\Vol}{Vol}

\DeclareMathOperator{\tr}{tr}

\DeclareMathOperator{\Li}{Li}
\DeclareMathOperator{\csch}{csch}

\def\Im{\mathop{\rm Im}}
\def\Re{\mathop{\rm Re}}
\def\Res{\mathop{\rm Res}}

\newcommand{\pslash}{\ensuremath\diagup\!\!\!\!\!{+}}

\begin{document}

\title{Wilson loops in 5d long quiver gauge theories}

\author{Christoph F.~Uhlemann} 
\email{uhlemann@umich.edu}
  
\affiliation{Leinweber Center for Theoretical Physics, Department of Physics
	\\
	University of Michigan, Ann Arbor, MI 48109-1040, USA}

\preprint{LCTP-20-10}

\begin{abstract}

Quiver gauge theories with a large number of nodes host a wealth of Wilson loop operators. 
Expectation values are obtained, using supersymmetric localization, for Wilson loops in the antisymmetric representations associated with each individual gauge node, for a sample of 5d long quiver gauge theories whose  UV fixed points have holographic duals in Type IIB. The sample includes the $T_N$ theories and the results are uniformly given in terms of Bloch-Wigner functions. The holographic representation of the Wilson loops is identified. It comprises, for each supergravity solution, a two-parameter 
family of D3-branes which exactly reproduce the field theory results and identify points in the internal space with the faces of the associated 5-brane web.
The expectation values of (anti)fundamental Wilson loops exhibit an enhanced scaling for many operators, which matches between field theory and supergravity.
\end{abstract}

\maketitle

\tableofcontents

\baselineskip 14pt

\section{Introduction}

Extended objects play a prominent role in quantum field theory.
They are crucial in classifying quantum field theories and understanding relations between them,
they can serve as order parameters, and can be used to define lower-dimensional theories from higher-dimensional parents.

Wilson loop operators are natural objects to study in gauge theories.
Extensive work on Wilson loops, e.g.\ in 4d $\cN=4$ SYM and 3d Chern-Simons-matter theories, has led to a wealth of exact results and revealed rich physical and mathematical structures  (see e.g.\ the 3d road map paper \cite{Drukker:2019bev}).
It is natural to embark on similar explorations in 5d. Gauge theories in 5d are perturbatively non-renormalizable, but many theories with 8 supercharges are believed to describe well-defined quantum field theories that flow to strongly-coupled fixed points in the UV.
The UV fixed points are superconformal field theories (SCFTs) with no supersymmetric marginal couplings (classification attempts include \cite{Intriligator:1997pq,DelZotto:2017pti,Jefferson:2018irk,Bhardwaj:2018yhy,Bhardwaj:2018vuu,Apruzzi:2018nre,Closset:2018bjz,Apruzzi:2019opn,Apruzzi:2019enx,Bhardwaj:2019fzv,Saxena:2019wuy,Apruzzi:2019kgb,Bhardwaj:2020gyu,Bhardwaj:2020kim}), whose spectrum of loop operators includes the gauge theory Wilson loops.
Wilson loops have been studied for 5d $USp(N)$ theories in \cite{Assel:2012nf,Alday:2014bta,Alday:2015jsa} and aspects of S-duality were studied in \cite{Assel:2018rcw}.
Aspects of higher-form symmetries were discussed recently in \cite{Morrison:2020ool,Albertini:2020mdx}.
The purpose of the present paper is to initiate the study of Wilson and more general loop operators in a class of 5d gauge theories that are described by quiver diagrams with a large number of nodes, and which flow to strongly-coupled 5d SCFTs with holographic duals in Type IIB supergravity.

Quiver gauge theories with a large number of nodes host a wealth of Wilson loop operators, which can be characterized by a representation with respect to each of the gauge nodes.
The focus here will be on Wilson loops associated with individual gauge nodes, characterized by a representation with respect to a single node. 
For each choice of representation this is a large family of Wilson loops. 
We will study Wilson loops in fundamental and antifundamental as well as general antisymmetric representations of individual gauge nodes, for a sample of 5d long quiver gauge theories.
The sample includes the $+_{N,M}$ theories introduced in \cite{Aharony:1997bh} and the 5d $T_N$ theories \cite{Benini:2009gi}, which are 5d uplifts of the 4d $T_N$ theories of \cite{Gaiotto:2009we}.
They are complemented by the $Y_N$ and $X_{N}$ theories, discussed previously in \cite{Bergman:2018hin}, 
whose gauge theory descriptions have distinct features like unsaturated nodes and Chern-Simons terms.

In the limits considered here the gauge theory descriptions of the $+_{N,M}$, $T_N$ , $Y_N$ and $X_N$ theories each have a large number of $SU(\cdot)$ gauge nodes, with most of them having large rank.
Building on the supersymmetric localization results for long quiver gauge theories of \cite{Uhlemann:2019ypp}, exact results for the expectation values of Wilson loops in antisymmetric representations will be derived, in the limit where all gauge nodes are strongly coupled. 
The scaling can be expressed uniformly in terms of the length of the quivers.
The expectation values for large antisymmetric Wilson loops scale quadratically, and are uniformly given in terms of Bloch-Wigner dilogarithm functions, with complex arguments encoding the gauge node which the Wilson loop is associated with and the number of boxes in the Young tableau.
For (anti)fundamental Wilson loops the expectation values depend strongly on the gauge node they are associated with.
Depending on the type of gauge node, the scaling can be linear, logarithmically enhanced, or subleading.
These features are rooted in the properties of the matrix models and have interesting manifestations in the holographic duals.

The holographic representation of the Wilson loops will be identified.
The $+_{N,M}$, $T_N$ , $Y_N$ and $X_N$ theories can all be engineered by $(p,q)$ 5-brane webs in Type IIB string theory \cite{Aharony:1997ju,Aharony:1997bh},
and their UV fixed points have holographic duals in Type IIB supergravity \cite{DHoker:2016ujz,DHoker:2016ysh,DHoker:2017mds,DHoker:2017zwj,Uhlemann:2019lge}.
The supergravity solutions describe the conformal limit of the 5-brane webs.
The geometry is a warped product of $AdS_6$ and $S^2$ over a Riemann surface $\Sigma$, which has distinguished points on its boundary at which $(p,q)$ 5-branes emerge.
These are the external 5-branes of the associated 5-brane web and identify the dual field theory.
Antisymmetric Wilson loops are represented by a two-parameter family of D3-branes for each solution, one for each point of $\Sigma$.
This reflects that the gauge node and representation of the Wilson loop are two effectively continuous parameters. 
These D3-branes reproduce the field theory results exactly, and the precise identification will assign to each point of $\Sigma$ a face in the associated 5-brane web.
This connects the internal space of the supergravity solutions locally to the internal structure of the associated 5-brane webs, and deepens the relation between supergravity solutions, brane webs, and the field theories.
Fundamental and anti-fundamental Wilson loops will be understood in terms of D3-branes approaching points on the boundary of $\Sigma$ and in terms of fundamental strings at the limiting points.

5d gauge theories host further loop operators represented by D-strings, which are associated with instanton particles, and more general $(p,q)$ strings.
Their expectation values can be obtained directly from the holographic duals, and the results will be discussed in the context of dualities. 
$SL(2,\ZZ)$ dualities in 5d provide relations between gauge theories with a common UV fixed point, 
which will be used to relate $(p,q)$ loop operators to Wilson loops in S-dual gauge theory deformations.
Moreover, although the $+_{N,M}$, $T_N$, $Y_N$ and $X_{N}$ theories are distinct theories, which are not related to each other by $SL(2,\ZZ)$, there are large-$N$ dualities between the $Y_N$ and $T_{N}$ and between the $X_N$ and $+_{N,N}$ theories, which provide further relations that will be discussed.

\smallskip

\underline{Outline}:
The general class of gauge theories is introduced in sec.~\ref{sec:Wilson-loc}, along with relevant aspects of supersymmetric localization for long quivers.
The Wilson loop expectation values for the $+_{N,M}$, $T_N$, $Y_N$ and $X_{N}$ theories are obtained. 
In sec.~\ref{sec:brane-web-Wilson} the representation of line defects in 5-brane webs is discussed, to guide the identification of Wilson loops in the holographic duals.
In sec.~\ref{sec:AdSCFT} the holographic duals for the theories of sec.~\ref{sec:Wilson-loc} are introduced, brane and string embeddings are discussed and Wilson loop expectation values are obtained holographically.
The results are discussed in sec.~\ref{sec:discussion}. 
Various technical results are derived in appendices.

\section{Wilson loops in long quiver theories}\label{sec:Wilson-loc}

The gauge theories of interest here are 5d linear quiver gauge theories, with $SU(\cdot)$ gauge nodes connected by bifundamental hypermultiplets.
The nodes may have non-trivial Chern-Simons terms and/or additional fundamental hypermultiplets, such that a generic quiver takes the form
\begin{align}\label{eq:SU-fund-quiver}
SU(&N_0)_{c^{}_0} - SU(N_1)_{c^{}_1} - \ldots  - SU(N_L)_{c^{}_L}
 \nonumber\\[-1mm]
 &\vert \hskip 20mm \vert \hskip 29mm \vert
  \\[-1mm] \nonumber
  & \!\!\!\! [k_0] \hskip 15mm [k_1] \hskip 24.5mm [k_L]
\end{align}
The ranks of the gauge groups are encoded in $N_t$ with $t=0,\ldots L$, the Chern-Simons levels are encoded in $c_t$, and the numbers of fundamental hypermultiplets for each node are encoded in $k_t$.
A natural description for long quiver gauge theories is obtained by introducing an effectively continuous coordinate $z\in[0,1]$ along the quiver, and replacing the discrete data
$\lbrace N_t, k_t, c_t \vert\, t=0,..,L\rbrace$ by continuous data
\begin{align}\label{eq:cont-quiver}
  z&=\frac{t}{L}~, &   N(z)&= N_{z L}~,  & k(z)&= k_{zL}~, & c(z)&=c_{zL}~.
\end{align}
For the theories considered here $N(z)$ is a piecewise-linear concave function, and fundamental hypermultiplets or Chern-Simons terms can appear at nodes where $N(z)$ has a kink.
This ensures that the gauge theories flow to strongly-coupled fixed points in the UV, described by 5d SCFTs.

The loop operators of interest in this section are $\tfrac{1}{2}$-BPS supersymmetric Wilson loops of the general form
\begin{align}\label{eq:Wilson-gen}
 W_R^{(t)}&=
 \frac{1}{{\rm dim} R} \tr_R\mathcal P \exp\left\lbrace\int \left(iA^{(t)}+\sigma^{(t)}ds \right)\right\rbrace~.
\end{align}
$t=0,1\ldots ,L$ refers to the gauge node that the Wilson loop is associated with,
$A^{(t)}$ is the gauge field associated with the $t^{\rm th}$ gauge node and $\sigma^{(t)}$ is the real scalar in the vector multiplet.
$R$ is a representation of the gauge group at the node $t$ and $\mathcal P$ denotes path ordering. 
A Wilson loop of the form (\ref{eq:Wilson-gen}) on a great circle of $S^5$ preserves half the supersymmetries,
and the $SU(2)$ $R$-symmetry \cite{Assel:2012nf}. 
The preserved symmetries of the UV SCFT fit into $D(2,1;2)\oplus SU(2)$, which is a sub-superalgebra of $F(4)$ with bosonic symmetry $SO(2,1)\oplus SU(2)^3$ \cite{Frappat:1996pb}. The remaining $SU(2)$ factors are associated with the preserved $SO(4)$ isometries.

For gauge theories of the form (\ref{eq:SU-fund-quiver}) with $L$ large, Wilson loops are labeled by the effectively continuous coordinate $z$ along the quiver, defined in (\ref{eq:cont-quiver}).
For each choice of representation this yields an effectively continuous family of operators.
The focus here will be on the fundamental and anti-fundamental representations, denoted by $f$ and $\bar f$, 
and on general $k$-fold antisymmetric representations, denoted by $\wedge^k$.
To uniformly label the antisymmetric representations it is convenient to introduce a parameter $\mathds{k}$ valued in $[0,1]$ and defined for the $t^{\rm th}$ gauge node by $\mathds{k}\equiv k/N_t$.
A concise notation for the Wilson loops then is
\begin{align}\label{eq:W-z-kk-def}
 W_{f/\bar f}(z)&\equiv W_{f/\bar f}^{(zL)}~, &
 W_\wedge(z,\mathds{k})&\equiv W_{\wedge^{\mathds{k}N(z)}}^{(zL)}~.
\end{align}
These are families of loop operators labeled, respectively, by one and two effectively continuous parameters.
In this section expectation values for these operators will be obtained in the regime where all gauge couplings become infinitely strong, for a sample of 5d long quiver gauge theories.
The expectation values for fundamental and anti-fundamental Wilson loops will differ only for the $Y_N$ theories which involve a Chern-Simons term.

Relevant aspects of supersymmetric localization for long quiver theories and general formulae for Wilson loop expectation values are discussed in sec.~\ref{sec:long-loc}.
The $+_{N,M}$, $T_N$, $Y_N$ and $X_{N}$ quiver gauge theories are discussed explicitly in sections \ref{sec:plus-loc} -- \ref{sec:XNN-loc}.

\subsection{Localization in long quivers}\label{sec:long-loc}

The general form of the zero-instanton part of the partition function of 5d gauge theories on (squashed) spheres was derived in \cite{Kallen:2012va,Kim:2012ava,Imamura:2013xna,Lockhart:2012vp}, and the saddle points for a number of gauge theories of the form (\ref{eq:SU-fund-quiver}) were found in \cite{Uhlemann:2019ypp}.
In this section general expressions for Wilson loops are derived.

The partition function for theories of the form (\ref{eq:SU-fund-quiver}), after supersymmetric localization, is given by an integral over the Cartan algebra,
\begin{align}\label{eqn:generalpartfunc}
\cZ_{S^5}  &= 
\left[ \prod_{t=1}^{L}
\prod_{i=1}^{N_t-1} \int_{-\infty}^{\infty} d \lambda_i^{(t)} \right]e^{-\mathcal F}~,
\end{align}
where $\lambda_i^{(t)}$ are the eigenvalues associated with the $t^{\rm th}$ gauge node. 
The precise form of the integrand $\exp\lbrace -\mathcal F\rbrace$  will not be needed, it can be found for theories of the form (\ref{eq:SU-fund-quiver}) in \cite{Uhlemann:2019ypp}.
The Wilson loop operator (\ref{eq:Wilson-gen}) evaluated on the localization locus (where $A=0$ and $\sigma$ is constant) reduces to
\begin{align}\label{eq:WR-gen}
  W_R^{(t)}&=\frac{1}{{\rm dim} R}\tr_R e^{2\pi\lambda^{(t)}}~.
\end{align}
The expectation value of the Wilson loop can be obtained by inserting this factor into the matrix model
\begin{align}\label{WR-exp-gen}
 \big\langle W_R^{(t)}\big\rangle&=\frac{1}{\cZ_{S^5}}
 \left[ \prod_{t^\prime=1}^{L}
\prod_{i=1}^{N_t-1} \int_{-\infty}^{\infty} d \lambda_i^{(t^\prime)} \right]
 \,\frac{1}{{\rm dim} R}(\tr_R e^{2\pi\lambda^{(t)}})\,e^{-\cF}~.
\end{align}
For Wilson loops in sufficiently small representations the insertion is usually subleading with respect to $\exp\lbrace -\cF\rbrace$ and does not affect the saddle point equations. The scaling for the theories of the form (\ref{eq:SU-fund-quiver}) will be discussed below. If the saddle point equations are unaffected by the insertion, the Wilson loop expectation value reduces to
\begin{align}\label{WR-exp-gen-saddle}
 \big\langle W_R^{(t)}\big\rangle&=\frac{1}{{\rm dim} R}\tr_R e^{2\pi\lambda^{(t)}}\Big\vert_{\rm saddle}~.
\end{align}

The saddle points for a number of long quiver theories were determined in \cite{Uhlemann:2019ypp}.
One introduces a family of eigenvalue distributions $\lbrace \rho_t(\lambda),t=0,..,L\rbrace$, one for each gauge node.
In the continuum formulation this family of eigenvalue distributions is replaced by a single function of two variables
\begin{align}\label{eq:cont-rho}
 \rho(z,\lambda)&=\rho_{zL}(\lambda)~.
\end{align}
For each $z$ it gives the eigenvalue distribution for the gauge node labeled by $t=zL$.
Non-trivial solutions to the saddle point equations exist if the eigenvalues scale with the length of the quiver, such that the eigenvalues and the eigenvalue distributions can be parametrized by\footnote{The factors of $3$ are included for compatibility with \cite{Uhlemann:2019ypp}. The rescaling was used there also to absorb the dependence on the squashing parameters. For the round $S^5$ this results in factors of $3$.}
\begin{align}\label{eq:lambda-scaling}
 \lambda&=3Lx~,
 &
 \hat\rho(z,x)&=3L\rho(z,3L x)~,
\end{align}
with $x$ of order one.
The saddle point equations then demand that $N(z)\hat\rho(z,x)$ be a harmonic function.
It is the solution to an electrostatics problem for a charge configuration with certain boundary conditions that encode the features of the gauge theory under consideration. 

For Wilson loops in the fundamental or anti-fundamental representation of the $t^{\rm th}$ gauge node, (\ref{WR-exp-gen-saddle}) becomes, respectively,
\begin{align}\label{eq:W-f-d}
 \big\langle W_{f}^{(t)}\big\rangle&=\frac{1}{N_t}\sum_{i=1}^{N_t} e^{2\pi\lambda_i^{(t)}}  
 =
 \int d\lambda\, \rho_t(\lambda) e^{2\pi\lambda}~,
 &
 \big\langle W_{\bar f}^{(t)}\big\rangle&=
 \int d\lambda\, \rho_t(\lambda) e^{-2\pi\lambda}~.
\end{align}
Switching to the continuous quiver coordinate $z$ and to $\rho(z,\lambda)$ defined in (\ref{eq:cont-rho}), and further implementing the scaling in (\ref{eq:lambda-scaling}),
with the notation for the Wilson loop in (\ref{eq:W-z-kk-def}), leads to
\begin{align}\label{eq:W-f}
 \big\langle W_{f}(z)\big\rangle
 &=\int dx \,\hat\rho\left(z,x\right) e^{6\pi L x}~,
 &
\big\langle W_{\bar f}(z)\big\rangle
 &=\int dx \,\hat\rho\left(z,x\right) e^{-6\pi L x}~.
\end{align}
The scaling of the expectation value at leading order is determined by the largest eigenvalue for the fundamental and by the smallest eigenvalue for the anti-fundamental Wilson loop, and due to (\ref{eq:lambda-scaling}) naively proportional to the length of the quiver.

We now turn to the $k$-fold antisymmetric representation of $SU(N_t)$, denoted by $\wedge^k$, with $k$ of $\mathcal O(N_t)$.
The Wilson loop expectation value (\ref{WR-exp-gen-saddle}) becomes
\begin{align}\label{eq:W-wedge-d}
 \big\langle W_{\wedge ^k}^{(t)}\big\rangle&= \binom{N_t}{k}^{-1} \sum_{j_1<j_2<\ldots<j_k} e^{2\pi\sum_{\ell=1}^k \lambda_{j_\ell}^{(t)}}~.
\end{align}
Due to the strict inequalities between the $j_\ell$ in the sum, the leading-order scaling is determined by the $k$ distinct largest eigenvalues
(similar discussions can be found in \cite{Yamaguchi:2006tq,Assel:2012nf}).
To determine the $k$ largest eigenvalues from the eigenvalue distributions, we define $b_{t,\ell}$ by
\begin{align}\label{eq:b-ell-t}
 N_t\int_{b_{t,\ell}}^\infty d\lambda \,\rho_t(\lambda)&=\ell~.
\end{align}
The eigenvalue distribution has to be integrated from $b_{t,\ell}$ to infinity to capture a fraction $\ell/N_t$ of the largest eigenvalues.
At leading order, the Wilson loop expectation value is then given by
\begin{align}\label{eq:W-wedge-disc}
 \ln\big\langle W_{\wedge^k}^{(t)}\big\rangle &= 
 2\pi \sum_{\ell=1}^k  b_{t,\ell}~.
\end{align}
For $k$ of $\mathcal O(N_t)$ the sum can be converted to an integral.
The scaling of the eigenvalues can be taken into account by switching to the electrostatics potential $\hat\rho(z,x)$ and the parametrization in terms of $x$ in (\ref{eq:lambda-scaling}).
The expectation value (\ref{eq:W-wedge-disc}) with (\ref{eq:b-ell-t}) then becomes
\begin{align}\label{eq:W-wedge-k}
 \ln\big\langle W_{\wedge^k}^{(t)}\big\rangle &= 
 2\pi N_t \int_{0}^{k/N_t} dy\, b_{t,yN_t}~,
 &
 \int^{\infty}_{b_{t,y}/(3L)} dx \,\hat\rho \big(\tfrac{t}{L},x\big) &=y~.
\end{align}
It is convenient to replace the $\lbrace b_{t,\ell}\rbrace$ by a function of two variables $b(z,y)$ and absorb a rescaling.
In terms of the continuous quiver coordinate $z$ of (\ref{eq:cont-quiver}) and the label $\mathds{k}$ for the antisymmetric representation defined in (\ref{eq:W-z-kk-def}), the Wilson loop expectation value then becomes
\begin{align}\label{eq:W-wedge-k-2}
\ln \langle W_\wedge(z,\mathds{k})\rangle &= 6\pi L N(z)\int_0^{\mathds{k}}dy\,b(z,y)~,
&
\int^{\infty}_{b(z,y)} dx \,\hat\rho (z,x) &=y~.
\end{align}
It will often be convenient to use the inverse function $y(b)$, to express the expectation value as
\begin{align}\label{eq:W-wedge-k-3}
 \ln\big\langle W_{\wedge}(z,\mathds{k})\big\rangle&=
 6\pi L\,N(z) \int_{b(0,z)}^{b(\mathds{k},z)} y'(b) db\, b
 =-6\pi L\,N(z) \int_{b(0,z)}^{b(\mathds{k},z)} b\, \hat\rho(z,b)\, db~.
\end{align}
Fundamental and antisymmetric Wilson loops will be evaluated for specific theories using the expressions in (\ref{eq:W-f}) and (\ref{eq:W-wedge-k-2}), (\ref{eq:W-wedge-k-3}) in the following subsections.

\smallskip

We close this part with comments on the scaling of the Wilson loop expectation values and general features of the saddle point eigenvalue distributions.
Since the eigenvalues were scaled linearly with $L$ in (\ref{eq:lambda-scaling}), the (anti)fundamental and $k$-fold antisymmetric Wilson loops in (\ref{eq:W-f-d}) and (\ref{eq:W-wedge-d}) naively scale like $\exp(L)$ and $\exp(kL)$, respectively.
This is indeed the scaling at nodes where the saddle point eigenvalue distribution $\hat\rho(z,x)$ has compact support.
However, $\hat\rho(z,x)$ typically has support for $x$ on the entire real line for the theories in (\ref{eq:SU-fund-quiver}), with exponentially decaying tails for large $|x|$.
The tails signals that, after absorbing the linear scaling of the eigenvalues with $L$ as in (\ref{eq:lambda-scaling}), the rescaled eigenvalues $x$ still spread out onto the entire real line: 
the ``bulk'' of the eigenvalues scales linearly with $L$, but the largest eigenvalues have a logarithmically enhanced scaling.
This will be seen more explicitly below.
For quantities which involve an $\mathcal O(1)$ fraction of the eigenvalues, such as Wilson loops in large antisymmetric representations, the tails do not affect the leading-order behavior.
But for quantities with stronger sensitivity to individual eigenvalues, such as (anti)fundamental Wilson loops, the leading-order behavior can change.
The scaling will be enhanced logarithmically for (anti)fundamental Wilson loops at nodes where $\hat\rho(z,x)$ has unbounded support,
and remain unmodified at nodes where $\hat\rho(z,x)$ has bounded support.
Since the free energy for the theories in (\ref{eq:SU-fund-quiver}) scales like $L^4$  \cite{Uhlemann:2019ypp}, 
the scaling of all Wilson loops discussed here is subleading with respect to the ``background'' set by the gauge theory, which was assumed for (\ref{WR-exp-gen-saddle}).

\subsection{\texorpdfstring{$+_{N,M}$}{+[N,M]} theory}\label{sec:plus-loc}

The $+_{N,M}$ theories, discussed in \cite{Aharony:1997bh} and named in \cite{Bergman:2018hin} for the shape of the 5-brane junctions that realizes them (fig.~\ref{fig:plus}), are the UV fixed points of the quiver gauge theories
\begin{align}\label{eq:D5NS5-quiver}
[N]-\underbrace{SU(N)-\ldots -SU(N)}_{SU(N)^{M-1}}-[N] ~.
\end{align}
All Chern-Simons levels are zero and $L=M-1$.
The ``rank function'' $N(z)$ (with a slight abuse of notation) is given by $N(z)=N$.
The S-dual quiver has $N$ and $M$ exchanged.
All nodes are saturated, in the sense that the effective number of flavors, including bifundamentals, is twice the number of colors.
The saddle point eigenvalue distributions for general theories of this class were found in \cite{Uhlemann:2019ypp}.
For the $+_{N,M}$ theory with $N$ and $M$ large the saddle point is given by
\begin{align}\label{eq:rho-sol-plus}
 \hat\rho_s(z,x)&=\frac{4\sin (\pi  z) \cosh \left(2 \pi x\right) }{\cosh \left(4 \pi x\right)-\cos (2 \pi  z)}~.
\end{align}
The eigenvalue distributions have exponential tails for large $|x|$ at all interior nodes, 
and they collapse to $\delta$-functions in $x$ at the boundary nodes corresponding to $z\rightarrow 0$ and $z\rightarrow 1$.

\smallskip

We start the discussion of Wilson loops with large antisymmetric representations, for which the expectation values can be obtained from (\ref{eq:W-wedge-k-2}), (\ref{eq:W-wedge-k-3}).
To this end the function $b(z,y)$ defined in (\ref{eq:W-wedge-k-2}) has to be determined.
The required integral is
\begin{align}\label{eq:plus-y}
 y=\int_{b(z,y)}^\infty dx\,\hat\rho(z,x)&=\frac{1}{2}-\frac{1}{\pi}\tan ^{-1}(\sinh (2 \pi b(z,y)) \csc (\pi  z))~,
\end{align}
which leads to
\begin{align}\label{eq:plus-b}
 b(z,y)&=\frac{1}{2\pi}\sinh^{-1}\left(\cot\left(\pi y\right)\sin(\pi z)\right)~.
\end{align}
This function is monotonic for $y\in(0,1)$, with logarithmic divergences at the end points which are integrable and will be discussed shortly.
With $b$ in hand, the Wilson loop expectation values follow straightforwardly from (\ref{eq:W-wedge-k-3}).
This leads to
\begin{align}
 \ln\big\langle W_{\wedge}(z,\mathds{k})\big\rangle&=
 \frac{3MN}{\pi}\left[
 D\Big(e^{2\pi b(\mathds{k},z)+i\pi z}\Big)+D\Big(e^{2\pi b(\mathds{k},z)+i\pi(1-z)}\Big) \right]~,
 \label{eq:plus-Wwedge}
\end{align}
where $b(z,y)$ is given in (\ref{eq:plus-b}) and $D$ is the Bloch-Wigner function,\footnote{
$D(z)$ is single-valued and real analytic on $\CC$ except at $z=0,1,\infty$, where it is continuous but not differentiable. 
It satisfies $D(z)=-D(1/z)=D(1-1/z)=D(1/(1-z))=-D(1-z)=-D(z/(z-1))$ and the 5-term relation $D(x)+D(y)+D(1-xy)+D(\frac{1-x}{1-xy})+D(\frac{1-y}{1-xy})=0$.\label{foot:D-rel}}
\begin{align}\label{eq:D-def}
 D(u)&=\Im\left(\Li_2(u)\right)+\arg(1-u)\ln |u|~,
\end{align}
with $\arg$ denoting the branch of the argument that lies between $-\pi$ and $\pi$.

The expectation value in (\ref{eq:plus-Wwedge}) is invariant under $z\rightarrow 1-z$, reflecting the symmetry of the quiver in (\ref{eq:D5NS5-quiver}),
and under $\mathds{k}\rightarrow 1-\mathds{k}$, corresponding to charge conjugation.
It approaches one for $\mathds{k}\rightarrow 0$ and $\mathds{k}\rightarrow 1$, where the representation becomes trivial, 
and it also approaches one for $z\rightarrow \lbrace 0,1\rbrace$, corresponding to the boundary nodes where the eigenvalue distributions are $\delta$-functions.
The maximal value overall is obtained for $\mathds{k}=z=\frac{1}{2}$, 
\begin{align}\label{eq:plus-max}
 \ln\big\langle W_{\wedge}\big(\tfrac{1}{2},\tfrac{1}{2}\big)\big\rangle&=\frac{6C}{\pi}MN ~,
\end{align}
where $C\approx 0.915966$ is Catalan's constant.
The overall factor in (\ref{eq:plus-Wwedge}) depends on $N$ and $M$ through the S-duality invariant combination $MN$.
Perhaps less obviously, the expression is invariant under exchange of $\mathds{k}$ and $z$. 
This will be discussed further in sec.~\ref{sec:plus-hol}.

\smallskip

Before moving on to (anti)fundamental Wilson loops, it is instructive to discuss the function $b(z,y)$ in (\ref{eq:plus-b}).
By definition, $b(z,y)$ encodes that a fraction $y$ of the largest (rescaled) eigenvalues $x$ at the node $z$ is contained in the interval $(b(z,y),\infty)$. 
The interval $(b(z,1-y),b(z,y))$ thus includes all eigenvalues except for a fraction $y$ of the largest and smallest eigenvalues.
Since $b(z,y)$ is finite for $y\in(0,1)$, the ``bulk'' of the rescaled eigenvalues $x$ (any fixed fraction of them) for each node is contained in a finite interval, and the corresponding $\lambda$ scale linearly with $L$.
However, at the interior nodes 
\begin{align}\label{eq:plus-b-asympt}
 b(z,y)\big\vert_{y\rightarrow 0}&\sim -\frac{1}{2\pi}\ln y~, &
 b(z,y)\big\vert_{y\rightarrow 1}&\sim \frac{1}{2\pi}\ln(1-y)~.
\end{align}
This signals the logarithmically enhanced scaling of the largest eigenvalues discussed below (\ref{eq:W-wedge-k-3}):
An order one number of the largest eigenvalues, corresponding to $y=\mathcal O(1/N(z))$, is contained in an interval with lower bound $b(z,y)=\mathcal O(\ln N(z))$.
Similarly, an order one number of the smallest eigenvalues is contained in an interval with upper bound of $\mathcal O(-\ln N(z))$.
The divergences in $b(z,y)$ for $y\rightarrow \lbrace 0,1\rbrace$ are an artifact: 
when $b(z,y)$ becomes large enough that the fraction of eigenvalues in the interval $(b(z,y),\infty)$ becomes small compared to $1/N(z)$, which would correspond to a single actual eigenvalue, there is effectively no eigenvalue larger than $b(z,y)$ and the description in terms of a continuous distribution ultimately fails.
This introduces a physical cut-off. 
Combining these observations shows that the largest and smallest eigenvalues have a logarithmically enhanced scaling compared to the ``bulk'' of the eigenvalues which scale linearly with $L$.

This enhanced scaling is inconsequential for the leading-order behavior of Wilson loops in large antisymmetric representations.
Intuitively speaking, the expectation values are determined by the $k$ distinct largest eigenvalues, almost all of which scale linearly with $L$.
The contribution from the $\mathcal O(1)$ largest eigenvalues in (\ref{eq:W-wedge-d}) is enhanced logarithmically, but it remains subleading with respect to the contribution of the ``bulk eigenvalues''.
This is reflected in the divergences of $b(z,y)$ being integrable, which led to (\ref{eq:plus-Wwedge}).

\smallskip

We now turn to (anti)fundamental Wilson loops.
At the boundary nodes at $z\in\lbrace 0,1\rbrace$, where $\hat\rho_s(0,x)=\hat\rho(1,x)=\delta(x)$, the leading-order expectation values obtained from (\ref{eq:W-f}) are
\begin{align}\label{eq:Wf-plus}
 \big\langle W_{f}(0)\big\rangle=\big\langle W_{\bar f}(0)\big\rangle&=1~,
 \nonumber\\
 \big\langle W_{f}(1)\big\rangle=\big\langle W_{\bar f}(1)\big\rangle&=1~.
\end{align}
For the (anti)fundamental Wilson loops at interior nodes the enhanced scaling of the largest eigenvalues comes into play.
When the saddle point eigenvalue distribution has unbounded support, the integral in (\ref{eq:W-f}) is divergent, since the argument of the exponential scales with $L$.
In parallel to the discussion above, this reflects that a continuous eigenvalue distribution ceases to be an effective description when $x$ is large.
The integration in (\ref{eq:W-f}) should be cut off when the fraction of eigenvalues larger than $x$ becomes smaller than $\epsilon/N(z)$ for some $\epsilon<1$, i.e.\ when there are effectively no eigenvalues left.
Since the divergence is logarithmic the leading-order result is independent of~$\epsilon$.
An operationally equivalent approach is to use that the $k$-fold antisymmetric representations of $SU(N_t)$ with $k=1$ and $k=N_t-1$ are, respectively, the fundamental and anti-fundamental representations.
This corresponds to $\mathds{k}=1/N(z)$ and $\mathds{k}=1-1/N(z)$ and leads to
\begin{align}\label{eq:W-wedge-f}
 \ln \langle W_f(z)\rangle &= \ln\langle W_\wedge (z,1/N(z))\rangle~,
 &
 \ln \langle W_{\bar f}(z)\rangle &= \ln\langle W_\wedge (z,1-1/N(z))\rangle~.
\end{align}
With (\ref{eq:plus-Wwedge}) one obtains
\begin{align}\label{eq:Wf-plus-int}
 &&
 \ln \langle W_f(z)\rangle&=\ln \langle W_{\bar f}(z)\rangle=3M \ln N~,
 &
 z&\in (0,1)~.
\end{align}
The logarithmically enhanced scaling reflects that the scaling of the fundamental/anti-fundamental Wilson loops in (\ref{eq:W-f-d}) is determined by the largest/smallest eigenvalue. 
The scaling and the coefficients in (\ref{eq:Wf-plus-int}), which are independent of $z$, 
are consistent with results extracted from a numerical study of Wilson loops in the $+_{N,M}$ matrix model for finite $N$ and $M$.

\subsection{\texorpdfstring{$T_N$}{T[N]} theory}

The 5d $T_N$ theories were proposed in \cite{Benini:2009gi} as uplift of the 4d $T_N$ theories.
The 5d $T_N$ theories can be defined as the UV fixed points of the quiver gauge theories \cite{Bergman:2014kza,Hayashi:2014hfa}
\begin{align}
\label{TNquiver}
[2]-SU(2)-SU(3)-\ldots-SU(N-1)-[N] ~,
\end{align}
with all Chern-Simons levels zero.
The S-dual gauge theories are identical.
Various aspects were studied recently in \cite{Eckhard:2020jyr}.
The quivers (\ref{TNquiver}) are characterized by $L=N-2$ and the rank function $N(z)=N z$.
The saddle-point eigenvalue distributions are encoded in (sec.~4.2 of \cite{Uhlemann:2019ypp})
\begin{align}\label{eq:TN-saddle}
 \hat\rho_s(z,x)&=\frac{\sin (\pi  z)}{z}\frac{1}{\cosh \left(2\pi x\right)+\cos (\pi  z)}~.
\end{align}
The distribution obtained for $z\rightarrow 0$ is finite and non-vanishing, with unbounded support. For $z\rightarrow 1$ it approaches a $\delta$-function.

The expectation values of antisymmetric Wilson loops can, again, be obtained from (\ref{eq:W-wedge-k-2}).
To this end, we note that\footnote{%
The integral simplifies with the expression of sec.~3 of \cite{Uhlemann:2019ypp}, in which 
$\hat\rho_s=\frac{i}{z}\left(\frac{1}{u+1}-\frac{1}{\bar u+1}\right)$ with $u=\exp(2\pi x+i\pi z)$.}
\begin{align}
 \int_{b(z,y)}^\infty dx\, \hat\rho(z,x)&=1+\frac{i}{2\pi z}\ln\left(\frac{1+e^{2\pi b(z,y)+i\pi z}}{1+e^{2\pi b(z,y)-i\pi z}}\right)\stackrel{!}{=}y~,
\end{align}
such that
\begin{align}\label{eq:b-TN}
 b(z,y)&=\frac{1}{2\pi}\ln\left[\sin(\pi(1-y)z)\csc(\pi y z)\right]~.
\end{align}
The expectation value, via (\ref{eq:W-wedge-k-2}) or (\ref{eq:W-wedge-k-3}), evaluates to
\begin{align}\label{eq:W-wedge-TN}
 \ln\left\langle W_{\wedge}(z,\mathds{k})\right\rangle
 &=\frac{3}{\pi}N^2 D\Big(e^{2\pi b(\mathds{k},z)+i\pi (1-z)}\Big)~,
\end{align}
where $D$ is the Bloch-Wigner function defined in (\ref{eq:D-def}) and $b$ is defined in (\ref{eq:b-TN}).\footnote{
The Bloch-Wigner function can be expressed in terms of its argument on the unit circle (Clausen functions), through
$2D(z)=D(\frac{z}{\bar z})+D(\frac{1-\bar z}{1-z})+D(\frac{\bar z (1-z)}{z(1-\bar z)})$. This leads to $\ln\left\langle W_{\wedge}(z,\mathds{k})\right\rangle=\frac{3}{2\pi}N^2\big[D\big(e^{2\pi i \mathds{k} z}\big)-D\big(e^{2\pi i(\mathds{k}-1)z}\big)-D\big(e^{2\pi i z}\big)\big]$.
}

The expectation values in (\ref{eq:W-wedge-TN}) approach one for $z\rightarrow 1$ and for $z\rightarrow 0$ (the eigenvalue distribution at $z=0$ is non-trivial, but the gauge groups become small), and also for $\mathds{k}\rightarrow 0$ and $\mathds{k}\rightarrow 1$.
The maximal expectation value is obtained for $z=\tfrac{2}{3}$ and $k=\tfrac{1}{2}$. It is given by
\begin{align}
 \ln\big\langle W_{\wedge}\big(\tfrac{2}{3},\tfrac{1}{2}\big)\big\rangle
 &=
 \frac{3}{\pi}N^2 D\big(e^{\frac{i\pi}{3}}\big)~,
\end{align}
where $D(e^{i\pi/3})\sim 1.0149416$ is the maximum of the Bloch-Wigner function on the unit circle.

The leading-order expectation values of (anti)fundamental Wilson loops can be obtained using (\ref{eq:W-f}) for the boundary node at $z=1$,
which leads to
\begin{align}\label{eq:W-f-TN}
 \langle W_f(1)\rangle=\langle W_{\bar f}(1)\rangle&=1~.
\end{align}
At all other nodes the eigenvalue distributions have unbounded support and the scaling is enhanced.
Similarly to the discussion leading to (\ref{eq:Wf-plus-int}), the leading-order expectation values of (anti)fundamental Wilson loops can be obtained 
from (\ref{eq:W-wedge-TN}) using the relations in (\ref{eq:W-wedge-f}). This leads to
\begin{align}\label{eq:Wf-TN-int}
 \ln \langle W_f(z)\rangle&=\ln \langle W_{\bar f}(z)\rangle=3N \ln N~, & z&\in[0,1)~.
\end{align}
The scaling of the expectation values is again enhanced uniformly, from linear scaling with $N$ to $N\ln N$,
with the coefficient independent of the gauge node.

\subsection{\texorpdfstring{$Y_N$}{Y[N]} theory}\label{sec:YN-loc}

The $Y_N$ theory, named in \cite{Bergman:2018hin} for the shape of the associated 5-brane junction (fig.~\ref{fig:YN}), 
is the UV fixed point of the quiver gauge theory
\begin{align}\label{eq:YN-quiver-2}
SU&(2)-SU(3)-\ldots-SU(N\,{-}\,1)-SU(N)^{}_{\pm 1}-SU(N\,{-}\,1)-\ldots-SU(3)-SU(2)
\nonumber\\
&\ \vert \hskip 130mm \vert
\\  \nonumber
&[2] \hskip 127mm [2]
\end{align}
The central node has an effective number of flavors (including bifundamentals) which is less than twice the number of colors, and involves a Chern-Simons term.
The S-dual gauge theory deformation is given by
\begin{align}\label{eq:YN-quiver-S}
SU(2)-SU(4)-SU(6)-\ldots-SU(2N-2)-[2N] ~.
\end{align}
For the gauge theory in (\ref{eq:YN-quiver-2}), $L=2N-3$ and the saddle point eigenvalue distributions are encoded in (sec.~4.3 of \cite{Uhlemann:2019ypp})
\begin{align}\label{eq:varrho-YN}
 \hat\rho_s(z,x)&=\frac{2N}{N(z)\sqrt{-1-4e^{-4\pi c_{N-1}x+2\pi iz}}}+\rm{c.c.}
 &
 N(z)&=N(1-|1-2z|)~.
\end{align}
The eigenvalue distributions (\ref{eq:varrho-YN}) are finite for $z\rightarrow 0$ and $z\rightarrow 1$, with unbounded support. 

The expectation values of antisymmetric Wilson loops are obtained from (\ref{eq:W-wedge-k-2}).
To determine $b(z,y)$ we assume $z\leq \frac{1}{2}$; the results for $z\geq \tfrac{1}{2}$ follow from the symmetry of the quiver under $z\rightarrow 1-z$.
The required integral is
\begin{align}
 y\stackrel{!}{=}\int_{b(z,y)}^{\infty} dx\, \hat\rho_s(z,x)
 &=\frac{iN}{\pi N(z)}\ln\left(\frac{1+\sqrt{1+4e^{-4\pi b(z,y)-2\pi iz}}}{1+\sqrt{1+4e^{-4\pi b(z,y)+2\pi iz}}}\right)~,
\end{align}
which leads to
\begin{align}\label{eq:b-YN}
 b(z,y)&=\frac{1}{4\pi}\ln\left[\sin^2\left(2\pi z(1-y)\right)\csc\left(\pi z(2-y)\right)\csc(\pi y z)\right]~.
\end{align}
The expectation value, via (\ref{eq:W-wedge-k-2}) or (\ref{eq:W-wedge-k-3}), becomes\footnote{%
The result can also be expressed as 
$  \ln\langle W_{\wedge}(z,\mathds{k})\rangle = 
  \frac{3}{\pi}N^2\left[
  D\big(e^{2\pi i(\mathds{k}-2)z}\big)+D\big(e^{2\pi i \mathds{k}z}\big)-D\big(e^{4\pi i(\mathds{k}-1)z}\big)
  \right]$.
}
\begin{align}\label{eq:Wwedge-YN}
  \ln\langle W_{\wedge}(z,\mathds{k})\rangle &= 
  \frac{6}{\pi}N^2\, D\Big(1+e^{2\pi i (\mathds{k}-1)z}\sin(\pi \mathds{k}z)\csc(\pi(\mathds{k}-2)z)\Big)~.
\end{align}
This expression is valid for $z\leq 1/2$; the result for $z>1/2$ follows from symmetry under $z\rightarrow 1-z$, which can be implemented by replacing $z$ with $(1-|1-2z|)/2$ on the right hand side.

\begin{figure}
\subfigure[][]{\label{fig:YN-plot-1}
 \includegraphics[width=0.4\linewidth]{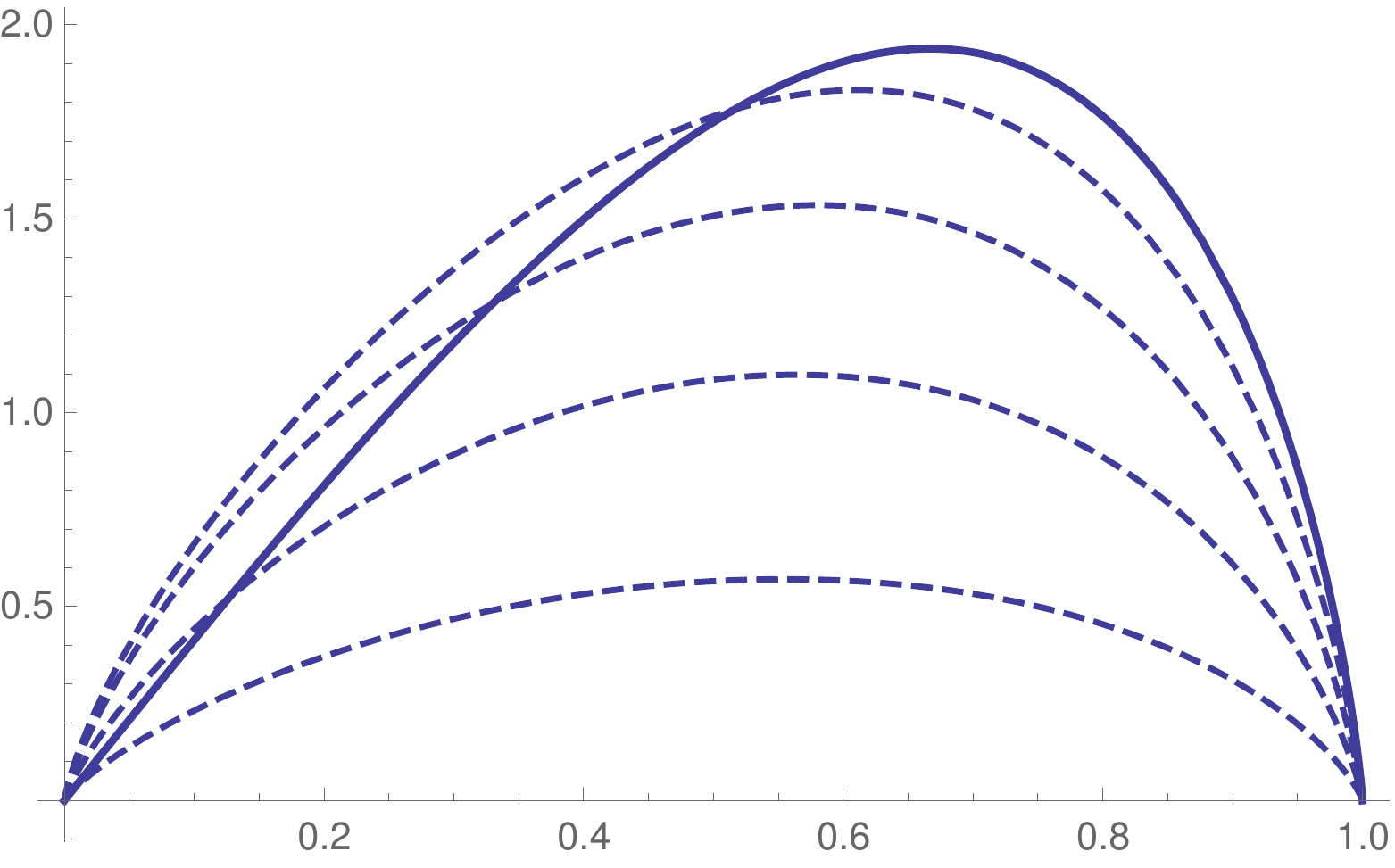}
 \put (3,1) {$\mathds{k}$}
 \put (-200,120) {\small $\ln\langle W_\wedge\rangle/N^2$}
}\hskip 10mm
\subfigure[][]{
 \includegraphics[width=0.4\linewidth]{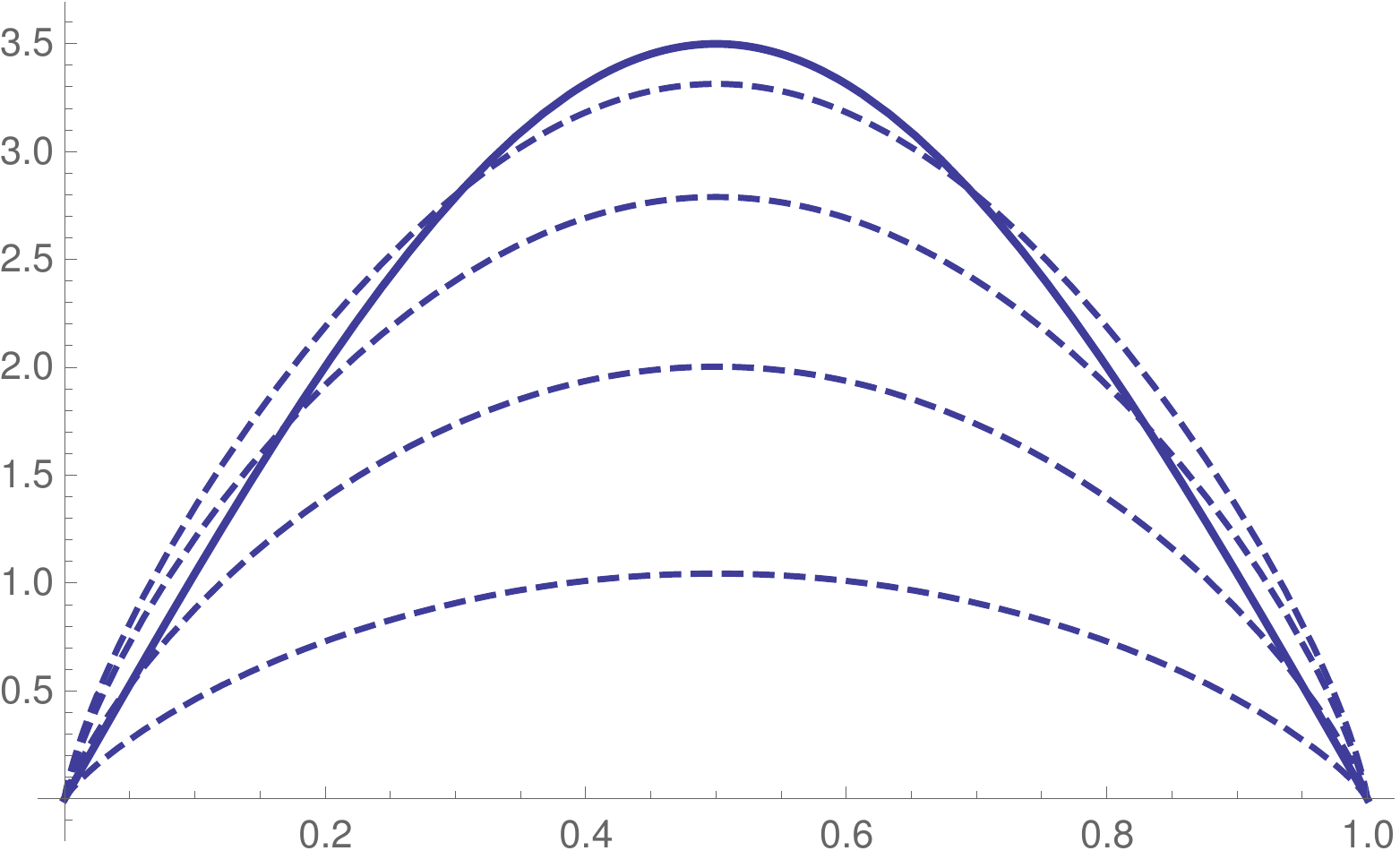}
 \put (3,1) {$\mathds{k}$}
 \put (-200,120) {\small $\ln\langle W_\wedge\rangle/N^2$}
}
 \caption{Left: Expectation value of the $\wedge^k$ Wilson loop in the $Y_N$ theory. The dashed curves from bottom to top are for $z\in\lbrace \tfrac{1}{10},\ldots,\tfrac{4}{10}\rbrace$, the solid curve is for $z=\tfrac{1}{2}$.
 Right: Analogous plot for the $X_{N}$ theory.}
\end{figure}

The expectation values in (\ref{eq:Wwedge-YN}) vanish for $\mathds{k}\rightarrow 0$ and $\mathds{k}\rightarrow 1$.
They are not invariant under $\mathds{k}\rightarrow 1-\mathds{k}$, which results from the eigenvalue distribution not being symmetric under reflection due to the Chern-Simons term. 
A plot is shown in fig.~\ref{fig:YN-plot-1}. 
At the central node (\ref{eq:Wwedge-YN}) reduces to 
\begin{align}
\ln\big\langle W_{\wedge}\big(\tfrac{1}{2},\mathds{k}\big)\big\rangle &=
\frac{6}{\pi}N^2D\big(e^{i\pi(1-\mathds{k})}\big)~.
\end{align}
The global maximum is attained for $\mathds{k}=\tfrac{2}{3}$ at $z=\tfrac{1}{2}$ and is related by a factor two to the maximum for the $T_N$ theory.

We now turn to (anti)fundamental Wilson loops.
For $c_{N-1}=1$ the support of the eigenvalue distribution at the central node is bounded from above.
The expectation value of the fundamental Wilson loop at the central node thus scales linearly and via (\ref{eq:W-f}) evaluates to
\begin{align}\label{eq:YN-fund-loc}
\big\langle W_{f}\big(\tfrac{1}{2}\big)\big\rangle
 &=\int_{-\infty}^{\ln 2/(2\pi)} dx \,\hat\rho_s\left(\tfrac{1}{2},x\right) e^{6\pi L x}
 =\frac{ \Gamma \left(3N+\tfrac{1}{2}\right)}{\sqrt{\pi } \Gamma
   \left(3N+1\right)}\,2_{}^{\,6N}~.
\end{align}
The scaling at leading order is given by $\ln\langle W_{f}(\tfrac{1}{2})\rangle = 6N\ln 2+\mathcal O(\ln N)$
and reflects the largest eigenvalue $x_0=\ln 2/(2\pi)$.
The expectation value of the anti-fundamental Wilson loop has enhanced scaling since the eigenvalue distribution is unbounded from below.
For $c_{N-1}=-1$ the situation is reversed.
The expectation values of the remaining (anti)fundamental Wilson loops can be obtained from (\ref{eq:Wwedge-YN}) using  the relations in (\ref{eq:W-wedge-f}).
This leads to
\begin{align}\label{eq:Wf-YN-int}
  \ln \langle W_f(z)\rangle&=3N \ln N~,  & z&\neq \tfrac{1}{2}~,
  \nonumber\\
  \ln \langle W_{\bar f}(z)\rangle&=6N\ln N~,  & z&\in [0,1]~.
\end{align}
As in the previous examples, the scaling is uniformly enhanced to $N\ln N$.
Since the $Y_N$ theory has a Chern-Simons term, the results for the fundamental and anti-fundamental representation differ.

The S-dual quiver (\ref{eq:YN-quiver-S}) is similar to the quiver for the $T_N$ theory: compared to (\ref{TNquiver}) all gauge nodes are scaled up by a factor two and there are no fundamentals on the left end. The function $\hat \rho(z,x)$ encoding the saddle point eigenvalue distributions is identical to that of the $T_N$ theory \cite{Uhlemann:2019ypp}.
The $\wedge^k$ Wilson loop expectation value in the S-dual quiver (\ref{eq:YN-quiver-S}) is therefore given by that of the $T_N$ theory rescaled by a factor $2$, originating in the fact that $N(z)$ is larger by a factor $2$.
The expectation value of the fundamental Wilson loop associated with the $SU(N-2)$ node in (\ref{eq:YN-quiver-S}) can be obtained from (\ref{eq:W-f}),
and is one, in parallel with the discussion for the $T_N$ theory.

\subsection{\texorpdfstring{$X_{N}$}{X[N]} theory}\label{sec:XNN-loc}

The $X_{N}$ theories are a special case of the $X_{N,M}$ theories discussed in \cite{Bergman:2018hin}, named for the shape of the associated 5-brane junction (fig.~\ref{fig:X-web}).
The quiver gauge theory is
\begin{align}\label{eq:XNN-quiver}
 SU(2)-SU(4)-\ldots -SU(2N\,{-}\,2) - SU(2N) - SU(2N\,{-}\,2) - \ldots -SU(4) - SU(2)~,
\end{align}
and $L=2N-1$.
There is no Chern-Simons term at the central node and the S-dual quiver is identical.
The electrostatics potential encoding the saddle point eigenvalue distributions is derived in app.~\ref{sec:XNN-FS5}, and leads to
\begin{align}
\label{eq:XNN-rho}
 \hat\rho_s(z,x)&=
 \frac{2N}{N(z)}
 \frac{1}{\sqrt{1-2\coth^2 (2 \pi  x+i \pi  z)}}+\mathrm{c.c.}
&
N(z)&=2N(1-|1-2z|)~.
\end{align}
Similarly to the saddle point for the $Y_N$ theory it has an interior node at $z=\frac{1}{2}$ where the eigenvalue distribution has bounded (and in this case compact) support; the distribution is non-vanishing only for  $\cosh(2\pi x)\leq \sqrt{2}$.
At all other nodes the eigenvalue distributions have exponential tails.

The expectation values for the antisymmetric Wilson loops are obtained from (\ref{eq:W-wedge-k-2}).
To determine $b(z,y)$ from the expression for $\hat\rho(z,x)$ in (\ref{eq:XNN-rho}),
we assume $z\leq \tfrac{1}{2}$. The results for $z\geq \frac{1}{2}$ again follow from the symmetry of the quiver under $z\rightarrow 1-z$.
The required integral is
\begin{align}\label{eq:X-y}
 y\stackrel{!}{=}\int_{b(z,y)}^{\infty} dx\, \hat\rho_s(z,x)
 &=\frac{1}{4}-\frac{1}{4 \pi i z}\sinh^{-1} \left(\cosh (2\pi b(z,y)+i\pi z)\right)+\rm{c.c.}
\end{align}
and solving for $b$ leads to 
\begin{align}\label{eq:b-XNN}
 b(z,y)&=\frac{1}{2 \pi }\sinh ^{-1}\left(\sin (\pi z (1-2 y)) \sqrt{\frac{\cos ^2(\pi z (1-2 y) )+\cos ^2(\pi  z)}{\cos ^2(\pi z (1-2 y) )-\cos
   ^2(\pi  z)}}\right)~.
\end{align}
The Wilson loop expectation value can be obtained from (\ref{eq:W-wedge-k-3}), which leads to 
\begin{align}\label{eq:W-wedge-XNN}
 \ln\big\langle W_{\wedge}(z,\mathds{k})\big\rangle&=
 \frac{6}{\pi}N^2\left[ D(u)+D(1+u) \right],
&
u&=e^{ 2\pi i \mathds{k}z - \sinh^{-1}\left(\sin(2\pi\mathds{k}z)\cot(2\pi(1-\mathds{k})z)\right) }\,,
\end{align}
with $b$ as given in (\ref{eq:b-XNN}).
This expression is valid for $z\leq \tfrac{1}{2}$, and extends to $z\in[0,1]$ by symmetry of the quiver under $z\rightarrow 1-z$.
The maximum occurs at the central node for $z=\mathds{k}=\tfrac{1}{2}$, 
\begin{align}
 \ln\big\langle W_{\wedge}\big(\tfrac{1}{2},\tfrac{1}{2}\big)\big\rangle=\frac{12}{\pi}CN^2~,
\end{align}
where $C$ is Catalan's constant. This differs from the maximal expectation value for the $+_{N,M}$ theory in (\ref{eq:plus-max}) with $M=N$ by a factor $2$, 
which will be discussed further in sec.~\ref{sec:AdSCFT}.

At the central node the fundamental and antifundamental Wilson loop expectation values both scale linearly with $N$.
They have identical expectation values due to charge conjugation symmetry of the quiver. 
From (\ref{eq:W-f}) with (\ref{eq:XNN-rho}) and the substitution $x=(2\pi)^{-1}\ln(t+3)$,
\begin{align}\label{eq:XNN-fund-loc}
\big\langle W_{f}\big(\tfrac{1}{2}\big)\big\rangle=
 \big\langle W_{\bar f}\big(\tfrac{1}{2}\big)\big\rangle
 &=\int_{-2\sqrt{2}}^{2\sqrt{2}}dt\, \frac{(t+4)(t+3)^{\frac{3L}{2}-1}}{2\pi\sqrt{8-t^2}}
 \approx
 \big(1+\sqrt{2}\big)^{6N}~.
\end{align}
The integral can be evaluated explicitly in terms of hypergeometric functions. 
The leading-order behavior is given by the last expression.
The remaining (anti)fundamental Wilson loops have enhanced scaling since the eigenvalue distributions have unbounded support,
and can be obtained from (\ref{eq:W-wedge-XNN}) using (\ref{eq:W-wedge-f}).
This leads to
\begin{align}\label{eq:Wf-X-int}
&&
 \ln \langle W_f(z)\rangle&=\ln \langle W_{\bar f}(z)\rangle=3N \ln N~, & z&\neq \tfrac{1}{2}~.
\end{align}
As in the previous examples the scaling is enhanced logarithmically and uniformly for the nodes where the saddle point eigenvalue distribution has unbounded support.

\section{Line defects in 5-brane webs}\label{sec:brane-web-Wilson}

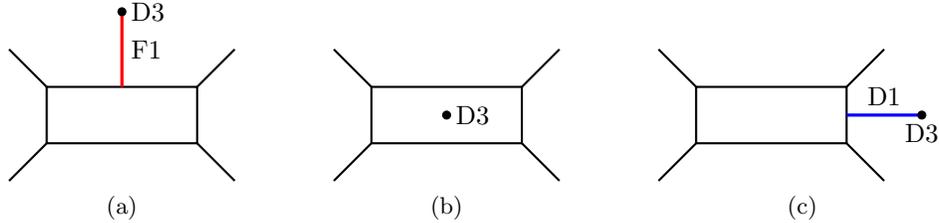
\begin{figure}
 \subfigure[][]{\label{fig:SU2-F1}
   \begin{tikzpicture}
     \draw[thick] (0,0) -- (2,0) -- (2,0.75) -- (0,0.75) -- (0,0);
     \draw[thick] (0,0) -- (-0.5,-0.5);
     \draw[thick] (2,0) -- (2.5,-0.5);
     \draw[thick] (2,0.75) -- (2.5,1.25);
     \draw[thick] (-0.5,1.25) -- (0,0.75);
     \draw[very thick, red] (1,0.75) -- (1,1.75);
     \node [anchor=west] at (1,1.25) {\small F1};
     \draw [fill=black] (1,1.75) circle (1.5pt);
     \node [anchor=west] at (1,1.75) {\small D3};
   \end{tikzpicture}
 }\qquad\quad
 \subfigure[][]{\label{fig:SU2-D3}
    \begin{tikzpicture}
     \draw[thick] (0,0) -- (2,0) -- (2,0.75) -- (0,0.75) -- (0,0);
     \draw[thick] (0,0) -- (-0.5,-0.5);
     \draw[thick] (2,0) -- (2.5,-0.5);
     \draw[thick] (2,0.75) -- (2.5,1.25);
     \draw[thick] (-0.5,1.25) -- (0,0.75);
     \draw [fill=black] (1,0.375) circle (1.5pt);
     \node [anchor=west] at (1,0.375) {\small D3};
   \end{tikzpicture}
 }\qquad\quad
 \subfigure[][]{\label{fig:SU2-D1}
    \begin{tikzpicture}
     \draw[thick] (0,0) -- (2,0) -- (2,0.75) -- (0,0.75) -- (0,0);
     \draw[thick] (0,0) -- (-0.5,-0.5);
     \draw[thick] (2,0) -- (2.5,-0.5);
     \draw[thick] (2,0.75) -- (2.5,1.25);
     \draw[thick] (-0.5,1.25) -- (0,0.75);
     \draw[very thick, blue] (2,0.375) -- (3,0.375);
     \node [anchor=south] at (2.5,0.375) {\small D1};
     \draw [fill=black] (3,0.375) circle (1.5pt);
     \node [anchor=north] at (3,0.375) {\small D3};
   \end{tikzpicture}
 }
 \caption{5-brane web for an $SU(2)$ gauge theory, with a fundamental Wilson loop realized by a fundamental string on the left. In the center a configuration with a D3-brane in the closed face of the web. On the right a line operator realized by a D1-brane.
 The configurations are related by Hanany-Witten transitions.\label{fig:SU2}}
\end{figure}

The gauge theories discussed in the previous section can be engineered by $(p,q)$ 5-brane webs in Type IIB string theory \cite{Aharony:1997bh}.
This allows to  identify their holographic duals, to be discussed in the next section, and will guide the identification of Wilson loops in the supergravity duals.
Line defects can be realized in various ways by adding branes and strings to 5-brane webs.
In \cite{Assel:2018rcw} (see also \cite{Kim:2016qqs}) line defects realized by D3-branes in 5-brane webs were studied and used to extract Wilson loop expectation values. 
A basic picture will be sufficient here, to guide the identification of Wilson loops in the holographic duals for the theories of sec.~\ref{sec:Wilson-loc}.

An $SU(2)$ gauge theory, realized on a pair of D5-branes suspended between NS5-branes, is shown in fig.~\ref{fig:SU2}. 
Line defects preserving the same symmetries as the $\tfrac{1}{2}$-BPS Wilson loops discussed in the previous section can be realized by adding $(p,q)$ strings and D3-branes with the following orientations
\begin{center}
\def\arraystretch{0.6}
\begin{tabular}{cccccc|cc|ccc}
\toprule
& \ 0 \ & \ 1 \ & \ 2 \ & \ 3 \ & \ 4 \ & \ 5 \ & \ 6 \ & \ 7 \ & \ 8 \ & \ 9 \ \\
\hline
 D5 & $\times$ & $\times$ & $\times$ & $\times$ & $\times$ & $\times$ & \\
 NS5 & $\times$ & $\times$ & $\times$ & $\times$ & $\times$ &  & $\times$ \\
\hline
 F1 & $\times$ & & & & & & $\times$ \\
 D1 & $\times$ & & & & & $\times$ & \\
 D3 & $\times$ & & & & & & & $\times$ & $\times$ & $\times$\\
\bottomrule
\end{tabular}
\end{center}
Generic $(p,q)$ 5-branes are at angles in the 5-6 plane reflecting their D5 and NS5 charges, and $(p,q)$ strings are similarly at angles reflecting their F1 and D1 charges.
Generic $(p,q)$ strings can end on $(p,q)$ 5-branes, to which they are perpendicular \cite{Aharony:1996xr,Aharony:1997bh}.

A Wilson loop in the fundamental representation is realized by a fundamental string extending from the D5-branes in the vertical direction.
The string may be terminated on a D3-brane, which wraps the directions orthogonal to the D5 and NS5 branes, fig.~\ref{fig:SU2-F1}. 
The D3-brane is in a Hanany-Witten orientation with respect to the D5 and the NS5 branes:
upon crossing a D5/NS5 brane a fundamental string/D-string is eliminated or created.
The D3-brane on which the fundamental string ends in fig.~\ref{fig:SU2-F1} may be moved into the closed face of the web,
which eliminates the fundamental string and leads to the picture in fig.~\ref{fig:SU2-D3}.
D-strings can be connected to the brane web in a way similar to the F1, fig.~\ref{fig:SU2-D1}, and describe line operators related to instanton particles.
For the $SU(2)$ theory these operators are related to Wilson loops by the Hanany-Witten transitions in fig.~\ref{fig:SU2}, and by S-duality, which 
corresponds to a ninety-degree rotation of the brane web.

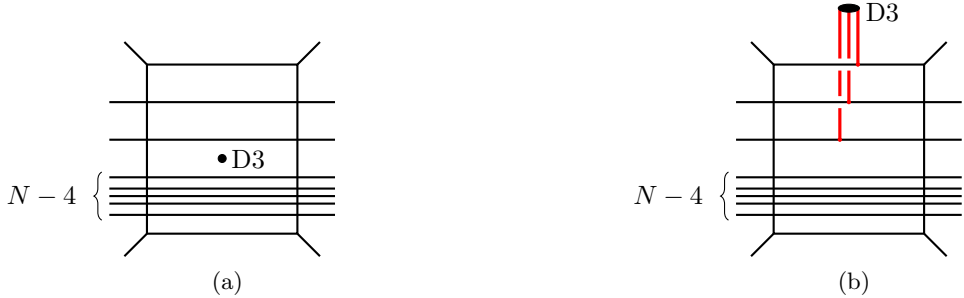
\begin{figure}
 \subfigure[][]{\label{fig:SUN-web-1}
   \begin{tikzpicture}
     \draw[thick] (-1,-1) rectangle (1,1.25);
     
     \draw[thick] (1,1.25) -- +(0.3,0.3);     
     \draw[thick] (-1,1.25) -- +(-0.3,0.3);
     \draw[thick] (1,-1) -- +(0.3,-0.3);     
     \draw[thick] (-1,-1) -- +(-0.3,-0.3);
     
     \foreach \i in {-0.75,-0.25,0.25,0.75}{
       \draw[thick] (-1.5,\i) -- (1.5,\i);
     }
     
     \foreach \i in {1,...,3}{
      \draw[thick] (-1.5,-0.3-0.1*\i) -- +(3,0);
     }
     
     \draw [decorate,decoration={brace,amplitude=3pt}] (-1.6,-0.82) -- (-1.6,-0.18) node [black,midway,xshift=-0.8cm] {\small $N-4$};
     \node at (3,0) {};
     
     \draw [fill=black] (0,0) circle (1.5pt);
     \node [anchor=west] at (0,0) {\small D3};
   \end{tikzpicture}
 }\hskip 20mm
 \subfigure[][]{\label{fig:SUN-web-2}
   \begin{tikzpicture}
\draw[thick] (-1,-1) rectangle (1,1.25);
     
     \draw[thick] (1,1.25) -- +(0.3,0.3);     
     \draw[thick] (-1,1.25) -- +(-0.3,0.3);
     \draw[thick] (1,-1) -- +(0.3,-0.3);     
     \draw[thick] (-1,-1) -- +(-0.3,-0.3);
     
     \foreach \i in {-0.75,-0.25,0.25,0.75}{
       \draw[thick] (-1.5,\i) -- (1.5,\i);
     }
     
     \foreach \i in {1,...,3}{
      \draw[thick] (-1.5,-0.3-0.1*\i) -- +(3,0);
     }
     
     \draw [decorate,decoration={brace,amplitude=3pt}] (-1.6,-0.82) -- (-1.6,-0.18) node [black,midway,xshift=-0.8cm] {\small $N-4$};
     \node at (3,0) {};
     
     \draw[fill=red,red] (-0.12,0.25) circle (0.6pt);
     \draw[very thick,red] (-0.12,0.25) -- (-0.12,0.67);
     \draw[very thick,red] (-0.12,0.83) -- (-0.12,1.17);
     \draw[very thick,red] (-0.12,1.33) -- (-0.12,2);
     
     \draw[fill=red,red] (0,0.75) circle (0.6pt);
     \draw[very thick,red] (0,0.75) -- (0,1.17);
     \draw[very thick,red] (0,1.33) -- (0,2);
     
     \draw[fill=red,red] (0.12,1.25) circle (0.6pt);
     \draw[very thick,red] (0.12,1.25) -- (0.12,2);
     
     \draw [fill=black] (0,2) ellipse (4.0pt and 1.5pt);
     \node [anchor=west] at (0.1,1.95) {\small D3};
   \end{tikzpicture} }
 \caption{Left: $SU(N)$ brane web with a D3-brane in a face separated by $k=3$ D5-branes from the asymptotic region in the upward direction and separated by $N-k$ D5-branes from the asymptotic region in the downward direction. Right: configuration with $k=3$ fundamental strings resulting from Hanany-Witten transitions moving the D3-brane out in the upward direction. Avoided intersections are shown as broken lines.\label{fig:SUN-web}}
\end{figure}

For larger rank gauge groups and brane webs realizing quiver gauge theories a relation between D3-branes and Wilson loop operators can be found along similar lines. Fig.~\ref{fig:SUN-web} shows a representative brane web for an $SU(N)$ gauge theory (or the part representing an $SU(N)$ gauge node in a larger brane web realizing a quiver theory).
A D3-brane can be placed in any face of the web.
Suppose the D3-brane is separated from the asymptotic region along the vertical axis by $k$ D5-branes. 
Moving it out of the web through Hanany-Witten transitions creates a fundamental string for each D5-brane crossed, leading to a total of $k$ fundamental strings, fig.~\ref{fig:SUN-web-2}.
Since the strings end on the same D3-brane, they are constrained by the $s$-rule \cite{Hanany:1996ie}, which limits the number of strings between the D3 and each D5 to at most one. 
This leads to the avoided intersections shown as broken lines in fig.~\ref{fig:SUN-web-2}.
The configuration with $k$ fundamental strings describes a Wilson loop in a $k$-fold tensor product of the fundamental representation. Due to the $s$-rule forcing the strings to end on distinct D5-branes it is the $k$-fold antisymmetric representation $\wedge^k$.
Moving the D3-brane out of the 5-brane web in the opposite direction leads, upon taking into account the orientation of the strings that are created, to the $\overline{\wedge^{N-k}}$ representation.

Labeling the faces of the web in the vertical direction by $0,\ldots N$, the resulting vertical ``coordinate'' of the D3-brane in the web thus encodes the representation of the Wilson loop that it represents.
For quiver theories with multiple gauge nodes one can similarly introduce a discrete coordinate in the horizontal direction of the brane web (see e.g.~fig.~\ref{fig:plus}).
The horizontal coordinate then encodes the gauge node the Wilson loop is associated with.

The UV fixed points of gauge theories realized by 5-brane webs are obtained in the limit where all length scales in the brane web vanish, leading to a junction of $(p,q)$ 5-branes at a point.
These junctions are described by the supergravity solutions discussed in the next section.
The positions of strings and branes in the 5-brane web remain meaningful in the conformal limit, and this will be used to identify the Wilson loops represented by strings and branes in the supergravity solutions.

Dualities between gauge theories can be understood from their brane web realization as Type IIB $SL(2,\ZZ)$ dualities transforming the charges of $(p,q)$ 5-branes and $(p,q)$ strings.
D3-branes are mapped to D3-branes by $SL(2,\ZZ)$. If the transformation of a brane web realizing a gauge theory leads again to a brane web with a gauge theory description, the Wilson loops realized by D3-branes that are mapped into each other are related by the duality.
Strings attached to the brane web are mapped to appropriately $SL(2,\ZZ)$-transformed strings, which will be used to relate Wilson loops realized by fundamental strings to loop operators realized by more general $(p,q)$ strings.

\section{Wilson loops in \texorpdfstring{$AdS_6\times S^2\times\Sigma$}{AdS6xS2xSigma}}
\label{sec:AdSCFT}

Type IIB supergravity solutions describing $(p,q)$ 5-brane junctions were constructed in \cite{DHoker:2016ysh,DHoker:2017mds,DHoker:2017zwj,Uhlemann:2019lge}.
The geometry is a warped product of $AdS_6$ and $S^2$ over a Riemann surface $\Sigma$, which is a disc or equivalently the upper half plane.
The $S^2$ collapses at the boundary of the disc, to close off the space smoothly. 
Each solution is defined by a pair of locally holomorphic functions $\cA_\pm$ on $\Sigma$.
The differentials of $\cA_\pm$ have poles at isolated points on the boundary of $\Sigma$, at which $(p,q)$ 5-branes emerge with charges given by the residues.
This allows to identify the associated 5-brane junction.
In this section Wilson loops will be identified as probe strings and branes embedded into these solutions.

The general expressions for the Einstein-frame metric, complex two-form $C_{(2)}$, and axion-dilaton scalar $B=(1+i\tau)/(1-i\tau)$ in terms of $\cA_\pm$ are as follows,
\begin{align}\label{eqn:ansatz}
 ds^2 &= f_6^2 \, ds^2 _{\mathrm{AdS}_6} + f_2^2 \, ds^2 _{\mathrm{S}^2} 
 + 4\rho^2\, |dw|^2~,
 &
 C_{(2)}&=\cC \vol_{S^2}~,
 \nonumber\\
 B &=\frac{\partial_w \cA_+ \,  \partial_{\bar w} \cG - R \, \partial_{\bar w} \bar \cA_-   \partial_w \cG}{
R \, \partial_{\bar w}  \bar \cA_+ \partial_w \cG - \partial_w \cA_- \partial_{\bar w}  \cG}~,
\end{align}
where $w$ is a complex coordinate on $\Sigma$,
$\vol_{S^2}$ and $ds^2_{S^2}$ are the volume form and line element for unit-radius $S^2$, 
and $ds^2_{AdS_6}$ is the line element of unit-radius $AdS_6$.
The metric functions are
\begin{align}\label{eq:metric-functions}
 f_6^2&=\sqrt{6\cG T}~, & f_2^2&=\frac{1}{9}\sqrt{6\cG}\,T ^{-\tfrac{3}{2}}~, & \rho^2&=\frac{\kappa^2}{\sqrt{6\cG}} T^{\tfrac{1}{2}}~.
\end{align}
The function $\cC$ appearing in the complex two-from field is given by
\begin{align}
\cC&=\frac{2i}{3}\left(
 \frac{\partial_{\bar w}\cG\partial_w\cA_++\partial_w \cG \partial_{\bar w}\bar\cA_-}{3\kappa^{2}T^2} - \bar{\mathcal{A}}_{-} - \mathcal{A}_{+}  \right)~.
\end{align}
The composite quantities $\cG$, $\kappa^2$, $T$ and $R$ in these expressions are defined in terms of $\cA_\pm$ as follows,
\begin{align}\label{eq:kappa2-G}
 \kappa^2&=-|\partial_w \cA_+|^2+|\partial_w \cA_-|^2~,
 &
 \partial_w\cB&=\cA_+\partial_w \cA_- - \cA_-\partial_w\cA_+~,
 \nonumber\\
 \cG&=|\cA_+|^2-|\cA_-|^2+\cB+\bar{\cB}~,
 &
  T^2&=\left(\frac{1+R}{1-R}\right)^2=1+\frac{2|\partial_w\cG|^2}{3\kappa^2 \, \cG }~.
\end{align}
Constructing $\cG$ for given $\cA_\pm$ needs an additional integration, to determine $\cB$.

The general form of the functions $\cA_\pm$ for solutions representing $(p,q)$ 5-brane junctions is reviewed in app.~\ref{sec:cG-gen}, where also a general expression for $\cG$ in terms of Bloch-Wigner functions is derived.
Explicit $\cA_\pm$ realizing holographic duals for the theories of sec.~\ref{sec:Wilson-loc} will be given below.

It will be convenient to choose a form for the $AdS_6$ metric that makes the symmetries preserved by the Wilson loops of interest manifest,
\begin{align}
 ds_{AdS_6}^2&=d\rho^2+\cosh^2\!\rho\,ds^2_{AdS_2}+\sinh^2\!\rho\,ds^2_{S^3}~.
\end{align}
To realize the defect conformal symmetry, probe branes or strings have to wrap $AdS_2$.
To realize the $SO(4)$ isometries they may either wrap the entire $S^3$ or be localized at $\rho=0$.
Finally, to realize the $SU(2)$ R-symmetry they have to either wrap the entire $S^2$ or be localized on the boundary of $\Sigma$, where the $S^2$ collapses.
$(p,q)$ strings can thus be located on $\partial\Sigma$, wrapping $AdS_2$ in $AdS_6$, and D3-branes can be in the interior of $\Sigma$, wrapping $AdS_2$ and the $S^2$.

In sec.~\ref{sec:pq-string} and \ref{sec:D3-branes} $\tfrac{1}{2}$-BPS embeddings of $(p,q)$ strings and D3-branes are discussed.
In sec.~\ref{sec:plus-hol} -- \ref{sec:X-sol} Wilson loops are discussed in the holographic duals for the theories of sec.~\ref{sec:Wilson-loc}.

\subsection{\texorpdfstring{$(p,q)$}{(p,q)} strings}\label{sec:pq-string}

We start with $(p,q)$ strings wrapping $AdS_2$ in $AdS_6$.
The action, with the dilaton convention $\tau=\chi+i e^{-2\phi}$ and with $(p,q)=(1,0)$ corresponding to F1 and $(0,1)$ to D1, is \cite{Schwarz:1995dk,Bergshoeff:2006gs}
\begin{align}
 S_{(p,q)}&=-T \int d^2\xi \sqrt{e^{2\phi}(p-\chi q)^2+e^{-2\phi}q^2}\sqrt{-\det(g)}+T \int \left(p B_2+q C_{(2)}^{\rm RR}\right)~,
\end{align}
where $T=(2\pi\alpha^\prime)^{-1}$ and $g$ is the induced Einstein-frame metric.
The pullback of $B_2$ and $C_{(2)}^{\rm RR}$ to the string worldvolume vanishes.
The induced Einstein-frame metric is
\begin{align}
 g&=f_6^2\,ds^2_{AdS_2}~.
\end{align}
The $(p,q)$ string action thus reduces to
\begin{align}\label{eq:S-pq-gen}
 S_{(p,q)}&=-T \Vol_{AdS_2} f_6^2\sqrt{e^{2\phi}(p-\chi q)^2+e^{-2\phi}q^2}~,
\end{align}
with $\Vol_{AdS_2}$ the renormalized volume of $AdS_2$. 
For global AdS$_2$, appropriate for the description of circular Wilson loops, 
the renormalized volume is given by
\begin{align}
 {\rm Vol}_{AdS_2}=-2\pi~.
\end{align}
For Poincar\'e $AdS_2$, describing a straight Wilson line, the renormalized volume vanishes.

The condition for a $(p,q)$ string of this form to preserve half the supersymmetries is derived in app.~\ref{sec:F1-BPS}.
The analysis shows that the string has to be located either at a pole or at a point where
\begin{align}\label{eq:BPS-pq-string}
 \left(p+iq\right)\partial_w\cA_+&=\left(p-iq\right)\partial_w\cA_-~.
\end{align}
This condition implies $\kappa^2=0$, which for regular solutions only occurs on the boundary of $\Sigma$.
$\tfrac{1}{2}$-BPS string embeddings are thus restricted to the boundary of $\Sigma$, as expected to preserve the $SU(2)$ R-symmetry.\footnote{Some care is needed in treating the point at infinity on the upper half plane, where the differentials fall off. Spurious solutions can be avoided by demanding the ratio of the differentials to approach the value dictated by (\ref{eq:BPS-pq-string}).}
The action (\ref{eq:S-pq-gen}) is evaluated for such embeddings in app.~\ref{sec:F1-BPS}, which leads to
\begin{align}\label{eq:Spq-dSigma}
 S_{(p,q)}&=-T\Vol_{AdS_2}\left|(p+iq)(\cA_+-\bar \cA_-)+(p-iq)(\bar\cA_+-\cA_-)\right|~.
\end{align}

For solutions with $K$ poles in the differentials and no monodromy, the condition (\ref{eq:BPS-pq-string}) leads to a polynomial of degree $K-2$
(this can be seen explicitly from the product form of $\partial_w\cA_\pm$ in (3.7) of \cite{DHoker:2017mds}).
For supergravity solutions with $K$ poles, describing 5-brane junctions of $K$ (stacks of) 5-branes, there can thus be $K-2$ supersymmetric $(p,q)$-string embeddings for each $(p,q)$ at generic points of $\partial\Sigma$.
At the poles of $\partial_w\cA_\pm$, the $(p,q)$ string action (\ref{eq:Spq-dSigma}) diverges unless the pole corresponds to 5-branes of type $(p,q)$ (this can be seen e.g.\ from (\ref{eq:5-brane-pq})).

\subsection{D3-branes}\label{sec:D3-branes}

D3-brane embeddings that preserve the desired symmetries wrap $AdS_2$ in $AdS_6$ and the $S^2$ realizing the R-symmetry,
and are localized in the interior of $\Sigma$.
The action, with $T_{\rm Dp}^{-1}=(2\pi)^{p}\sqrt{\alpha^\prime}^{\:p+1}$ and the dilaton convention $\tau=\chi+ie^{-2\phi}$, reads \cite{Bergshoeff:1996tu,Simon:2011rw}
\begin{align}\label{eq:S-D3-gen}
 S_{\rm D3}&=-T_{\rm D3} \int d^4\xi e^{-2\phi}\sqrt{-\det\left(\tilde g_{ab}+\cF_{ab}\right)}
 +T_{\rm D3}\int e^\cF\wedge\sum_q C^{\rm RR}_{(q)}~,
\end{align}
where $\cF=F-B_2$ and  $\tilde g=e^\phi g$ is the induced string-frame metric.  $2\pi\alpha^\prime$ has been set to one and we will do so from now on. 
The induced Einstein-frame metric is given by
\begin{align}
 g&=f_6^2ds^2_{AdS_2}+f_2^2ds^2_{S^2}~.
\end{align}
The general worldvolume gauge field compatible with the desired symmetries takes the form
\begin{align}\label{eq:D3-F}
 F&=\mathfrak{f}_{\rm el} \vol_{AdS_2}+\mathfrak{f}_{\rm m}\vol_{S^2}~,
\end{align}
where $\vol_{AdS_2}$ and $\vol_{S^2}$ are the canonical volume forms on unit-radius $AdS_2$ and $S^2$, respectively.
With the identification $C_{(2)}=B_2+i C_{(2)}^{\rm RR}$, the D3-brane action then becomes
\begin{align}\label{eq:S-D3-1}
 S_{\rm D3}&=T_{\rm D3}\Vol_{AdS_2}\Vol_{S^2}\left[
 -L_{\rm DBI} +\mathfrak{f}_{\rm el} \big(\Im(\cC)+\chi (\mathfrak{f}_{\rm m}-\Re(\cC))\big)
 \right]~,
 \nonumber\\
 L_{\rm DBI}&=\sqrt{\left(f_6^4-e^{-2\phi}\mathfrak{f}_{\rm el}^2\right)\left(f_2^4+e^{-2\phi}\left(\mathfrak{f}_{\rm m}-\Re(\cC)\right)^2\right)}~.
\end{align}

The conditions for the D3-brane to preserve half the supersymmetries are discussed in app.~\ref{sec:D3-susy}.
They fix the worldvolume fluxes in terms of the position of the D3-brane.
The result is
\begin{align}\label{eq:D3-fluxes}
 \mathfrak{f}_{\rm el}&=-2\nu f_6 e^\phi \Re\left(e^{-i\theta_\kappa}\bar\alpha\beta\right)~, &
 \mathfrak{f}_{\rm m}-\Re(\cC)&=\frac{2}{3}\nu f_2e^\phi \Im\left(e^{-i\theta_\kappa}\bar\alpha\beta\right)~,
\end{align}
where
$\alpha$, $\beta$ are Killing spinor components given in terms of $\cA_\pm$ and $B$ in (\ref{eq:alphabeta-def}) and $\theta_\kappa$ is defined in terms of $B$ in (\ref{eq:theta-k}).
The expressions will not be needed here.
The $\tfrac{1}{2}$-BPS requirement does not constrain the location of the D3-brane on $\Sigma$, so this is a $2_\RR$-parameter family of solutions.

The D3-brane carries F1 charge determined by the electric components of the gauge field, $\mathfrak{f}_{\rm el}$, and D1 charge determined by the magnetic components, $\mathfrak{f}_{\rm m}$.
The flux on $S^2$ is subject to the quantization condition
\begin{align}\label{eq:ND1}
 N_{\rm D1}&=\frac{1}{2\pi}\int_{S^2} F=2\left[\Re(\cC)+\frac{2}{3}\nu f_2e^\phi \Im\left(e^{-i\theta_\kappa}\bar\alpha\beta\right)\right]~,
\end{align}
which determines the D1 charge.
The conserved charge density associated with the electric components of the gauge field, integrated over the $S^2$, determines the F1 charge and is given by
\begin{align}\label{eq:NF1}
 N_{\rm F1}&=-\int_{S^2}\frac{\delta S_{\rm D3}}{\delta \mathfrak{f}_{\rm el}}
 \nonumber\\
 &=
 -T_{\rm D3}\Vol_{S^2}\left[\frac{f_2^4+e^{-2\phi}\left(\mathfrak{f}_{\rm m}-\Re(\cC)\right)^2}{L_{\rm DBI}}\,e^{-2\phi} \mathfrak{f}_{\rm el}+\Im(\cC)+\chi(\mathfrak{f}_{\rm m}-\Re(\cC))\right].
\end{align}
The expressions are evaluated explicitly in app.~\ref{sec:D3-action-charge}, which leads to the remarkably simple result
\begin{align}\label{eq:NF1-ND1-cApm}
  N_{\rm F1}+i N_{\rm D1}&=\frac{4}{3}\left(\cA_+ +\bar \cA_-\right)~.
\end{align}
The real and imaginary parts of $\cA_++\bar\cA_-$ are harmonic functions, so $N_{\rm F1}$ and $N_{\rm D1}$ take their maxima and minima on the boundary of $\Sigma$.\footnote{In the M-theory perspective on the $AdS_6$ solutions discussed in \cite{Kaidi:2018zkx}, the combinations of $\cA_\pm$ appearing in $N_{\rm F1}$ and $N_{\rm D1}$ correspond to the coordinates on the M-theory torus.}

The expectation value of the Wilson loop represented by the D3-brane is given by a Legendre transform of $S_{\rm D3}$ \cite{Drukker:2005kx,Drukker:2006zk}.
The explicit expressions are again evaluated in app.~\ref{sec:D3-action-charge}, and lead to 
\begin{align}\label{eq:Whol-G}
\ln\langle W_\wedge\rangle
&=
 S_{\rm D3} - \mathfrak{f}_{\rm el} \frac{\delta S_{\rm D3}}{\delta \mathfrak{f}_{\rm el}}
 =
 -\frac{2}{3}T_{\rm D3} {\rm Vol}_{AdS_2}{\rm Vol}_{S^2} \,\cG~,
\end{align}
where $\cG$ is defined in (\ref{eq:kappa2-G}) and $T_{\rm D3} {\rm Vol}_{AdS_2}{\rm Vol}_{S^2}=-4\pi/(2\pi \alpha^\prime)^2$.
The function $\cG$ that featured prominently in the construction of the supergravity solutions thus directly encodes the (Legendre-transformed) on-shell action for supersymmetric D3-branes embedded into the $AdS_6$ solutions.

\subsection{\texorpdfstring{$+_{N,M}$}{+[N,M]} solution}\label{sec:plus-hol}

\begin{figure}
\subfigure[][]{\label{fig:plus}
\begin{tikzpicture}[scale=0.6]
 \foreach \i in {-1.5,-1.0,-0.5,0,0.5,1,1.5}{
 \draw[thick] (\i,-3) -- (\i,3);
 }
 \foreach \j in {-0.75,-0.25,0.25,0.75}{
 \draw[thick] (3.3,\j) -- (-3.3,\j);
 }
 
%

 \node at (0,-3.5) {$M$\,NS5};
 \node at (-4.2,0) {$N$\,D5};

\end{tikzpicture}
}\hskip 4mm
\subfigure[][]{\label{fig:plus-disc}
 \begin{tikzpicture}
  \node at (0,0) {\includegraphics[width=3.8cm]{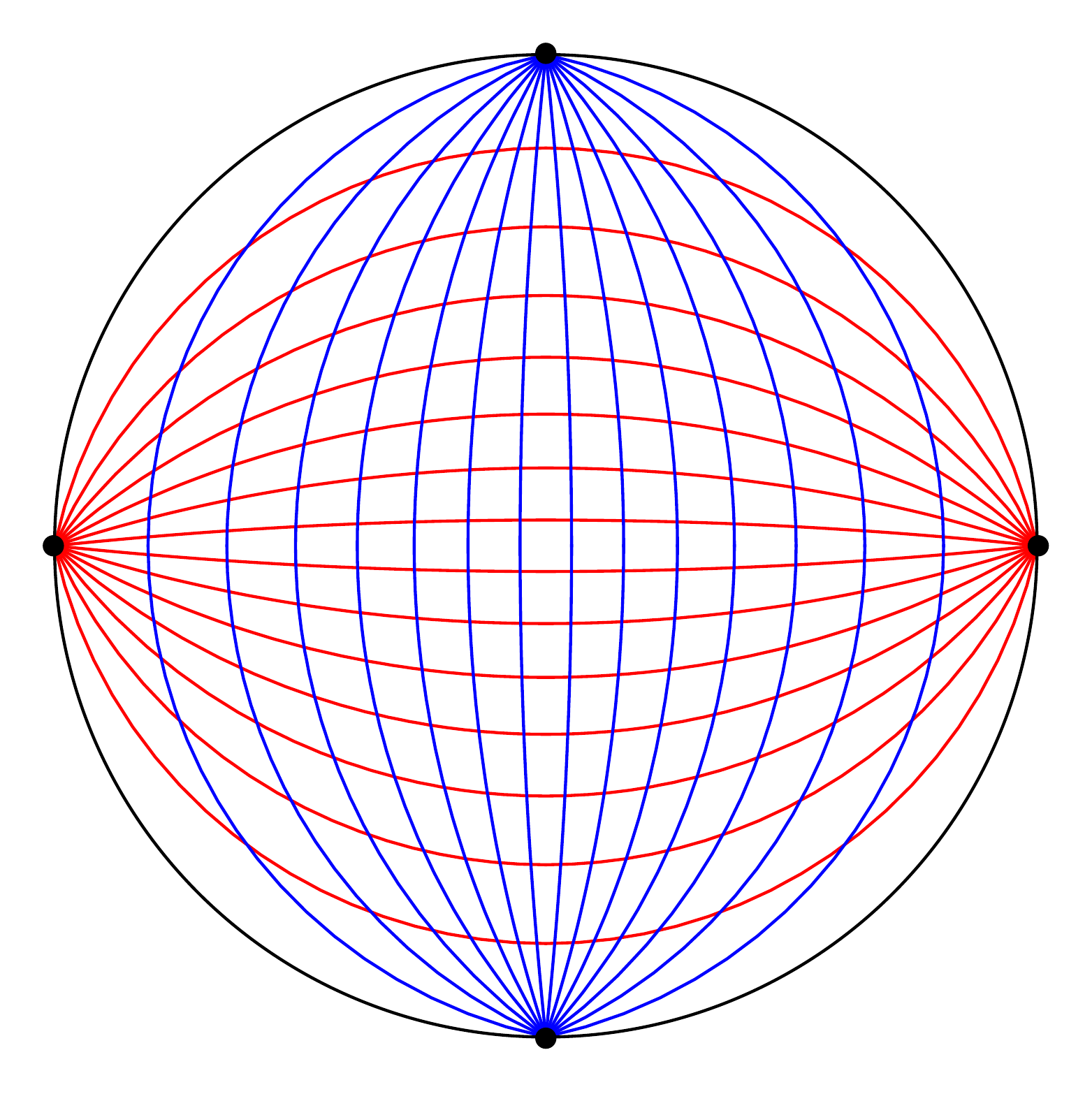}};
  \draw[thick] (0,0) circle (1.7);
  

  \draw[very thick] (0,1.7) -- (0,1.85);
\draw[very thick] (0,-1.7) -- (0,-1.85);
\draw[very thick] (1.7,0) -- (1.85,0);
\draw[very thick] (-1.7,0) -- (-1.85,0);

\node at (0,2.0) {NS5};
\node at (0,-2.0) {NS5};
\node at (2.1,0) {D5};
\node at (-2.1,0) {D5};
\draw[thick,fill=red] (1.7,0) circle (0.08);
\node at (1.3,0) {\small \bf F1};

\draw[thick,fill=red] (-1.7,0) circle (0.08);
\node at (-1.3,0) {\small \bf F1};

\draw[thick,fill=blue] ({sin(135)*1.7},{cos(135)*1.7}) circle (0.08);
\node at ({sin(135)*2.1},{cos(135)*2.1}) {\scriptsize $(1,1)$};
\draw[thick,fill=blue] ({sin(45)*1.7},{cos(45)*1.7}) circle (0.08);
\node at ({sin(45)*2.12},{cos(45)*2.12}) {\scriptsize $(1,-1)$};
\draw[thick,fill=blue] ({sin(-45)*1.7},{cos(-45)*1.7}) circle (0.08);
\draw[thick,fill=blue] ({sin(-135)*1.7},{cos(-135)*1.7}) circle (0.08);

 \end{tikzpicture}
 }\hskip 4mm
\subfigure[][]{\label{fig:plus-cont}
 \begin{tikzpicture}
  \node at (0,0) {\includegraphics[width=0.24\linewidth]{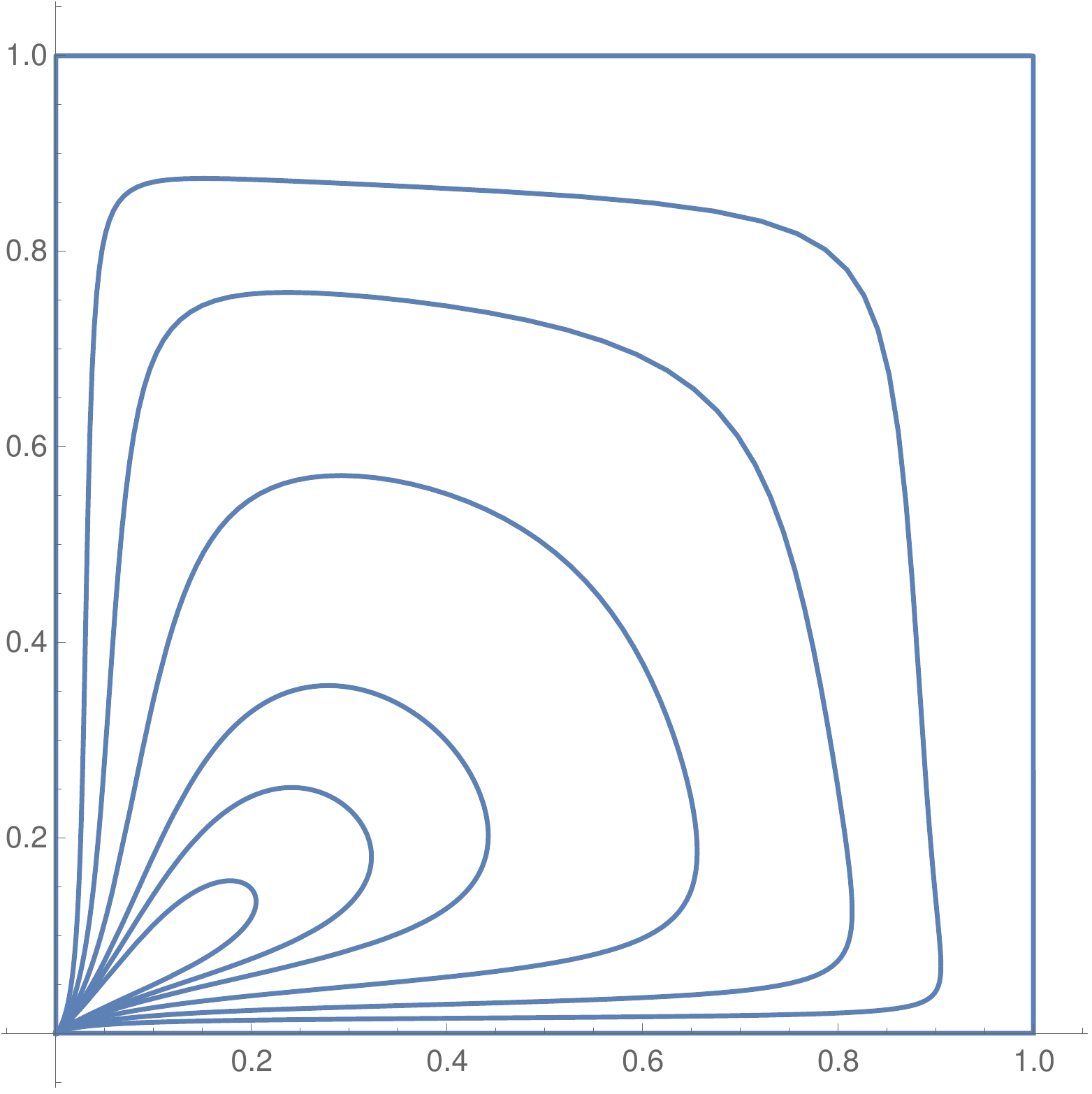}};
  \node[rotate=90] at (-2.15,0) {\scriptsize $N_{\rm F1}/N$};
  \node at (0,-2.1) {\scriptsize $N_{\rm D1}/M$};
 \end{tikzpicture}
}
\caption{Left: brane web realizing the gauge theory (\ref{eq:D5NS5-quiver}). 
Center: $+_{N,M}$ supergravity solution on the disc, with the 5-brane poles and some of the string embeddings. 
Along the red/blue curves the D3-brane has constant F1/D1 charge.
Right: $(N_{\rm D1},N_{\rm F1})$ curves, from outer to inner for $\Im(w)\in\lbrace 10^{-6},\tfrac{1}{20},\tfrac{1}{10},\tfrac{1}{5},\tfrac{2}{5},\tfrac{3}{5},1\rbrace$.
The $\Im(w)=10^{-6}$ curve traces the boundary of the disc in fig.~\ref{fig:plus-disc}.
Each segment of the boundary in fig.~\ref{fig:plus-cont} corresponds to a pole in fig.~\ref{fig:plus-disc}.
The F1/D1 charges naturally identify the vertical/horizontal coordinate of the face in which the  D3-brane is located in the brane web.}
\end{figure}
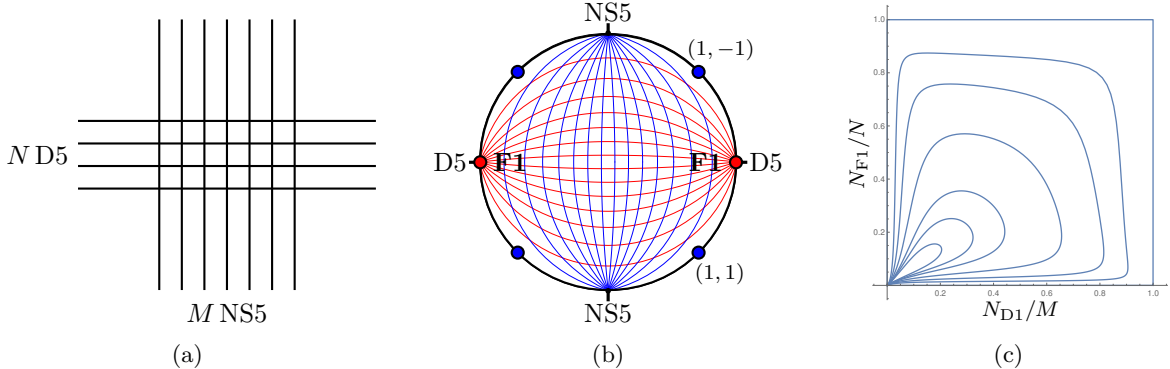

The $+_{N,M}$ theories are realized by intersections of $N$ D5-branes and $M$ NS5-branes, fig.~\ref{fig:plus}.
Functions $\cA_\pm$ for supergravity solutions with the appropriate 5-brane charges were discussed in \cite{Gutperle:2017tjo,Bergman:2018hin}.
Their explicit form on the upper half plane is (sec.~4.2 of \cite{Bergman:2018hin} with $2\pi\alpha^\prime=1$)
\begin{align}\label{eq:cA-plus}
 \cA_\pm&=\frac{3}{8\pi}\left[iN\left(\ln(2w-1)-\ln(w-1)\right)\pm M \left(\ln (3w-2)-\ln w\right)\right]~.
\end{align}
The corresponding $\partial_w\cA_\pm$ have four poles, reflecting the external 5-branes.
The 5-brane charges at a general pole are given by \cite{Bergman:2018hin}
\begin{align}\label{eq:5-brane-pq}
 \Res(\partial_w\cA_\pm)&=\frac{3}{8\pi}(\pm N_{\rm NS5}+i N_{\rm D5})~.
\end{align}
For (\ref{eq:cA-plus}) there are D5-brane poles at $w\in\lbrace \tfrac{1}{2},1\rbrace$ and NS5 poles at $w\in\lbrace 0,\tfrac{2}{3}\rbrace$.
The free energy obtained holographically from this solution was first matched to a field theory computation in \cite{Fluder:2018chf}.
Aspects of strings in these solutions were previously studied in \cite{Kaidi:2017bmd}.
The function $\cG$ defined in (\ref{eq:kappa2-G}), using the expression in (\ref{eq:cG-gen}) and the identities in footnote~\ref{foot:D-rel}, 
can be expressed as
\begin{align}
 \cG&=\frac{9}{8\pi^2}NM\left[D\left(\frac{3w-2}{w}\right)+D\left(\frac{w}{2-3w}\right)\right]~,
\end{align}
where $D$ is the Bloch-Wigner function defined in (\ref{eq:D-def}).

Wilson loops in antisymmetric representations can be realized in 5-brane webs by D3-branes, as discussed in sec.~\ref{sec:brane-web-Wilson}.
The embeddings into the supergravity solutions are discussed in sec.~\ref{sec:D3-branes}.
The expectation values of antisymmetric Wilson loops, computed from the action of D3-branes embedded into this solution at a point $w$ of the upper half plane, via (\ref{eq:Whol-G}), are given by
\begin{align}\label{W-wedge-plus-hol}
 \ln\langle W_{\wedge}\rangle &= \frac{3}{\pi}NM\left[D\left(\frac{3w-2}{w}\right)+D\left(\frac{w}{2-3w}\right)\right]~.
\end{align}
To make this result meaningful the gauge node that the Wilson loop is associated with and the representation have to be identified.
To this end we note that the F1 and D1 charges of the D3-brane embedded at a point $w$ of $\Sigma$ are given, via (\ref{eq:NF1-ND1-cApm}), by
\begin{align}\label{eq:NFND-plus}
 N_{\rm F1}&=\frac{N}{\pi}\Im\ln\left(\frac{w-1}{2w-1}\right)~,
 &
 N_{\rm D1}&=\frac{M}{\pi}\Im\ln\left(\frac{3w-2}{w}\right)~.
\end{align}
$N_{\rm F1}$ is valued in $[0,N]$, and $N_{\rm D1}$ in $[0,M]$.
Curves of constant F1 and D1 charge are shown in fig.~\ref{fig:plus-disc}, for which the upper half plane has been conformally mapped to the unit disc.
Curves of constant $N_{\rm F1}$/$N_{\rm D1}$ connect the D5/NS5 poles.
At the NS5-brane poles $N_{\rm F1}\vert_{w=2/3}=N$ and $N_{\rm F1}\vert_{w=0}=0$,
while at the D5-brane poles $N_{\rm D1}\vert_{w=1}=0$ and $N_{\rm D1}\vert_{w=1/2}=M$.

The space of F1 and D1 charges carved out by the D3-brane embeddings as their position is varied on $\Sigma$ is also shown in fig.~\ref{fig:plus-cont}.
The figure shows the charges along curves with constant $\Im(w)$ on the upper half plane. 
The outer curve in fig.~\ref{fig:plus-cont} corresponds to small $\Im(w)$, and traces the boundary of the upper half plane.
Each boundary segment of the square carved out by the F1 and D1 charges in fig.~\ref{fig:plus-cont} corresponds to a pole in fig.~\ref{fig:plus-disc}: $N_{\rm D1}$ and $N_{\rm F1}$ are both constant along the regular boundary segments between poles, and jump at the poles.
Since $N_{\rm D1}$ and $N_{\rm F1}$ are harmonic functions, they take their maxima and minima on the boundary of $\Sigma$.

Fig.~\ref{fig:plus-cont} shows that the space of F1 and D1 charges carved out by the D3-brane is precisely the grid diagram associated with the brane web in fig.~\ref{fig:plus}: 
The grid diagram is defined on an integer lattice. It is obtained as the dual graph to the brane web, by placing a point in each face of the web (open and closed) such that points in adjacent faces are connected by a line perpendicular to the 5-brane that separates them \cite{Aharony:1997bh}. 
The grid diagram encodes the structure of faces and vertices in the brane web. It is independent of the size of the faces and remains meaningful in the conformal limit where the size of the faces vanishes.
For the $+_{N,M}$ web in fig.~\ref{fig:plus} the grid diagram comprises the points $\lbrace (i,j)\,\vert\, i=0,\ldots, M, \ j=0,\ldots N\rbrace$,
where the origin for the integer coordinates is chosen arbitrarily and can be shifted.
The F1 and D1 charges in fig.~\ref{fig:plus-cont} precisely carve out the points of this grid diagram.

The D1/F1 charge of a D3-brane at a point of $\Sigma$ is thus naturally identified with the horizontal/vertical coordinate of the face in which the D3-brane is located in the $+_{N,M}$ brane web.
Denoting the horizontal and vertical coordinates of the face by $n_x$ and $n_y$, respectively, such that the lower left face corresponds to $n_x=n_y=0$,
the identification is
\begin{align}\label{eq:ND1-NF1-plus-0}
 n_x&=N_{\rm D1}~, & n_y&=N_{\rm F1}~.
\end{align}
With the expressions for $N_{\rm F1}$ and $N_{\rm D1}$ in terms of $w$ in (\ref{eq:NFND-plus}), 
this explicitly associates to each point of $\Sigma$ a face of the $+_{N,M}$ brane web.
Through the grid diagram the field theory data $L$ and $N(z)$ is encoded directly in the supergravity solution: 
$L$ is the difference between the maximal and minimal values of $N_{\rm D1}$, which is $M$,
while $N(z)$ is the difference between the maximal and minimal values of $N_{\rm F1}$ for fixed $N_{\rm D1}=Lz$, which is $N$ for all $z$.

Following the discussion of sec.~\ref{sec:brane-web-Wilson}, the D3-brane with coordinates $n_x$, $n_y$ given by (\ref{eq:ND1-NF1-plus-0}) describes a Wilson loop in the $k$-fold antisymmetric representation of the $t^{\rm th}$ gauge node with $k=N-n_y$ and $t=n_x$.
This identifies the field theory parameters $z= t/L$ and $\mathds{k}= k/N(z)$, introduced in sec.~\ref{sec:Wilson-loc} to label the Wilson loops, for the D3-brane at a given point of $\Sigma$.
For the $+_{N,M}$ theory one obtains
\begin{align}\label{eq:ND1-NF1-plus}
 z&=\frac{N_{\rm D1}}{M}~, &  \mathds{k}&=1-\frac{N_{\rm F1}}{N}~.
\end{align}
With (\ref{eq:NFND-plus}) these relations can be solved for $w$ in terms of $\mathds{k}$ and $z$, which leads to
\begin{align}\label{eq:w-zk-plus}
 \frac{w}{3w-2}&=\exp\left\lbrace-i\pi z+\sinh^{-1}\left(\cot(\pi \mathds{k})\sin(\pi z)\right)\right\rbrace~.
\end{align}
With the identification (\ref{eq:w-zk-plus}) the holographic result (\ref{W-wedge-plus-hol}) becomes
\begin{align}
\ln\langle W_{\wedge}(z,\mathds{k})\rangle &= \frac{3}{\pi}NM\left[D\left(\frac{1}{\bar u}\right)+D\left(-\bar u\right)\right]~,
&
u&=e^{i\pi z+\sinh^{-1}\left(\cot(\pi \mathds{k})\sin(\pi z)\right)}~.
\end{align}
This exactly matches the field theory result for antisymmetric Wilson loops given by (\ref{eq:plus-Wwedge}) with (\ref{eq:plus-b}), noting the identities for the Bloch-Wigner function in footnote \ref{foot:D-rel}.

\smallskip

We now discuss loop operators represented by strings.
For the functions $\cA_\pm$ in (\ref{eq:cA-plus}), the BPS condition for $(p,q)$ strings in (\ref{eq:BPS-pq-string}) becomes
\begin{align}
 2M p (w-1)(2w-1) +N q w(3w-2)&=0~.
\end{align}
This is a quadratic equation with two real solutions for each $(p,q)$. 
The gauge node which the corresponding operators are associated with can be identified from the location of the embedding and the identification of points on $\Sigma$ with faces in the brane web in (\ref{eq:ND1-NF1-plus-0}).
For each $(p,q)$ one solution is associated with each of the boundary gauge nodes at $z=0$ and $z=1$.
With a straightforward generalization of the notation in (\ref{eq:W-z-kk-def}) the expectation values of the corresponding $(p,q)$ loop operators are given by
\begin{align}\label{eq:Wpq-plus}
 \langle W_{(p,q)}(0)\rangle=\langle W_{(p,q)}(1)\rangle&=e^{S_{(p,q)}}~,
\end{align}
where the action via (\ref{eq:Spq-dSigma}) evaluates to
\begin{align}\label{eq:Spq-plus}
 S_{(p,q)}&=
 3\left| N q\right|\, \sinh^{-1}\left|\frac{Mp}{Nq}\right|
 +3\left|M p\right|\, \sinh^{-1}\left| \frac{N q}{M p}\right|~.
\end{align}

In the 5-brane web in fig.~\ref{fig:plus} these $(p,q)$ strings are perpendicular to $(p,q)$ 5-branes, as discussed in sec.~\ref{sec:brane-web-Wilson}.
Consider the boundary segment in fig.~\ref{fig:plus-disc} between the NS5-branes pointing north and the D5-branes pointing east.
As one moves from the NS5 pole to the D5 pole, the string charges vary from $(0,-1)$ close to the NS5 pole, through the $(1,-1)$ string indicated in the figure, to $(1,0)$ at the D5 pole.
The $(1,-1)$ string connects to the brane web diagonally from north east. 
Towards the NS5 pole this turns into a D-string connecting to the brane web horizontally from the right, and towards the D5 pole it turns into a fundamental string connecting vertically to the brane web from above.
The discussion for the remaining boundary segments follows analogously.

For $(p,q)=(1,0)$ the embeddings obtained from (\ref{eq:Wpq-plus}) coincide with the D5-brane poles.
Following the discussion above, this embedding for each pole represents fundamental strings connecting to the brane web from above and from below, depending on which side the pole is approached from.
The fundamental strings at the D5-brane poles thus represent fundamental and anti-fundamental Wilson loops associated with each of the boundary nodes.
The action (\ref{eq:Spq-plus}) vanishes, leading to
\begin{align}\label{eq:Wf-plus-hol}
 \langle W_f(0)\rangle&=\langle W_{\bar f}(0)\rangle=1~,
 &
 \langle W_f(1)\rangle&=\langle W_{\bar f}(1)\rangle=1~.
\end{align}
This agrees with the field result (\ref{eq:Wf-plus}).
The (anti)fundamental Wilson loops associated with interior gauge nodes were found to have enhanced scaling in sec.~\ref{sec:plus-loc}. Their expectation values were determined from antisymmetric Wilson loops for $\mathds{k}=1/N(z)$ and $\mathds{k}=1-1/N(z)$. 
The same relations (\ref{eq:W-wedge-f}) can be used in the holographic representation. 
Since the antisymmetric Wilson loop expectation values obtained from D3-branes were already shown to exactly match the field theory results, this exactly reproduces the remaining (anti)fundamental Wilson loops in (\ref{eq:Wf-plus-int}).

The relations (\ref{eq:W-wedge-f}) have a geometric interpretation in the supergravity duals.
Changing $\mathds{k}$ for the antisymmetric Wilson loop represented by a D3-brane corresponds to changing the F1 charge while holding the D1 charge fixed. 
Small $\mathds{k}$ and $1-\mathds{k}$ each correspond to D3-branes approaching one of the NS5-brane poles in fig.~\ref{fig:plus-disc}, along one of the blue curves along which the D3-brane carries constant D1 charge (see also (\ref{eq:w-zk-plus})).
In fig.~\ref{fig:plus-cont} this corresponds to approaching the upper/lower horizontal boundaries along a vertical curve.
Close to the pole the D3-brane becomes equivalent to a fundamental string (in the brane web it is separated from the asymptotic region by a single D5-brane).
This yields a fundamental string for each curve of constant D1-charge ending at the pole, which is a total of $M$ fundamental strings for each pole and represents the (anti)fundamental Wilson loops at interior nodes.
A fundamental string at a generic 5-brane pole is indeed $\tfrac{1}{2}$-BPS, as discussed in app.~\ref{sec:F1-BPS},
but the action at a pole with NS5 charge is divergent. This is the holographic incarnation of the divergence discussed in sec.~\ref{sec:plus-loc} after (\ref{eq:Wf-plus}) and reflects the enhanced scaling.

More general $(p,q)$ strings can be discussed along similar lines.
D-strings with $(p,q)=(0,1)$ can be embedded at the NS5-brane poles at $w=0$ and $w=2/3$.
They connect to the brane web from the left/right depending on which side the poles are approached from.
They have vanishing action and the corresponding loop operators have expectation value one.
The expectation values (\ref{eq:Wpq-plus}) simplify for loop operators with $Mp=\pm Nq$, leading to
\begin{align}\label{eq:plus-pqstring-ops}
 \langle W_{(p,q)}(0)\rangle \big\vert_{Mp=\pm Nq}&=\langle W_{(p,q)}(1)\rangle \big\vert_{Mp=\pm Nq}=\big(1+\sqrt{2}\big)^{6|Nq|}~.
\end{align}
For $M=N$ and $(p,q)=(1,\pm 1)$ the expectation values agree with the field theory results (\ref{eq:XNN-fund-loc}) for the (anti)fundamental Wilson loops at the central node in the $X_{N}$ theory.
The relation between the $+_{N,N}$ and $X_{N}$ solutions will be discussed further in sec.~\ref{sec:X-sol}.
$(p,q)$ loop operators with enhanced scaling are realized by  $(p,q)$ strings at the poles, similar to the discussion for fundamental strings.

\smallskip

The S-dual quiver gauge theory, given by (\ref{eq:D5NS5-quiver}) with $N$ and $M$ exchanged, is obtained by a ninety-degree rotation of the brane web in fig.~\ref{fig:plus}.
It is described by a correspondingly transformed supergravity solution which has $N$ and $M$ exchanged (the transformation follows from (\ref{eq:dApm-SU11})).
This can be represented as a ninety-degree rotation of the disc in fig.~\ref{fig:plus-disc}.
S-duality exchanges the fundamental strings and D-strings, so that the $(0,1)$ loop operators represented by D-strings can be described as Wilson loops associated with the boundary nodes in the S-dual quiver. This explains their expectation value one.
The transformation exchanges the F1 and D1 charges of the D3-branes, and thus their horizontal and vertical coordinates in the brane web.
For the antisymmetric Wilson loops this exchanges the (rescaled) gauge node label $z$ and the (rescaled) representation label $\mathds{k}$. 
This explains the symmetry of the expectation value under this exchange observed in sec.~\ref{sec:plus-loc}.

\subsection{\texorpdfstring{$T_N$}{T[N]} solution}

\begin{figure}
\subfigure[][]{\label{fig:TN-web}
\begin{tikzpicture}[xscale=-0.6,yscale=-0.6]
     \draw[thick] (-4,0.75) -- (-0.5,0.75) -- (-0.5,-3);
     \draw[thick] (-0.5,0.75) -- +(1.75,1.75);
     
     \draw[thick] (-4,0.25) -- (0.25,0.25) -- (0.25,-3);
     \draw[thick] (0.25,0.25) -- +(1.75,1.75);
     
     \draw[thick] (-4,-0.25) -- (1.0,-0.25) -- (1.0,-3);
     \draw[thick] (1.0,-0.25) -- +(1.75,1.75);
     
     \draw[thick] (-4,-0.75) -- (1.75,-0.75) -- (1.75,-3);
     \draw[thick] (1.75,-0.75) -- +(1.75,1.75);
     
     \draw[thick] (-4,1.25) -- (-1.25,1.25) -- (-1.25,-3);
     \draw[thick] (-1.25,1.25) -- +(1.75,1.75);
     
     \node at (0,-3.5) {$N$ NS5};
    \node at (-5,0.25) {$N$ D5};
\end{tikzpicture}
}\hskip 4mm
\subfigure[][]{\label{fig:TN-disc}
 \begin{tikzpicture}
  \node at (0,0) {\includegraphics[width=3.8cm]{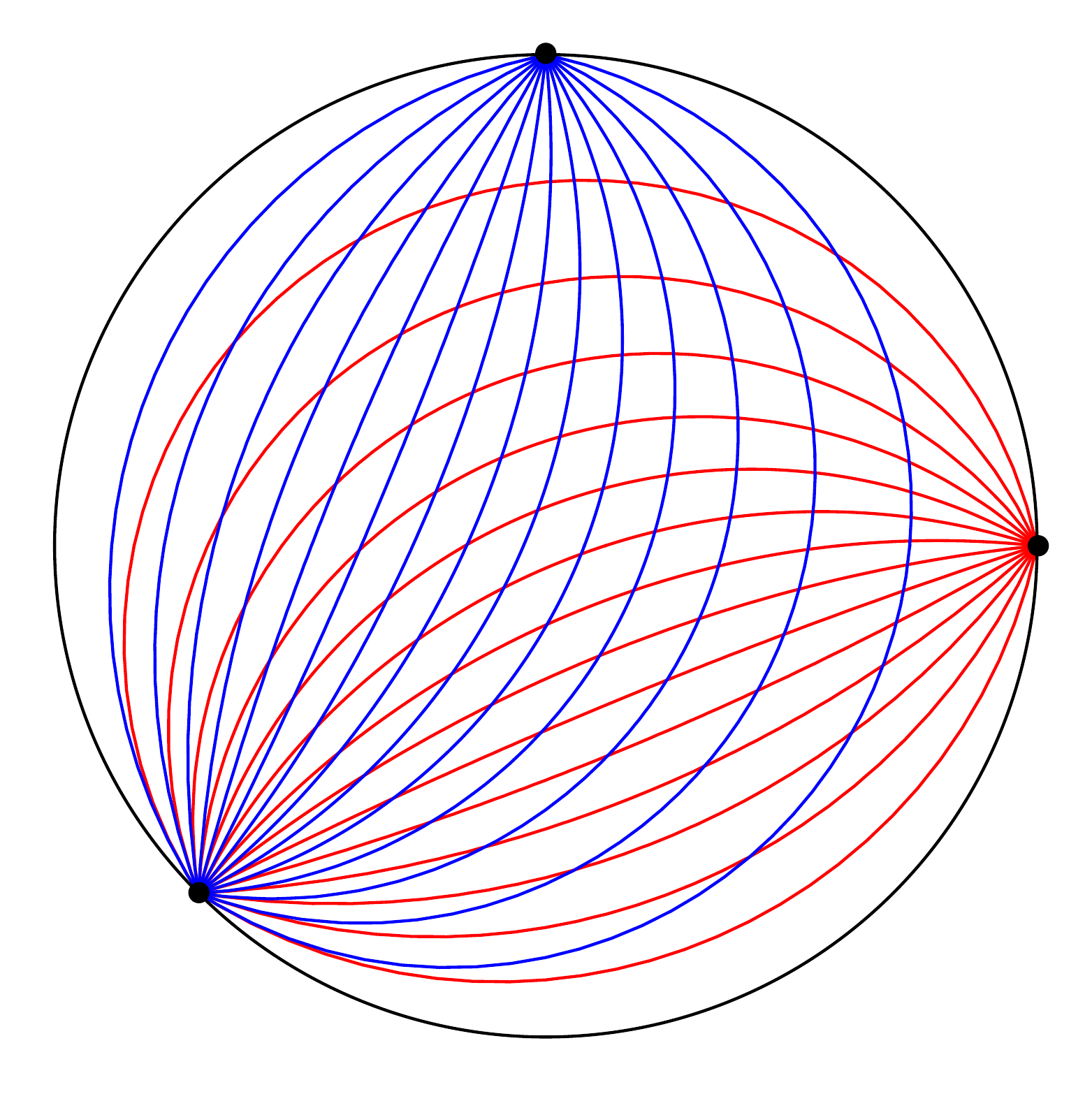}};
  \draw[thick] (0,0) circle (1.7);

\draw[very thick] ({-sin(45)*1.7},{-cos(45)*1.7}) -- ({-sin(45)*1.85},{-cos(45)*1.85});
\draw[very thick] (0,-1.7) -- (0,-1.85);
\draw[very thick] (1.7,0) -- (1.85,0);

\node at ({-sin(45)*2.15},{-cos(45)*2.15}) {$5_{[1,1]}$};
\node at (0,2.0) {NS5};
\node at (2.1,0) {D5};

\draw[thick,fill=red] (1.7,0) circle (0.08);
\node at (1.3,0) {\small \bf F1};

\node at ({sin(45)*2.15},{cos(45)*2.15}) {\scriptsize $(1,-1)$};
\draw[thick,fill=blue] ({sin(45)*1.7},{cos(45)*1.7}) circle (0.08);

\end{tikzpicture}
}\hskip 4mm
\subfigure[][]{\label{fig:TN-cont}
 \begin{tikzpicture}
  \node at (0,0) {\includegraphics[width=0.21\linewidth]{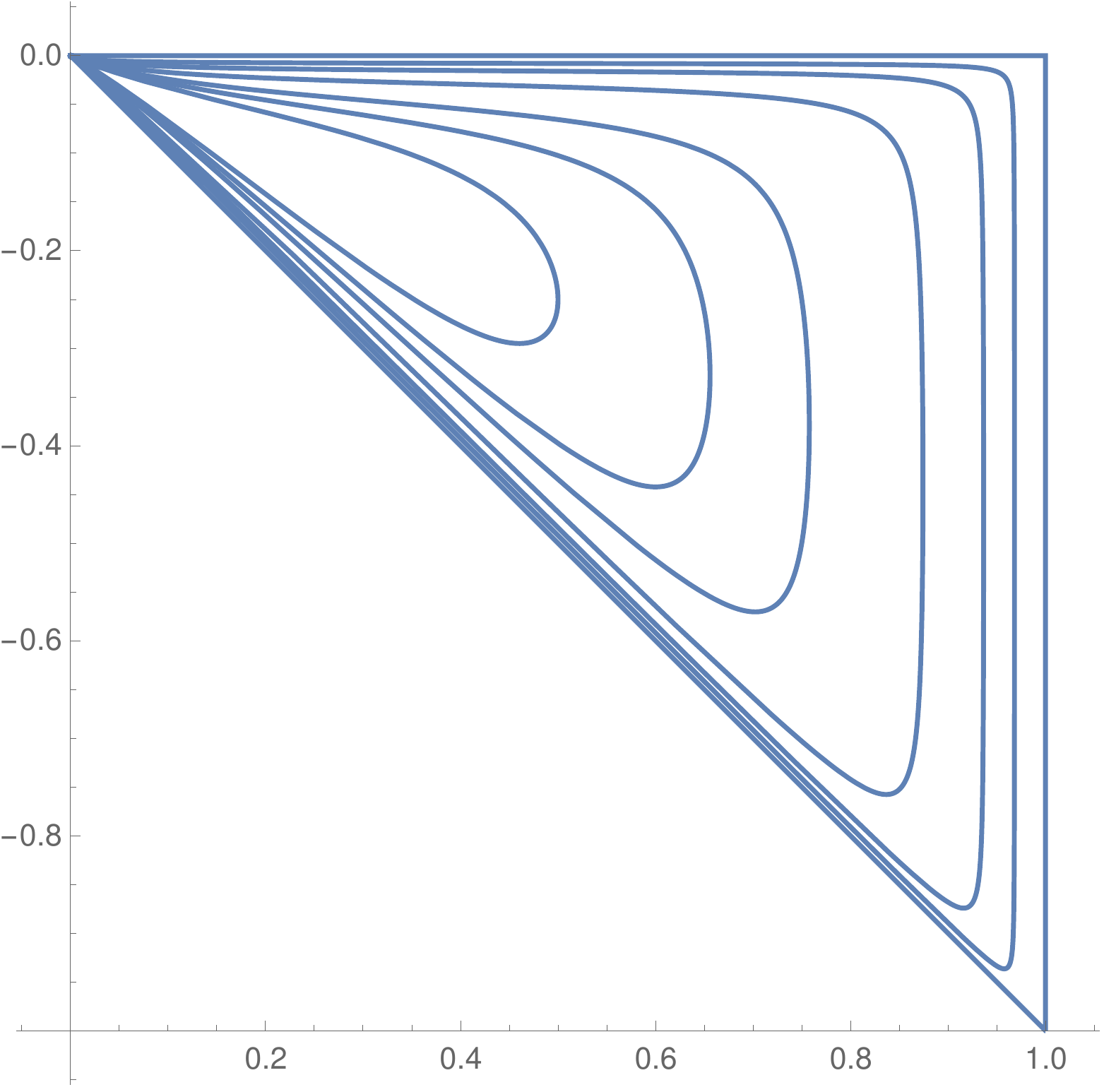}};
  \node[rotate=90] at (-2.05,0) {\scriptsize $N_{\rm F1}/N$};
  \node at (0,-2.0) {\scriptsize $N_{\rm D1}/N$};
 \end{tikzpicture}
}
\caption{Left: brane web for the gauge theory (\ref{TNquiver}). 
Center: $T_N$ supergravity solution on the disc, with the 5-brane poles and some of the string embeddings. 
Along the red/blue curves the D3-brane has constant F1/D1 charge.
Right: $(N_{\rm D1},N_{\rm F1})$ curves, from outer to inner for $\Im(w)\in\lbrace 10^{-6},\tfrac{1}{10},\tfrac{1}{5},\tfrac{1}{2},1 \rbrace$.
The $\Im(w)=10^{-6}$ curve follows the boundary of the disc in fig.~\ref{fig:TN-disc}.
The F1/D1 charges identify the vertical/horizontal coordinate of the face in which the  D3-brane is located in the brane web.}
\end{figure}
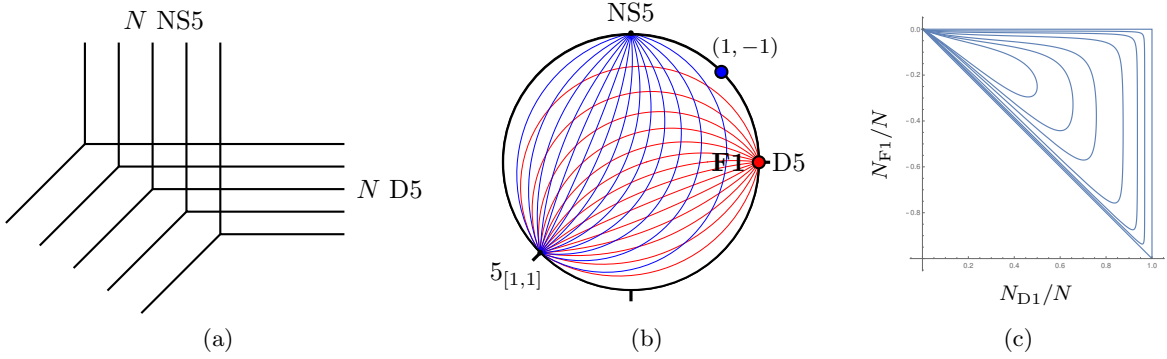

The $T_N$ theories are realized by junctions of $N$ D5-branes, $N$ NS5-branes and $N$ $(1,1)$ 5-branes, fig.~\ref{fig:TN-web}.
The supergravity solution for the $T_N$ junction correspondingly has three poles in $\partial_w\cA_\pm$, which can be placed at $\lbrace 0,\pm 1\rbrace$. 
The functions $\cA_\pm$ on the upper half plane are (sec.~4.3 of \cite{Bergman:2018hin})
\begin{align}\label{eq:Apm-TN}
 \cA_\pm&=\frac{3}{8\pi}N \left[\pm \ln(w-1)+i\ln(2w)+(\mp 1-i)\ln(w+1)\right]~.
\end{align}
For this theory the function $\cG$ defined in (\ref{eq:kappa2-G}), using the expression in (\ref{eq:cG-gen}) and the identities in footnote \ref{foot:D-rel}, can be expressed as
\begin{align}
 \cG&=\frac{9}{8\pi^2}N^2D\left(\frac{2w}{w+1}\right)~,
\end{align}
where $D$ is again the Bloch-Wigner function defined in (\ref{eq:D-def}).

The antisymmetric Wilson loop expectation values obtained from the action of D3-branes embedded into the solution, via (\ref{eq:Whol-G}), are given by
\begin{align}\label{W-wedge-TN-hol}
 \ln\langle W_{\wedge}\rangle &=
 \frac{3}{\pi}N^2D\left(\frac{2w}{w+1}\right)~.
\end{align}
To make this result meaningful the gauge node and representation of the Wilson loop represented by a D3-brane at a point $w$ have to be identified.
The F1 and D1 charges of the D3-brane at the point $w$ are obtained from (\ref{eq:NF1-ND1-cApm}) and given by
\begin{align}\label{eq:NFND-TN}
 N_{\rm F1}&=\frac{N}{\pi}\Im\ln\left(\frac{w+1}{w}\right)~,
 &
 N_{\rm D1}&=\frac{N}{\pi}\Im\ln\left(\frac{w-1}{w+1}\right)~.
\end{align}
$N_{\rm D1}$ varies between $0$ for $w\rightarrow\infty$ and $N$ at $w=0$;
$N_{\rm F1}$ varies between $0$ for $w\rightarrow\infty$ and $-N$ at $w=-\tfrac{1}{2}$.
Curves of constant D1/F1 charge are shown in fig.~\ref{fig:TN-disc}, for which the upper half plane has been mapped to the unit disc.
The curves of constant $N_{\rm D1}$ connect the NS5 and $(1,1)$ 5-brane poles, while the curves with constant $N_{\rm F1}$ connect the D5 and $(1,1)$ 5-brane poles.

As in the $+_{N,M}$ solution before, the D1 and F1 charges, shown in fig.~\ref{fig:TN-cont}, carve out the grid diagram of the brane web in fig.~\ref{fig:TN-web}, which now contains the points $\lbrace (i,-j)\,\vert\, i,j=0,\ldots, N, \ j\leq i\rbrace$.
The charges therefore naturally identify the face of the $T_N$ web in which the D3-brane is located.
The horizontal and vertical coordinates of the face, $n_x$ and $n_y$, such that the far left face of the brane web corresponds to $n_x=n_y=0$, is given by
\begin{align}
 n_x&=N_{\rm D1}~, & n_y&=N_{\rm F1}~.
\end{align}
With the expressions for $N_{\rm F1}$ and $N_{\rm D1}$ in (\ref{eq:NFND-TN}) this associates to each point of $\Sigma$ a face of the $T_N$ brane web.
The supergravity solution explicitly encodes the field theory data $L$ in the range of $N_{\rm D1}$, which leads to $L=N$, and the rank function $N(z)$ in the range of $N_{\rm F1}$ for fixed $N_{\rm D1}=zL$, which leads to $N(z)=Nz$. 
The coordinates of the D3-brane in the brane web identify the representation of the Wilson loop and the gauge node, following the logic of sec.~\ref{sec:brane-web-Wilson}, which leads to
\begin{align}\label{eq:kz-TN}
z&=\frac{N_{\rm D1}}{N}~,
&
N(z)&=N_{\rm D1}~,
&
 \mathds{k}&=-\frac{N_{\rm F1}}{N(z)}~.
\end{align}
From these expressions and (\ref{eq:NFND-TN}) a concrete identification of $w$ with $\mathds{k}$ and $z$ can be derived,
\begin{align}\label{eq:w-zk-TN}
 \frac{2w}{w+1}&=e^{i\pi \mathds{k}z}\csc(\pi(1-\mathds{k})z)\sin(\pi z)~.
\end{align}
This identifies for each D3-brane the properties of the associated Wilson loop.
The holographic result for the antisymmetric Wilson loop expectation values in (\ref{W-wedge-TN-hol}) becomes
\begin{align}
  \ln\langle W_{\wedge}\rangle &=
 \frac{3}{\pi}N^2D\left(e^{i\pi \mathds{k}z}\csc(\pi(1-\mathds{k})z)\sin(\pi z)\right)~.
\end{align}
Noting that $D(z)=D(1/(1-z))$ (see footnote~\ref{foot:D-rel}), this exactly matches the field theory result given by (\ref{eq:W-wedge-TN}) with (\ref{eq:b-TN}).

\smallskip

We now discuss loop operators represented by strings. 
The BPS condition for $(p,q)$ strings in (\ref{eq:BPS-pq-string}), with the functions $\cA_\pm$ in (\ref{eq:Apm-TN}), leads to a linear equation for $w$, with solution
\begin{align}
 w&=\frac{q}{q-2p}~.
\end{align}
For each choice of $(p,q)$ there is one $\tfrac{1}{2}$-BPS embedding of a $(p,q)$ string at a regular boundary point.
The gauge nodes which the corresponding loop operators are associated with can be identified from the location of the embedding and  (\ref{eq:kz-TN}).
For $|q|< |q-2p|$ this is the boundary node $z_{(p,q)}=1$, and otherwise the boundary node $z_{(p,q)}=0$.
The expectation values of the corresponding loop operators are given, via (\ref{eq:Spq-dSigma}), by
\begin{align}\label{eq:Spq-T}
 \langle W_{(p,q)}(z_{(p,q)})\rangle&=e^{S_{(p,q)}}~, &
 S_{(p,q)}&=\frac{3}{2}\left|Np\ln\frac{(p-q)^2}{p^2}-Nq\ln \frac{(p-q)^2}{q^2}\right|~.
\end{align}
In the $T_N$ brane web these strings correspond to $\tfrac{1}{2}$-BPS $(p,q)$ strings perpendicular to $(p,q)$ 5-branes, as discussed in sec.~\ref{sec:brane-web-Wilson} and below (\ref{eq:Spq-plus}).

Fundamental strings with $(p,q)=(1,0)$ can be embedded at the D5-brane pole at $w=0$, which corresponds to the boundary node at $z=1$. 
They connect to the 5-brane web from above or below, following a similar discussion as around (\ref{eq:Wf-plus-hol}),
and represent fundamental and anti-fundamental Wilson loops.
The action vanishes, leading to
\begin{align}
 \langle W_f(1)\rangle&=\langle W_{\bar f}(1)\rangle=1~.
\end{align}
This matches the field theory result (\ref{eq:W-f-TN}). 
The remaining Wilson loops are represented by fundamental strings embedded at the NS5 and (1,1) 5-brane poles, in parallel with the discussion for the $+_{N,M}$ solution. They can be recovered from D3-branes approaching the poles along lines where the D3-branes carry constant D1 charge (the blue curves in fig.~\ref{fig:TN-disc}). In fig.~\ref{fig:TN-cont} this corresponds to approaching the boundary of the triangle along vertical lines.
Since the antisymmetric Wilson loop expectation values were already shown to exactly match the field theory computation, the matching of the remaining (anti)fundamental Wilson loops to (\ref{eq:Wf-TN-int}) is automatic.

D-strings and $(1,1)$ strings can be embedded, respectively, at the NS5-brane pole at $w=1$ and at the $(1,1)$ 5-brane pole at $w=-1$ with vanishing action. 
A $(1,-1)$ string can be embedded at $w=1/3$, and represents a $(1,-1)$ loop operator at the $z=1$ boundary node
with expectation value
\begin{align}\label{eq:TN-1m1-loop}
\langle W_{(1,-1)}(1)\rangle &=2^{\,6N}~.
\end{align}
This coincides with the expectation value of the fundamental Wilson loop in the $Y_N$ theory, (\ref{eq:YN-fund-loc}).

\smallskip

S-duality rotates the brane web in fig.~\ref{fig:TN-web} by ninety degrees. The  gauge theory description is again given by (\ref{TNquiver}), up to a reversal of the coordinate along the quiver.
Fundamental strings and D-strings are exchanged, which relates the $(0,1)$ loop operators with expectation value one to (anti)fundamental Wilson loops in the S-dual quiver with expectation value one.
The $(1,-1)$ loop is mapped to a $(1,1)$ loop; a relation to the fundamental Wilson loop in the $Y_N$ theory will be discussed in the next section.
Antisymmetric Wilson loops are mapped by S-duality to antisymmetric Wilson loops, with transformed parameters $z$ and $\mathds{k}$.
From the rotation of the brane web and reversal of the coordinate along the quiver, 
the transformation is $(z,\mathds{k})\rightarrow (1-(1-\mathds{k})z,(1-z)/(1-(1-\mathds{k})z))$.
The results in (\ref{eq:W-wedge-TN}) and in (\ref{W-wedge-TN-hol}), (\ref{eq:w-zk-TN}) are invariant under this transformation.

\subsection{\texorpdfstring{$Y_N$}{Y[N]} solution}

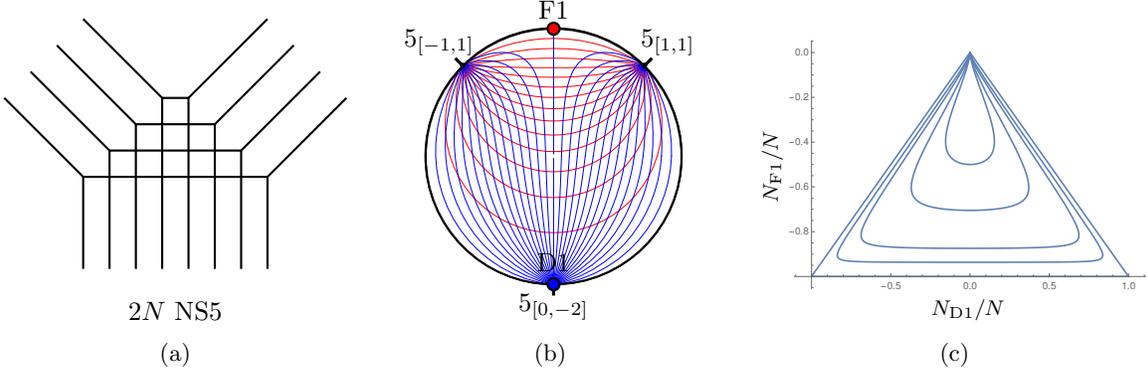
\begin{figure}
\subfigure[][]{\label{fig:YN}
\begin{tikzpicture}[scale=0.7]
     \draw[thick] (-0.25,-2.5) -- (-0.25,0.75) -- (0.25,0.75) -- (0.25,-2.5);
     \draw[thick] (-0.25,0.75) -- +(-1.5,1.5);
     \draw[thick] (0.25,0.75) -- +(1.5,1.5);
     \draw[thick] (-0.75,-2.5) -- (-0.75,0.25) -- (0.75,0.25) -- (0.75,-2.5);
     \draw[thick] (-0.75,0.25) -- +(-1.5,1.5);
     \draw[thick] (0.75,0.25) -- +(1.5,1.5);
     \draw[thick] (-1.25,-2.5) -- (-1.25,-0.25) -- (1.25,-0.25) -- (1.25,-2.5);
     \draw[thick] (-1.25,-0.25) -- +(-1.5,1.5);
     \draw[thick] (1.25,-0.25) -- +(1.5,1.5);
     \draw[thick] (-1.75,-2.5) -- (-1.75,-0.75) -- (1.75,-0.75) -- (1.75,-2.5);
     \draw[thick] (-1.75,-0.75) -- +(-1.5,1.5);
     \draw[thick] (1.75,-0.75) -- +(1.5,1.5);
     
     \node at (0,-3.3) {$2N$ NS5};
\end{tikzpicture}
}\hskip 4mm
\subfigure[][]{\label{fig:YN-disc}
 \begin{tikzpicture}
  \node at (0,0) {\includegraphics[width=3.8cm]{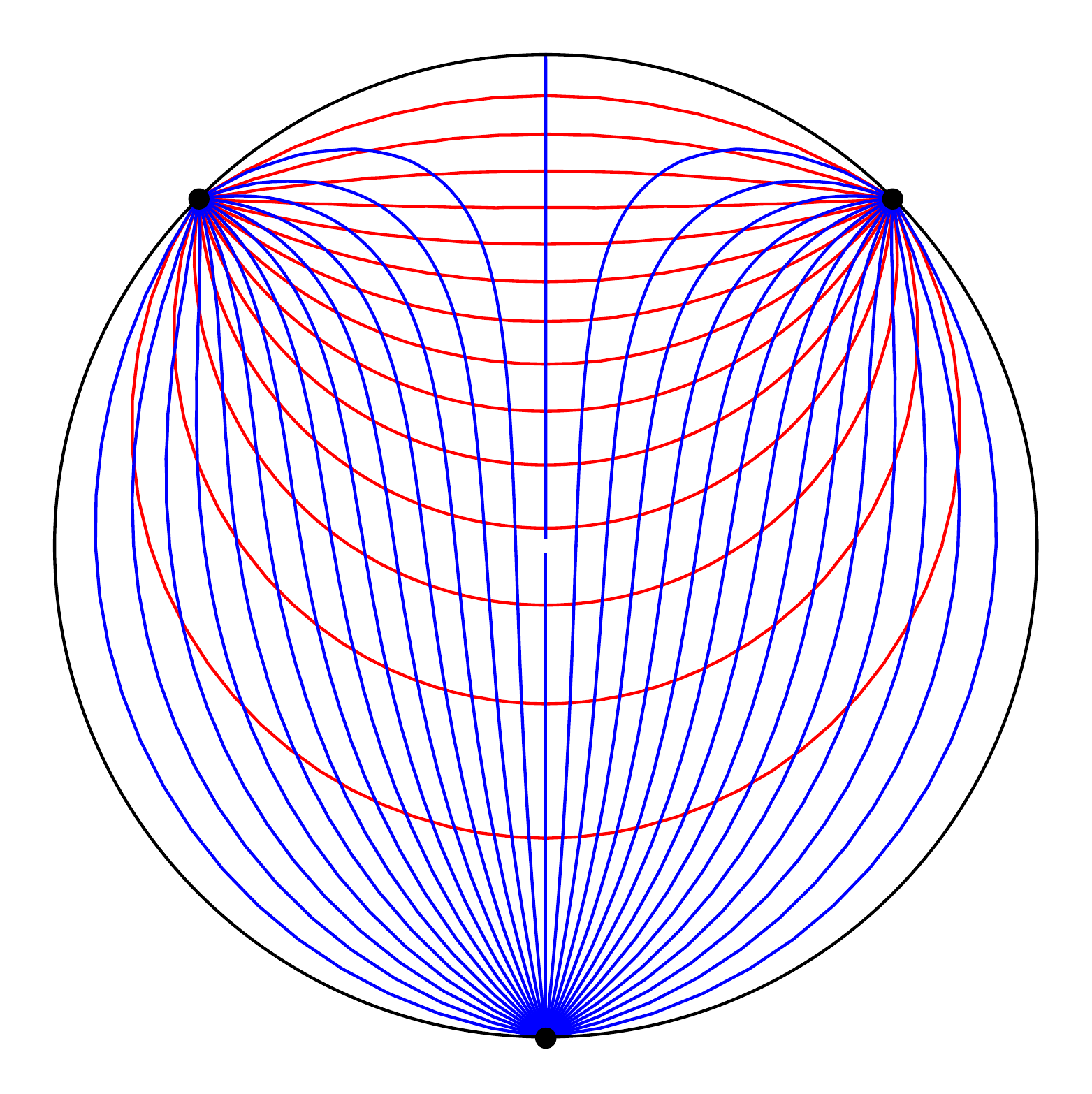}};
  \draw[thick] (0,0) circle (1.7);
  \foreach \j in {-45,180,45}{
    \draw[very thick] ({sin(\j)*1.7},{cos(\j)*1.7}) -- ({sin(\j)*1.85},{cos(\j)*1.85});
  }
  \node at ({sin(45)*2.15},{cos(45)*2.15}) {$5_{[1,1]}$};
  \node at ({sin(180)*2.0},{cos(180)*2.0}) {$5_{[0,-2]}$}; 
  \node at ({sin(-45)*2.15},{cos(-45)*2.15}) {$5_{[-1,1]}$}; 

  \draw[thick,fill=red] (0,1.7) circle (0.08);
  \node at (0,1.95) {F1};
  
  \node at (0,-1.4) {D1};
  \draw[thick,fill=blue] ({sin(0)*1.7},{-cos(0)*1.7}) circle (0.08);

 \end{tikzpicture}
}\hskip 4mm
\subfigure[][]{\label{fig:YN-cont}
 \begin{tikzpicture}
  \node at (0,0) {\includegraphics[width=0.29\linewidth]{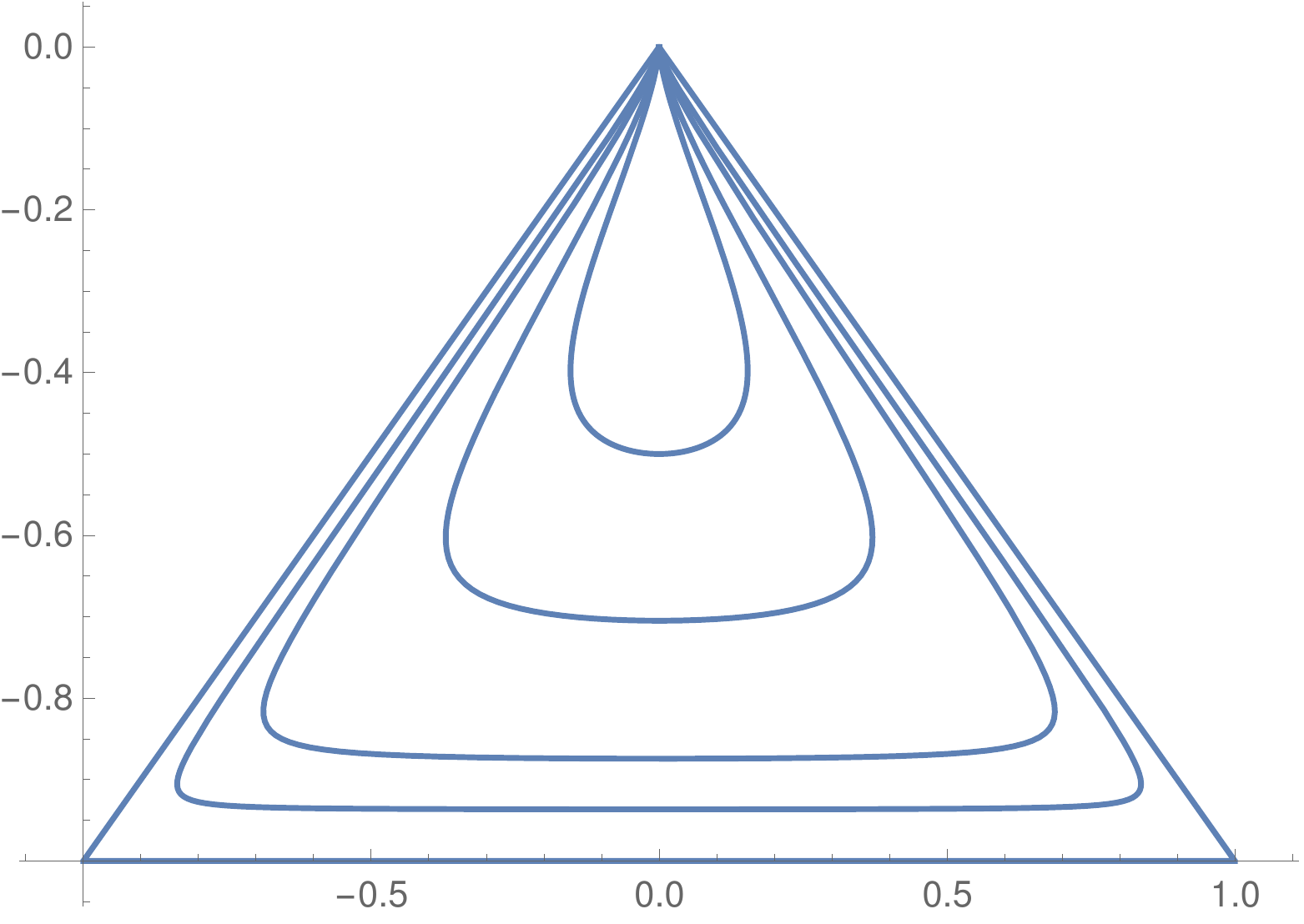}};
  \node[rotate=90] at (-2.6,0) {\scriptsize $N_{\rm F1}/N$};
  \node at (0,-1.9) {\scriptsize $N_{\rm D1}/N$};
 \end{tikzpicture}
}

\caption{Left: Brane web for the gauge theory (\ref{eq:YN-quiver-2}). 
Center: $Y_N$ supergravity solution on the disc, with the 5-brane poles and string embedding. 
Along the red/blue curves the D3-brane has constant F1/D1 charge.
Right: $(N_{\rm D1},N_{\rm F1})$ curves, from outer to inner curve for $\Im(w)\in\lbrace 10^{-6},\tfrac{1}{10},\tfrac{1}{5},\tfrac{1}{2},1 \rbrace$.
The F1/D1 charges carve out the grid diagram of the web in fig.~\ref{fig:YN} and identify the vertical/horizontal coordinate of the face in which the  D3-brane is located in the brane web.}
\end{figure}

The $Y_N$ theories are realized by junctions of $2N$ NS5-branes, $N$ $(1,1)$ 5-branes and $N$ $(-1,1)$ 5-branes, fig.~\ref{fig:YN}.
The supergravity solution for the $Y_N$ junction has $\partial_w\cA_\pm$ with three poles, which can be placed at $\lbrace 0,\pm 1\rbrace$. 
The functions $\cA_\pm$ on the upper half plane are
\begin{align}
 \cA_\pm&=\frac{3}{8\pi}N \left[(\pm 1+i)\ln(w-1)+(\pm 1-i)\ln(w+1)\mp 2\ln(2w)\right]~.
\end{align}
For this theory, from the general expression in (\ref{eq:cG-gen}) and the identities in footnote~\ref{foot:D-rel},
\begin{align}
 \cG&=\frac{9}{4\pi^2}N^2D\left(\frac{2w}{w+1}\right)~.
\end{align}

The antisymmetric Wilson loop expectation values obtained from D3-branes embedded into this solution,  via (\ref{eq:Whol-G}), are given by
\begin{align}\label{W-wedge-YN-hol}
 \ln\langle W_{\wedge}\rangle 
 & = \frac{6}{\pi}N^2D\left(\frac{2w}{w+1}\right)~.
\end{align}
Up to a factor 2 the expression for $\cG$ and (\ref{W-wedge-YN-hol}) are identical to the expressions for the $T_N$ solution.
The F1 and D1 charges of the D3-brane at a point $w$ on $\Sigma$, however, are different.
From (\ref{eq:NF1-ND1-cApm}),
\begin{align}
 N_{\rm D1}&=\frac{N}{\pi}\Im\ln\left(\frac{w^2-1}{w^2}\right)~,
 &
 N_{\rm F1}&=\frac{N}{\pi}\Im\ln\left(\frac{w+1}{w-1}\right)~.
\end{align}
$N_{\rm D1}$ is valued in $[-N,N]$ and $N_{\rm F1}$ in $[-N,0]$. 
This reflects that the gauge theory has $2N+\mathcal O(1)$ nodes and is different from the quiver deformation of the $T_N$ theory.
Curves of constant F1/D1 charge are shown in fig.~\ref{fig:YN-disc}, for which the upper half plane is mapped to the unit disc.
Half of the curves with constant $N_{\rm D1}$ connect the NS5 and $(1,1)$ 5-brane poles, and half of them connect the NS5 and $(1,-1)$ 5-brane poles.
Curves of constant $N_{\rm F1}$ extend from the $(-1,1)$ 5-brane pole to the $(1,1)$ 5-brane pole.

As in the previous examples, the F1 and D1 charges in fig.~\ref{fig:YN-cont} carve out the grid diagram of the brane web in fig.~\ref{fig:YN}.
They identify the face of the $Y_N$ brane web in which a D3-brane embedded at a point $w\in\Sigma$ is located.
The horizontal and vertical coordinates of the face in the brane web, $n_x$ and $n_y$, such that the lower left face in the web corresponds to $n_x=n_y=0$, 
for the D3-brane embedded at a point $w$ are given by
\begin{align}\label{eq:tk-YN}
 n_x&=N_{\rm D1}+N~,
 &
 n_y&=N_{\rm F1}+N~.
\end{align}
This associates with each point of $\Sigma$ a face in the $Y_N$ brane web.
The field theory data $L$ and $N(z)$ are encoded in the supergravity solution in the range of $n_x$ and in the range of $n_y$ for fixed $n_x=zL$, respectively.
The parameters of the Wilson loop represented by the D3-brane are found, following the discussion of sec.~\ref{sec:brane-web-Wilson}, as
\begin{align}\label{eq:zk-YN}
 z&=\frac{N_{\rm D1}+N}{2N}~,
 &
 N(z)&=N-\left|N_{\rm D1}\right|~,
 &
 \mathds{k}&=1-\frac{N_{\rm F1}+N}{N(z)}~.
\end{align}
The relations (\ref{eq:zk-YN}) can be solved for $w$ in terms of $z$ and $\mathds{k}$.
For $z\leq 1/2$ this leads to 
\begin{align}\label{eq:w-kz-YN}
 \frac{2w}{w+1}&=1+e^{2\pi i (\mathds{k}-1) z}\,\csc(\pi \mathds{k} z) \sin(\pi(\mathds{k}-2)z)~.
\end{align}
Using (\ref{eq:w-kz-YN}) in  (\ref{W-wedge-YN-hol}), the holographic result for the expectation values of antisymmetric Wilson loops becomes
\begin{align}
 \ln\langle W_{\wedge}\rangle 
 & = \frac{6}{\pi}N^2D\left(1+e^{2\pi i (\mathds{k}-1) z}\,\csc(\pi \mathds{k} z) \sin(\pi(\mathds{k}-2)z)\right)~.
\end{align}
This exactly matches the field theory result (\ref{eq:Wwedge-YN}), noting that $D(z)=D(\bar z/(\bar z-1))$.
Both expressions are valid for $z\leq \frac{1}{2}$ and extend to $z\geq \frac{1}{2}$ by symmetry of the quiver (and the supergravity solution) under $z\rightarrow 1-z$, providing a complete match.

\smallskip

We now discuss loop operators represented by strings.
The BPS condition in (\ref{eq:BPS-pq-string}) leads to a linear equation for $w$, 
\begin{align}
 w&=\frac{p}{q}~.
\end{align}
For each choice of $p$, $q$ there is one regular point on the boundary where a string with these charges is $\tfrac{1}{2}$-BPS.
The gauge node which the loop operator is associated with can be identified from the location of the embedding and (\ref{eq:zk-YN}).
It is the central node $z_{(p,q)}=\frac{1}{2}$ for $|p/q|>1$, the boundary node $z_{(p,q)}=1$ for $0<p/q<1$, and the boundary node $z_{(p,q)}=0$ for $-1<p/q<0$.
The expectation values of the corresponding $(p,q)$ loop operators, via (\ref{eq:Spq-dSigma}), evaluate to
\begin{align}\label{eq:Spq-Y}
 \langle W_{(p,q)}(z_{(p,q)})\rangle&=e^{S_{(p,q)}}~,
 &
 S_{(p,q)}&=3\left| Nq\ln\left|\frac{p-q}{p+q}\right|+Np\ln\left|\frac{4p^2}{p^2-q^2}\right|\right|~.
\end{align}
A D-string can be embedded at the NS5-pole at $w=0$ and has vanishing action. 
A fundamental string can be embedded at the point at infinity. 
The string represents a fundamental Wilson loop associated with the $z=\tfrac{1}{2}$ node, with expectation value
\begin{align}\label{eq:YN-fund-hol}
 \big\langle W_{f}\big(\tfrac{1}{2}\big)\big\rangle&=2^{\,6N}~.
\end{align}
This matches the field theory result (\ref{eq:YN-fund-loc}). The embedding of the string is shown in fig.~\ref{fig:YN-disc}.
Having only one $\tfrac{1}{2}$-BPS fundamental string at a regular boundary point reflects the field theory discussion in sec.~\ref{sec:YN-loc}: 
depending on the sign of the Chern-Simons term, only one of the fundamental and anti-fundamental Wilson loops scales linearly.

The fundamental Wilson loop with linear scaling at the central node is represented in terms of D3-branes by the single line of constant D1 charge for the D3-brane in fig.~\ref{fig:YN-disc} that ends at a regular boundary point.
The fundamental string at its end point represents the fundamental Wilson loop obtained for $k=1$.
The anti-fundamental Wilson loops for all nodes are represented by D3-branes approaching the NS5-brane pole along curves with constant D1 charge, in parallel with the discussion for the $+_{N,M}$ solution.
In fig.~\ref{fig:YN-cont} this corresponds to approaching the lower horizontal boundary along vertical lines.
The remaining fundamental Wilson loops are represented by D3-branes approaching the $(1,1)$ and $(1,-1)$ 5-brane poles along curves with constant D1 charge. They become equivalent to fundamental strings at these poles. 
In fig.~\ref{fig:YN-cont} this corresponds to approaching the upper boundary segments of the triangle along vertical lines.
Since the antisymmetric Wilson loop expectation values were already shown to exactly match the field theory computation, the matching of the remaining (anti)fundamental Wilson loops to (\ref{eq:Wf-YN-int}) is automatic.

\smallskip

The S-dual quiver deformation for the $Y_N$ theory, obtained from a ninety degree rotation of the brane web in fig.~\ref{fig:YN}, is given by (\ref{eq:YN-quiver-S}).
The supergravity solution for the dual web is obtained by a corresponding S-duality transformation, which can be obtained from (\ref{eq:dApm-SU11}) and changes the residues of $\partial_w\cA_\pm$ accordingly.
The result can be seen as a ninety degree rotation of fig.~\ref{fig:YN-disc}.
The D-string at the NS5 pole becomes a fundamental string at a D5 pole, with vanishing action. 
It describes a fundamental Wilson loop with expectation value one associated with a boundary node.
This matches the field theory discussion at the end of sec.~\ref{sec:YN-loc}.
The fundamental string in fig.~\ref{fig:YN-disc} becomes a D-string after S-duality, representing a $(0,1)$ loop with expectation value as in (\ref{eq:YN-fund-hol}).

\smallskip

The $Y_N$ supergravity solution is also related to the $T_N$ solution, by an $SL(2,\RR)$ transformation with $a=-b=c=d=1/\sqrt{2}$ combined with an overall rescaling of the 5-brane charges, $N\rightarrow N/\sqrt{2}$.
This transformation relates the results if the charges of probe branes are rescaled accordingly.
The $(1,-1)$ loop operator in the $T_N$ theory, with expectation value (\ref{eq:TN-1m1-loop}), is related by this transformation to the fundamental Wilson loop in the $Y_N$ theory, (\ref{eq:YN-fund-hol}), explaining the relation between their expectation values.
D3 branes are mapped to D3-branes with transformed $(p,q)$ string charges, which leads to a relation between antisymmetric Wilson loops.
It is
\begin{align}\label{eq:Y-T-W}
 \ln \langle W^{Y_N}_\wedge(z,\mathds{k})\rangle&= 2 \ln \Big\langle W^{T_N}_\wedge\Big(\hat z(2-\mathds{k})-1,\frac{\mathds{k}\hat z}{(2-\mathds{k})\hat z-1}\Big)\Big\rangle~,
\end{align}
where $\hat z=(1-|1-2z|)/2$.
This relation is satisfied by the expectation values in (\ref{eq:Wwedge-YN}) and (\ref{eq:W-wedge-TN}).

\subsection{\texorpdfstring{$X_{N}$}{X[N]} solution}\label{sec:X-sol}

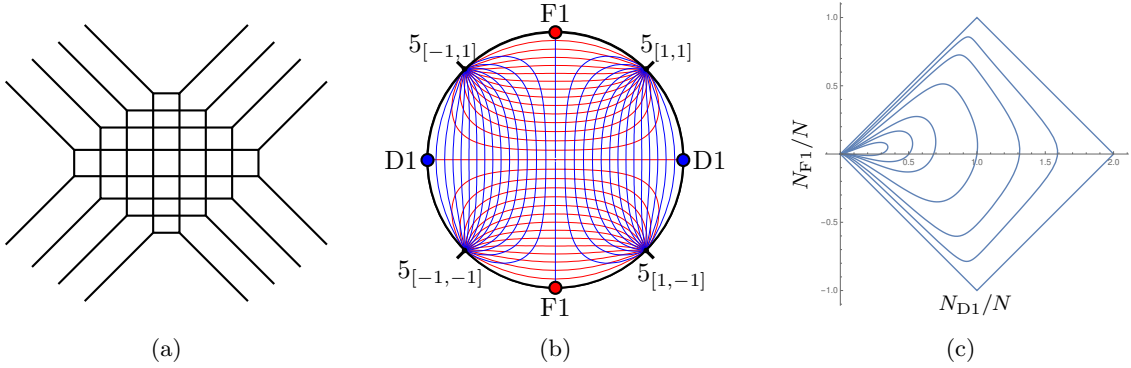
\begin{figure}
\subfigure[][]{\label{fig:X-web}
   \begin{tikzpicture}[scale=0.7]
     \draw[thick] (-1.75,-0.25) rectangle (1.75,0.25);
     \draw[thick] (-1.25,-0.675) rectangle (1.25,0.675);
     \draw[thick] (-0.75,-1) rectangle (0.75,1);
     \draw[thick] (-0.25,-1.325) rectangle (0.25,1.325);
     
     \draw[thick] (1.75,0.25) -- +(1.3,1.3);
     \draw[thick] (1.25,0.675) -- +(1.3,1.3);
     \draw[thick] (0.75,1) -- +(1.3,1.3);
     \draw[thick] (0.25,1.325) -- +(1.3,1.3);
     
     \draw[thick] (-1.75,0.25) -- +(-1.3,1.3);
     \draw[thick] (-1.25,0.675) -- +(-1.3,1.3);
     \draw[thick] (-0.75,1) -- +(-1.3,1.3);
     \draw[thick] (-0.25,1.325) -- +(-1.3,1.3);

     \draw[thick] (1.75,-0.25) -- +(1.3,-1.3);
     \draw[thick] (1.25,-0.675) -- +(1.3,-1.3);
     \draw[thick] (0.75,-1) -- +(1.3,-1.3);
     \draw[thick] (0.25,-1.325) -- +(1.3,-1.3);
     
     \draw[thick] (-1.75,-0.25) -- +(-1.3,-1.3);
     \draw[thick] (-1.25,-0.675) -- +(-1.3,-1.3);
     \draw[thick] (-0.75,-1) -- +(-1.3,-1.3);
     \draw[thick] (-0.25,-1.325) -- +(-1.3,-1.3);
     
     \node at (0,-2.9) {};
   \end{tikzpicture}
}\hskip 4mm
\subfigure[][]{\label{fig:XNM-disc}
 \begin{tikzpicture}
  \node at (0,0) {\includegraphics[width=3.8cm]{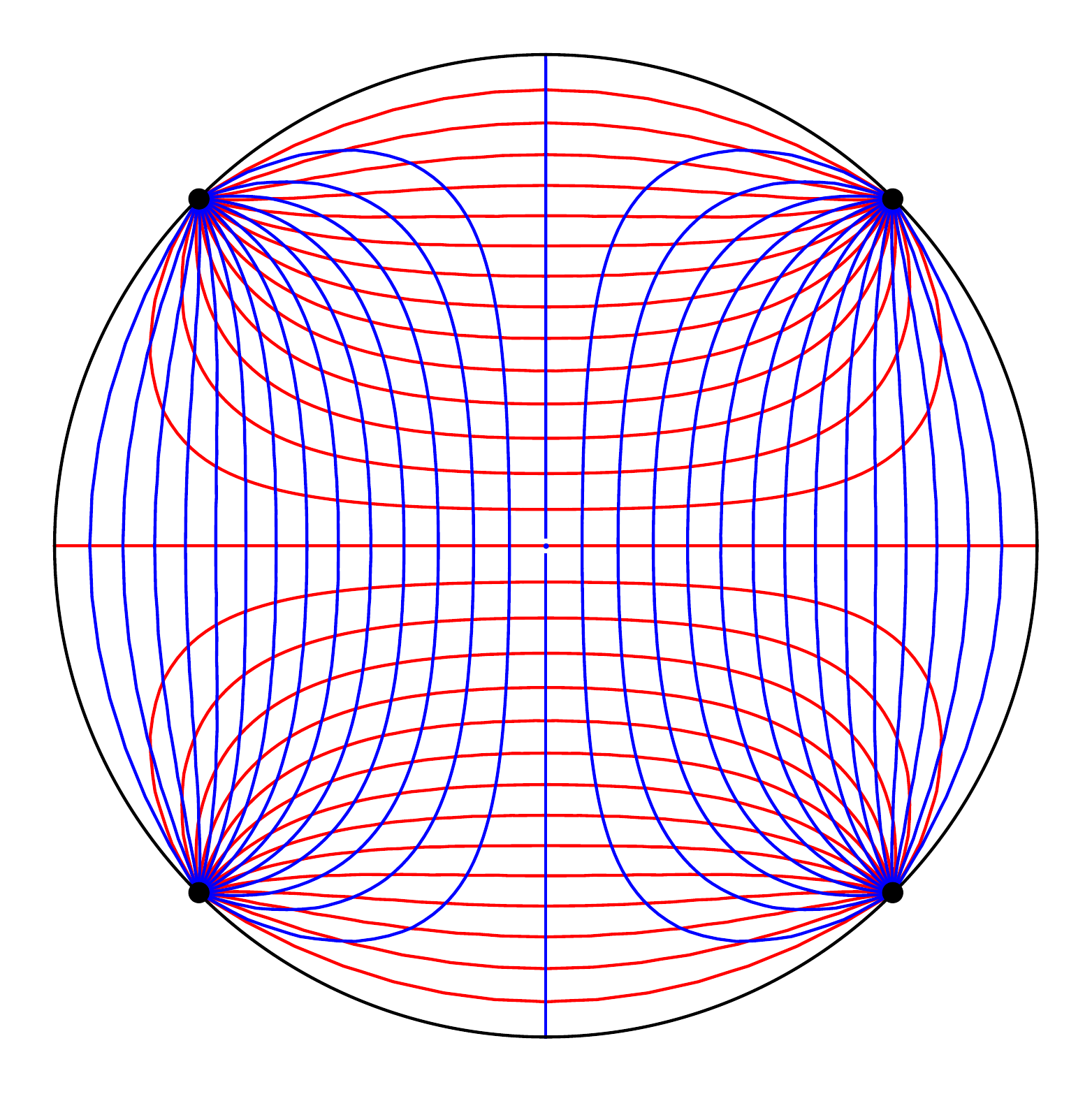}};
  \draw[thick] (0,0) circle (1.7);

\foreach \j in {-45,45,135,-135}{
 \draw[very thick] ({sin(\j)*1.7},{cos(\j)*1.7}) -- ({sin(\j)*1.85},{cos(\j)*1.85});
}
\node at ({sin(45)*2.1},{cos(45)*2.1}) {$5_{[1,1]}$};
\node at ({sin(-45)*2.1},{cos(-45)*2.1}) {$5_{[-1,1]}$}; 
\node at ({sin(135)*2.2},{cos(135)*2.2}) {$5_{[1,-1]}$};
\node at ({sin(-135)*2.15},{cos(-135)*2.15}) {$5_{[-1,-1]}$}; 

\draw[thick,fill=red] (0,1.7) circle (0.08);
\node at (0,1.95) {F1};

\draw[thick,fill=red] (0,-1.7) circle (0.08);
\node at (0,-1.95) {F1};

\draw[thick,fill=blue] (1.7,0) circle (0.08);
\node at (2.05,0) {D1};

\draw[thick,fill=blue] (-1.7,0) circle (0.08);
\node at (-2.05,0) {D1};

\end{tikzpicture}
}\hskip 4mm
\subfigure[][]{\label{fig:XN-cont}
 \begin{tikzpicture}
  \node at (0,0) {\includegraphics[width=0.25\linewidth]{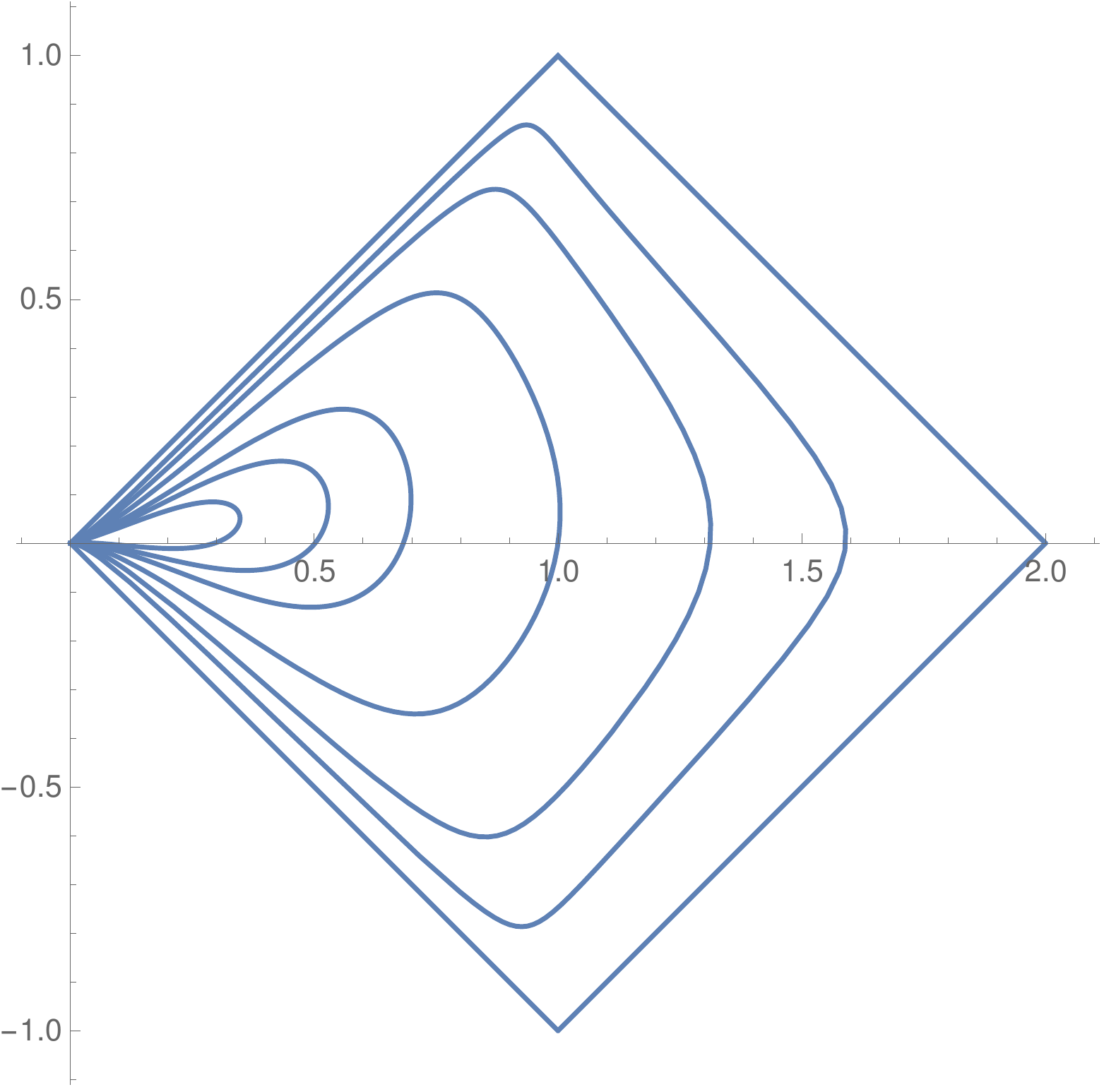}};
  \node[rotate=90] at (-2.3,0) {\scriptsize $N_{\rm F1}/N$};
  \node at (0,-2) {\scriptsize $N_{\rm D1}/N$};
 \end{tikzpicture}
}
\caption{Left: brane web realizing the gauge theory (\ref{eq:XNN-quiver}). 
Center: $X_{N}$ supergravity solution on the disc, with the 5-brane poles and some string embeddings. 
Along the red/blue curves the D3-brane has constant F1/D1 charge.
Right: $(N_{\rm D1},N_{\rm F1})$ curves, from outer to inner curve for $\Im(w)\in\lbrace 10^{-6},\tfrac{1}{20},\tfrac{1}{10},\tfrac{1}{5},\tfrac{2}{5},\tfrac{3}{5},1\rbrace$.
The $\Im(w)=10^{-6}$ curve traces the boundary of the disc in fig.~\ref{fig:XNM-disc}.
The F1/D1 charge identifies the vertical/horizontal coordinate of the face of the brane web in which the  D3-brane is.}
\end{figure}

The $X_{N}$ theories are realized by intersections of $N$ $(1,1)$ 5-branes and $N$ $(1,-1)$ 5-branes, fig.~\ref{fig:X-web}.
The $X_{N}$ supergravity solution is defined by (sec.~4.2 of \cite{Bergman:2018hin})
\begin{align}\label{eq:cA-X}
 \cA_\pm&=\frac{3}{8\pi}N\left[
 (\pm 1+i)\left(\ln(3w-2)-\ln w\right)+(\pm 1-i)\left(\ln(w-1)-\ln(2w-1)\right)
 \right]\,.
\end{align}
The pairs of poles with opposite-equal residues representing the pairs of external 5-branes are at $w\in\lbrace 0,\tfrac{2}{3}\rbrace$ and $w\in\lbrace \tfrac{1}{2},1\rbrace$.
The function $\cG$ following from (\ref{eq:cG-gen}) is given by
\begin{align}\label{eq:cG-X}
 \cG&=\frac{9}{4\pi^2}N^2\left[D\left(\frac{3w-2}{w}\right)+D\left(\frac{w}{2-3w}\right)\right]\,.
\end{align}
The antisymmetric Wilson loop expectation values obtained from D3-branes embedded at a point $w\in\Sigma$ in this solution, via (\ref{eq:Whol-G}), are given by
\begin{align}\label{eq:W-wedge-X-hol}
 \ln\langle W_{\wedge}\rangle&=\frac{6}{\pi}N^2\left[D\left(\frac{3w-2}{w}\right)+D\left(\frac{w}{2-3w}\right)\right]~.
\end{align}
The expressions in (\ref{eq:cG-X}), (\ref{eq:W-wedge-X-hol}) are related to those for the $+_{N,M}$ solution with $N=M$ by a factor 2, which is due to an $SL(2,\RR)$ relation similar to the one between the $T_N$ and $Y_N$ solutions.
The $(p,q)$ string charges of a D3-brane at $w$ are different; from (\ref{eq:NF1-ND1-cApm}) they are
\begin{align}\label{eq:NF-ND-X}
 N_{\rm D1}+N_{\rm F1}&=\frac{2}{\pi}N\Im \ln\left(\frac{w-1}{2w-1}\right)~,
 &
 N_{\rm D1}-N_{\rm F1}&=\frac{2}{\pi}N \Im\ln\left(\frac{3w-2}{w}\right)~.
\end{align}
The combinations $N_{\rm D1}+N_{\rm F1}$ and $N_{\rm D1}-N_{\rm F1}$ are valued in $[0,2N]$.
Curves of constant $N_{\rm D1}$ and $N_{\rm F1}$ are shown in fig.~\ref{fig:XNM-disc} for the solution on the unit disc.

The F1/D1 charges in fig.~\ref{fig:XN-cont} carve out the grid diagram of the brane web in fig.~\ref{fig:X-web}, and identify for each point of $\Sigma$ the face in which the D3-brane is located in the $X_{N}$ brane web.
Following the same logic as in the previous examples, the supergravity solution encodes the field theory data $L$ and $N(z)$ for the quiver (\ref{eq:XNN-quiver}), and the identification of the Wilson loop parameters for a D3-brane at a point $w$ is
\begin{align}\label{eq:zk-X}
 z&= \frac{N_{\rm D1}}{2N}~,
 &
 \mathds{k}&=\frac{1}{2}+\frac{N_{\rm F1}}{N(z)}
 ~,
 &
 N(z)&=2N-2|N_{\rm D1}-N|~.
\end{align}
From this relation between $z$, $\mathds{k}$ and $N_{\rm F1}$, $N_{\rm D1}$, with (\ref{eq:NF-ND-X}), one finds
\begin{align}
 \frac{2w-1}{w-1}&= e^{-2\pi i \mathds{k} z+\sinh^{-1}\left(\sin(2\pi\mathds{k}z)\cot(2\pi(1-\mathds{k})z)\right)}~.
\end{align}
Using the 5-term relation in footnote \ref{foot:D-rel} the expectation values in (\ref{eq:W-wedge-X-hol}) can be expressed as
\begin{align}\label{eq:W-wedge-X-hol-2}
 \ln\langle W_{\wedge}\rangle&=\frac{6}{\pi}N^2\left[D\left(u\right)+D\left(1+u\right)\right]~, & u&=e^{2\pi i \mathds{k} z-\sinh^{-1}\left(\sin(2\pi\mathds{k}z)\cot(2\pi(1-\mathds{k})z)\right)}~.
\end{align}
This exactly matches the field theory result in (\ref{eq:W-wedge-XNN}).

\smallskip

We now discuss loop operators realized by strings. 
The BPS condition (\ref{eq:BPS-pq-string}) for $(p,q)$ strings, with the functions $\cA_\pm$ in (\ref{eq:cA-X}), evaluates to
\begin{align}
 2 (w-1) (2 w-1) (p-q)+w (3 w-2) (p+q)&=0~.
\end{align}
It has two real solutions for each $(p,q)$.
The gauge nodes which the strings are associated with can be determined from the location of the embedding and (\ref{eq:zk-X}).
For $|p/q|>1$ they are both associated with the central node, $z_{(p,q)}=\frac{1}{2}$; for $|p/q|<1$ one embedding is associated with the boundary node at $z_{(p,q)}=0$ and the other with the node at $z_{(p,q)}=1$.
The expectation values of the corresponding $(p,q)$ loop operators are given by
\begin{align}\label{eq:W-pq-X}
 \langle W_{(p,q)}(z_{(p,q)})\rangle&= e^{S_{(p,q)}}~,
\end{align}
with the action resulting from (\ref{eq:Spq-dSigma})
\begin{align}\label{eq:Spq-X}
S_{(p,q)}&= 3\left|N(p+q)\right|\,\sinh^{-1}\left|\frac{p-q}{p+q}\right|+3\left|N(p-q)\right|\,\sinh^{-1}\left|\frac{p+q}{p-q}\right|~.
\end{align}

Fundamental strings and D-strings can be embedded, respectively, at $w_{\rm F1}=(4\pm \sqrt{2})/7$ and $w_{\rm D1}=2\pm \sqrt{2}$.
The expectation values of the corresponding loop operators are
\begin{align}\label{eq:XNN-fund-hol}
 \big\langle W_f\big(\tfrac{1}{2}\big)\big\rangle=
 \big\langle W_{\bar f}\big(\tfrac{1}{2}\big)\big\rangle&= (1+\sqrt{2})^{6N}~,
 \nonumber\\
 \langle W_D(0)\rangle=\langle W_D(1)\rangle &=(1+\sqrt{2})^{6N}~.
\end{align}
The fundamental strings are associated with the central node and represent the fundamental and anti-fundamental Wilson loops discussed in sec.~\ref{sec:XNN-loc}. The expectation values match (\ref{eq:XNN-fund-loc}).
The $X_N$ brane web is mapped to itself by S-duality, which exchanges the D-string and F-string operators.
The loop operators represented by D-strings are thus related to fundamental Wilson loops in the S-dual theory, which explains the expectation values in (\ref{eq:XNN-fund-hol}).

The fundamental and anti-fundamental Wilson loops with linear scaling in (\ref{eq:XNN-fund-hol}) can also be understood in terms of D3-branes, through the vertical curve of constant D1 charge in fig.~\ref{fig:XNM-disc} which has both end points at regular boundary points. The fundamental strings at the end points represent the fundamental and anti-fundamental Wilson loops obtained for $k=1$ and $k=2N-1$.
The D-strings in (\ref{eq:XNN-fund-hol}) can similarly be understood in terms of the horizontal curve of constant F1 charge in fig.~\ref{fig:XNM-disc} which connects two regular boundary points.
The curves of constant D1-charge corresponding to $z\neq \tfrac{1}{2}$ have both end points at poles.
The fundamental strings obtained at those end points represent the remaining (anti)fundamental Wilson loops with enhanced scaling.
Their expectation values can be obtained from the results for antisymmetric Wilson loops in (\ref{eq:W-wedge-X-hol-2}) using the relations in (\ref{eq:W-wedge-f}), which reproduces the field theory results in (\ref{eq:Wf-X-int}).

The $X_N$ solution is also related to the $+_{N,N}$ solution by an $SL(2,\RR)$ transformation combined with a rescaling, 
similar to the relation between the $Y_N$ and $T_N$ solutions. 
This relates the expectation values in (\ref{eq:XNN-fund-hol}) to those obtained for the $+_{N,M}$ theory in (\ref{eq:plus-pqstring-ops}) when $N=M$ and $(p,q)=(1,\pm 1)$. This connects the $(1,\pm 1)$ loop operators of the $+_{N,N}$ theories to the field theory computations of sec.~\ref{sec:XNN-loc}.
Antisymmetric Wilson loops are mapped to antisymmetric Wilson loops.
Using the action of $SL(2,\RR)$ on the $(p,q)$ string charges of the D3-brane and the identification of these charges with the field theory parameters for the two solutions leads to the relation
\begin{align}\label{eq:X-plus-W}
  \ln \langle W_\wedge^{X_{N}}(z,\mathds{k})\rangle&=2\ln\big\langle W_\wedge^{+^{}_{N,N}}\big(\mathds{k}(1-|1-2z|),(1-\mathds{k})(1-|1-2z|)\big)\big\rangle~.
\end{align}
It is satisfied by the expectation values (\ref{eq:W-wedge-XNN}) and (\ref{eq:plus-Wwedge}).

\section{Discussion}\label{sec:discussion}

Expectation values were obtained for $\tfrac{1}{2}$-BPS loop operators in 5d SCFTs with relevant deformations to long quiver gauge theories, using field theory methods and AdS/CFT.
The large number of Wilson loops in long quiver gauge theories makes them an excellent tool to study these theories in detail.
The results connect the local form of the internal space in the supergravity solutions
to the internal structure of the 5-brane webs realizing the 5d SCFTs and their gauge theory deformations, 
and to the local features of the saddle points of the field theory matrix models.

\smallskip

Wilson loops in (anti)fundamental and antisymmetric representations associated with individual nodes of the gauge theories were studied using supersymmetric localization.
The expectation values for large antisymmetric representations are uniformly given in terms of Bloch-Wigner functions,
whose complex arguments combine an effectively continuous coordinate $z\in[0,1]$ along the quiver, specifying the gauge node, and a parameter $\mathds{k}\in[0,1]$ specifying the number of boxes in the Young tableau as fraction of the maximal number for a given node.
For the $+_{N,M}$ and $T_N$ theories, which are, respectively, the UV fixed points of the quiver gauge theories (\ref{eq:D5NS5-quiver}) and (\ref{TNquiver}), the results are
\begin{align*}
 +_{N,M}:&&\ln\big\langle W_{\wedge}(z,\mathds{k})\big\rangle&=
 \frac{3}{\pi}NM\left[
 D(u)+D(1+u)\right], 
 \qquad u=e^{i\pi z+\sinh^{-1}\left(\cot\left(\pi \mathds{k}\right)\sin(\pi z)\right)}~,
\nonumber\\[1mm]
 T_N:&&\ln\left\langle W_{\wedge}(z,\mathds{k})\right\rangle
 &=\frac{3}{\pi}N^2 D\big(e^{i\pi (1-z)}\sin(\pi(1-\mathds{k})z)\csc(\pi \mathds{k} z)\big)\,,
\end{align*}
where $D$ is the single-valued Bloch-Wigner function $D(u)=\Im\left(\Li_2(u)\right)+\arg(1-u)\ln |u|$.
For the $Y_N$ and $X_{N}$ theories, whose gauge theory deformations (\ref{eq:YN-quiver-2}) and (\ref{eq:XNN-quiver}) have a central node with $N_f<2N_c$ and for $Y_N$ also a Chern-Simons term, the results are, with $\hat z=(1-|1-2z|)/2$,
\begin{align*}
   Y_N:&&\ln\langle W_{\wedge}(z,\mathds{k})\rangle &= 
   \frac{6}{\pi}N^2\, D\big(1+e^{2\pi i (\mathds{k}-1)\hat z}\sin(\pi \mathds{k}\hat z)\csc(\pi(\mathds{k}-2)\hat z)\big)\,,
 \nonumber\\[1mm]
  X_{N}:&&\ln\big\langle W_{\wedge}(z,\mathds{k})\big\rangle&=
  \frac{6}{\pi}N^2\left[ D(u)+D(1+u) \right],
 \qquad
  u=e^{ 2\pi i \mathds{k}\hat z - \sinh^{-1}\left(\sin(2\pi\mathds{k}\hat z)\cot(2\pi(1-\mathds{k})\hat z)\right) }\,.
\end{align*}
The expectation values scale quadratically with $N$ (and $M$), which is steeper than the $N^{3/2}$ scaling found for the 5d $USp(N)$ theories in \cite{Assel:2012nf}.
This is in line with the steeper scaling of the free energies, and can be understood from the matrix model perspective from the linear scaling of the majority of eigenvalues.
For all four theories the logarithms of the expectation values are positive for $(z,k)$ in the interior of the square $[0,1]^2$ and vanish on its boundary.
Their behavior as the boundary is approached varies, depending on the gauge node under consideration. 
This reflects that the behavior of fundamental and anti-fundamental Wilson loops, which are recovered for small $\mathds{k}$ and $1-\mathds{k}$, depends on the type of gauge node they are associated with.
The $X_N$ and $Y_N$ theories have (anti)fundamental Wilson loops with the naively expected linear scaling, at interior nodes with $N_f<2N_c$.
But the majority of (anti)fundamental Wilson loops exhibits a logarithmically enhanced scaling, which from the matrix model perspective originates in a small number of the largest/smallest eigenvalues having an enhanced scaling compared to the bulk of the eigenvalues.

\smallskip

The field theory results were reproduced exactly from the supergravity duals, which describe the near-horizon limit of 5-brane webs that realize the gauge theories and their UV fixed points. 
The geometry of the supergravity solutions is a warped product of $AdS_6$ and $S^2$ over a Riemann surface $\Sigma$, and each solution is defined by a pair of locally holomorphic functions $\cA_\pm$ on $\Sigma$.
Antisymmetric Wilson loops are represented by a two-parameter family of D3-branes for each solution, one for each point of $\Sigma$. 
The D3-branes carry $(p,q)$-string charges, which are given by the real and imaginary parts of $\cA_+ +\bar\cA_-$.
These charges were shown to carve out the grid diagram of the 5-brane web associated with the supergravity solution. 
They identify for each point of $\Sigma$ a face in the associated 5-brane web.
The internal structure of the 5-brane web and the data characterizing the gauge theory are thus directly encoded in each of the supergravity solutions.
For the D3-branes this identifies for each embedding the face which the D3-brane is located in in the brane web, 
which in turn identifies the representation and gauge node for the Wilson loop.
The expectation values obtained from the D3-brane actions with these identifications reproduce the field theory results exactly.

\smallskip

Fundamental and anti-fundamental Wilson loops were understood from the supergravity perspective in terms of D3-branes approaching boundary points of $\Sigma$ and in terms of fundamental strings embedded at the limiting points. 
The operators with linear scaling are identified with fundamental strings at regular boundary points, while those with non-linear scaling correspond to strings probing the points where 5-branes emerge.
The identification offers a field theory understanding of the points where $(p,q)$ 5-branes emerge in the supergravity solutions, and it would be interesting to explore this further.
Families of more general $\tfrac{1}{2}$-BPS $(p,q)$ string embeddings were identified for each solution, 
and the corresponding loop operators were related in various ways to Wilson loops.
It would be interesting to understand these operators more directly in field theory.

\smallskip

The D3-brane action from which Wilson loop expectation values are obtained was found to be given by the function $\cG$ defined in terms of $\cA_\pm$ in (\ref{eq:kappa2-G}).
This function plays a prominent role in the supergravity solutions, in that the geometry is specified entirely by $\cG$ and its derivatives, as can be seen from (\ref{eq:metric-functions}) with $\kappa^2=-\partial_w\partial_{\bar w}\cG$. 
Wilson loops thus provide an efficient way to (re)construct a significant part of the supergravity data from field theory.
It would be interesting to identify a set of field theory observables from which the entire holographic dual can be constructed.

\smallskip

For solutions describing 5-brane junctions involving only $(p,q)$ 5-branes, $\cG$ is generally given by sums of Bloch-Wigner functions, as shown in app.~\ref{sec:cG-gen}. The same is therefore true for the expectation values of antisymmetric Wilson loops in the corresponding gauge theories.
It would be interesting to explore possible geometric interpretations, noting that the Bloch-Wigner function encodes the volume of ideal hyperbolic tetrahedra \cite{Zagier2007}.
It would also be interesting to study theories engineered by 5-brane webs with additional ingredients like $[p,q]$ 7-branes \cite{DeWolfe:1999hj}.
The matrix models and saddle points for several examples were derived in \cite{Uhlemann:2019ypp} and supergravity duals were constructed in \cite{DHoker:2017zwj,Chaney:2018gjc}. 
Quiver theories with gauge nodes other than $SU(\cdot)$ can be realized by incorporating O7-planes into the brane webs \cite{Bergman:2015dpa} and the  corresponding supergravity solutions \cite{Uhlemann:2019lge}.
The expectation values of antisymmetric Wilson loops should contain higher polylogarithms for these theories.

\smallskip

Further interesting extensions include the study of Wilson loops in different representations, possibly involving an $\mathcal O(1)$ fraction of gauge nodes and relating to backreacted solutions along the lines of \cite{Yamaguchi:2006te,Lunin:2006xr,DHoker:2007mci},
and the identification of the operators described by the line defect solution of 6d supergravity \cite{Chen:2019qib} after uplifting to Type IIB \cite{Hong:2018amk,Malek:2018zcz}.
It would also be interesting to explore the defect CFTs defined by the Wilson loops and the corresponding AdS$_2$/CFT$_1$ dualities along the lines of \cite{Giombi:2017cqn}, and to explore long quiver theories in other dimensions, e.g.\ \cite{Apruzzi:2015wna,Nunez:2019gbg,Lozano:2019jza,Lozano:2019zvg,Faedo:2019cvr,vanGorsel:2019sbz,Bergman:2020bvi}.

\let\oldaddcontentsline\addcontentsline
\renewcommand{\addcontentsline}[3]{}

\begin{acknowledgments}
I am grateful to Oren Bergman and Jim Liu for illuminating discussions,
and to the attendants and organizers of the USC online particle physics seminar where part of the results were presented for interesting discussions and comments.
This work was supported, in part, by the US Department of Energy under Grant No.~DE-SC0007859
and by the Leinweber Center for Theoretical Physics.
\end{acknowledgments}

\let\addcontentsline\oldaddcontentsline

\appendix
\renewcommand\theequation{\Alph{section}.\arabic{equation}}

\section{Free energy for \texorpdfstring{$X_{N}$}{X[N]}}\label{sec:XNN-FS5}

In this appendix the saddle point eigenvalue distributions for the $X_{N}$ theories are derived,
and the free energy is computed.
The matrix model is given by (2.27) of \cite{Uhlemann:2019ypp}, and the notation and conventions of that reference will be used.
The quiver (\ref{eq:XNN-quiver}) is characterized by $L=2N$ and
\begin{align}
 N(z)&=2N(1-|1-2z|)~, 
\end{align}
and no Chern-Simons terms or fundamental flavors. This leads to
\begin{align}
 \cF_{X_{N}}&=L^2\int_0^1 dz \cL + 2NL^3\left[\mu_0\left(\int dx \hat\rho\left(\tfrac{1}{2},x\right)-1\right)+\tau_0 \int dx\,x\hat\rho\left(\tfrac{1}{2},x\right)\right]~.
\end{align}
The saddle point configuration is given by a harmonic function $\varrho(z,x)=N(z)\hat\rho(z,x)$ which satisfies 
$\varrho(0,x)=\varrho(1,x)=0$
and the junction condition (2.47) of \cite{Uhlemann:2019ypp}, which for this theory becomes
\begin{align}\label{eq:XNN-junction}
 \int dy \left[\partial_z\varrho(z,y)\right]_{z=\tfrac{1}{2}-\epsilon}^{z=\tfrac{1}{2}+\epsilon}F_H(x-y) +L\mu_0+L\tau_0 x&=0~.
\end{align}
Since the discontinuity in $N(z)$ at $z=\frac{1}{2}$ is not balanced by Chern-Simons terms or flavor contributions,
the junction condition leads to the following constraints,
\begin{align}
 \varrho\left(\tfrac{1}{2},x\right)&=0 \qquad \text{for $x\notin (x_0,x_1)$}~, 
 \nonumber\\
 \partial_z\varrho\left(\tfrac{1}{2},x\right)&=0 \qquad \text{for $x\in (x_0,x_1)$}~,
\end{align}
with some $x_0$ and $x_1$ to be determined. Due to the symmetry of the theory under charge conjugation, $x_0=-x_1$.
The function $\varrho$ can be constructed in analogy to the one for the $\pslash$ theory in sec.~4 of \cite{Uhlemann:2019ypp}.
This leads to
\begin{align}
 \varrho&=\frac{a(1-v)}{\sqrt{-v}}+\mathrm{c.c.} & v&=\frac{u e^{4\pi x_1}+1}{u+e^{4\pi x_1}}~, & u&=e^{4\pi x+2\pi i z}~.
\end{align}
Normalization fixes
\begin{align}
 a&=N\csch(2\pi x_1)~.
\end{align}
The $SU(N)$ constraint is satisfied automatically. Demanding terms quadratic in $x$ on the left of (\ref{eq:XNN-junction}) to vanish leads to 
\begin{align}
 2\pi x_1&=\cosh^{-1}\sqrt{2}~.
\end{align}
This leads to the saddle point configuration
\begin{align}
\label{eq:XNN-rho-app}
 \hat\rho_s(z,x)&=
 \frac{2N}{N(z)}
 \frac{1}{\sqrt{1-2\coth^2 (2 \pi  x+i \pi  z)}}+\mathrm{c.c.}
\end{align}
The junction condition (\ref{eq:XNN-junction}) is satisfied with
\begin{align}
 \mu_0&=\frac{7\zeta(3)N}{16\pi^2L}~.
\end{align}
With the saddle point eigenvalue distributions in hand, the free energy on squashed spheres can be obtained in parallel to the examples discussed in \cite{Uhlemann:2019ypp}, which leads to
\begin{align}
 F_{\vec{\omega}}&=\frac{7}{4\pi^2}\frac{\omega_{\rm tot}^3}{\omega_1\omega_2\omega_3}\zeta(3)N^4~.
\end{align}
This is four times the free energy of the $+_{N,N}$ theory, as expected from their $SL(2,\RR)$ relation at large $N$ (see sec.~\ref{sec:X-sol}).

\section{\texorpdfstring{$(p,q)$}{(p,q)} strings and D3-branes in \texorpdfstring{$AdS_6\times S^2\times\Sigma$}{AdS6xS2xSigma}}

In this appendix the supersymmetry conditions for the probe brane and string embeddings are discussed.
The preserved supersymmetries are generated by Killing spinors $\epsilon$ that satisfy \cite{Bergshoeff:1996tu, Cederwall:1996pv,Cederwall:1996ri}
\begin{align}\label{eq:D3-kappa-0}
 \Gamma_\kappa\epsilon&=\epsilon~.
\end{align}
In the following the background Killing spinors as derived in \cite{DHoker:2016ujz} are reviewed and then the BPS conditions and on-shell actions for $(p,q)$ strings and D3-branes are discussed.

A general Killing spinor is expanded in a basis of $AdS_6\times S^2$ Killing spinors as
\begin{align}
 \epsilon&=\sum_{\eta_1\eta_2}\chi^{\eta_1\eta_2}\otimes\zeta_{\eta_1\eta_2}~,
 &
 \chi^{\eta_1\eta_2}&=\epsilon_{AdS_6}^{\eta_1}\otimes \epsilon_{S^2}^{\eta_2}~,
\end{align}
where $\eta_1,\eta_2=\pm$. 
The explicit forms of the coefficient spinors $\zeta_{\eta_1\eta_2}$ are 
\begin{align}
 \zeta_{++}&=\begin{pmatrix} \bar\alpha\\ \beta \end{pmatrix}~,
 &
 \zeta_{--}&=\begin{pmatrix} -\bar\alpha\\ \beta \end{pmatrix}~,
 &
 \zeta_{\eta_1,-}&=i\nu \eta_1 \zeta_{\eta_1,+}~.
\end{align}
The complex conjugate spinor is given by
\begin{align}
 \cB^{-1}\epsilon^\star&=-i\sum_{\eta_1\eta_2}\chi^{\eta_1\eta_2}\otimes \eta_2\sigma_2\zeta_{\eta_1,-\eta_2}^\star~.
\end{align}
Thus,
\begin{align}
 \label{eq:eps}
 \epsilon&=\sum_{\eta}\left(\chi^{\eta,+}+i\nu \eta\chi^{\eta,-}\right)\otimes \zeta_{\eta,+}~,
 \\
 \cB^{-1}\epsilon^\star&=i\sum_{\eta}\left(\chi^{\eta,-}+i\nu \eta\chi^{\eta,+}\right)\otimes \sigma_2\zeta_{\eta,+}^\star~.
 \label{eq:eps-star}
\end{align}

The explicit form of the $S^2$ and $AdS_6$ Killing spinors has been derived in app.~B.1 and B.2 of \cite{DHoker:2016ujz}.
Metric and Killing spinors are, with constant spinors $\epsilon_{S^2,0}^{\eta_2}$, $\epsilon^{\eta_1}_{AdS_6,0}$,
\begin{align}\label{eq:S2-Killing}
 ds^2_{S^2}&=d\theta_2^2+\sin^2\!\theta_2\,d\theta_1^2~,
 &
 \epsilon_{S^2}^{\eta_2}&=\exp\left(\frac{i\eta_2}{2}\theta_2\sigma_2\right)\exp\left(-\frac{i}{2}\theta_1\sigma_3\right)\epsilon_{S^2,0}^{\eta_2}~,
\nonumber\\
ds^2_{AdS_6}&=dr^2+e^{2r}dx^\mu dx_\mu~,
 &
 \epsilon_{AdS_6}^{\eta_1}&=e^{\frac{\eta_1}{2}r\gamma_r}\left(1+\frac{1}{2}x^\mu\gamma_\mu\left(\eta_1-\gamma_r\right)\right)\epsilon^{\eta_1}_{AdS_6,0}~.
\end{align}

\let\oldaddcontentsline\addcontentsline
\renewcommand{\addcontentsline}[3]{}

\subsection{\texorpdfstring{$(p,q)$}{(p,q)} strings}\label{sec:F1-BPS}

The BPS condition for general $(p,q)$ strings can be derived by starting with a fundamental string, for which $\Gamma_\kappa=\Gamma_{(0)}\otimes \sigma^3$ \cite{Wulff:2016tju}, and then deducing the general form using $SL(2,\RR)$.
A phase has to be incorporated to accommodate the $SU(1,1)$ conventions, as discussed in \cite{Gutperle:2018vdd}.
Using the convention for complex notation of \cite{Karch:2015vra} the preserved supersymmetries are singled out by
\begin{align}\label{eq:theta-k}
 \Gamma_\kappa\epsilon&=e^{i\theta_\kappa}\Gamma_{(0)}\cB^{-1}\epsilon^\star~,
 &
 e^{2i\theta_\kappa}&=\frac{1+B}{1+\bar B}~.
\end{align}
For an embedding where the F1 wraps Poincar\'e AdS$_2$ with coordinates $(r,x^0$) in AdS$_6$,
\begin{align}
 \Gamma_{(0)}&=\Gamma^0\Gamma^1=\gamma_r\gamma_0\otimes \id_2\otimes \id_2~.
\end{align}
For $x^1=\ldots=x^4=0$, the AdS$_6$ Killing spinor reduces to
\begin{align}
 \epsilon_{AdS_6}^{\eta_1}&=e^{\frac{\eta_1}{2}r\gamma_r}\left(1+\frac{1}{2}x^0\gamma_0\left(\eta_1-\gamma_r\right)\right)\epsilon^{\eta_1}_{AdS_6,0}~.
\end{align}
Anticipating the form of the preserved supersymmetries, one can introduce a projector on the constant $AdS_6$ spinor such that
\begin{align}\label{eq:F1-proj}
 \gamma_r\gamma_0\,\epsilon^{\eta_1}_{0,AdS_6}&=\lambda \epsilon^{\eta_1}_{0,AdS_6}
 &
 &\Rightarrow&
 \gamma_r\gamma_0 \epsilon_{AdS_6}^{\eta_1} &= \lambda \epsilon_{AdS_6}^{-\eta_1}~,
\end{align}
with $\lambda^2=1$.
With the expression in (\ref{eq:eps-star}) this leads to
\begin{align}
 \Gamma_\kappa\epsilon&=i\lambda e^{i\theta_\kappa}\sum_\eta\left(\chi^{-\eta,-}+i\nu\eta\chi^{-\eta,+}\right)\otimes \sigma_2\zeta_{\eta,+}^\star
 \nonumber\\
 &=\sum_{\eta}\left(\chi^{\eta,+}+i\nu\eta\chi^{\eta,-}\right)\otimes \lambda\nu\eta e^{i\theta_\kappa}\sigma_2\zeta_{-\eta,+}^\star~,
\end{align}
where $\eta$ has been switched to $-\eta$ in the summation for the second line. Comparing to the expression for $\epsilon$ in (\ref{eq:eps}), 
the $\kappa$-symmetry constraint becomes
\begin{align}\label{eq:kappa-F1-5}
 0&=\Gamma_\kappa\epsilon-\epsilon=
 \sum_{\eta}\left(\chi^{\eta,+}+i\nu\eta\chi^{\eta,-}\right)\otimes\left(\lambda \nu\eta e^{i\theta_\kappa}\sigma_2\zeta_{-\eta,+}^\star - \zeta_{\eta,+}\right)~.
\end{align}
This expression vanishes without additional projectors if
\begin{align}
 \lambda \nu\eta e^{i\theta_\kappa}\sigma_2\zeta_{-\eta,+}^\star &= \zeta_{\eta,+}~.
\end{align}
The two equations for $\eta=\pm$ are conjugate to each other.
Moreover, the two components within each equation are conjugates of each other. 
This leads to the single  complex condition
\begin{align}\label{eq:susy-F1-0}
 \lambda e^{i\theta_\kappa}\alpha+\beta&=0~.
\end{align}
It constrains the location of the F1 on $\Sigma$.
From eq.~(4.9) of \cite{DHoker:2016ujz},
\begin{align}\label{eq:alphabeta-def}
 \rho\alpha^2&= f(\partial_{\bar w}\bar\cA_++\bar B\partial_{\bar w}\bar\cA_-)~,
 &
 \rho\beta^2&= f(B \partial_{\bar w}\bar\cA_++\partial_{\bar w}\bar\cA_-)~,
\end{align}
with $f^{-2}=1-|B|^2$.
The condition (\ref{eq:susy-F1-0}) becomes
\begin{align}\label{eq:susy-F1}
 \lambda \left(\frac{1+B}{1+\bar B}\right)^{1/2}\left(\frac{\partial_{\bar w}\bar\cA_++\bar B\partial_{\bar w}\bar\cA_-}{B\partial_{\bar w}\bar\cA_+ +\partial_{\bar w}\bar\cA_-}\right)^{1/2}
 &=-1~.
\end{align}
This is the BPS condition for fundamental strings wrapping AdS$_2$ to preserve half of the supersymmetries.
Upon squaring and multiplying by the denominators, (\ref{eq:susy-F1}) leads to
\begin{align}\label{eq:BPS-F1-2}
 (1-B\bar B)(\partial_{\bar w}\bar\cA_+-\partial_{\bar w}\bar\cA_-)&=0~.
\end{align}
The first factor is non-zero away from the poles.
For generic points on $\Sigma$ the condition thus is
\begin{align}\label{eq:BPS-F1-3}
 \partial_w\cA_+=\partial_w\cA_-~.
\end{align}
This condition can only be satisfied on the boundary of $\Sigma$, since it implies $\kappa^2=0$.
Using it in (\ref{eq:susy-F1}) fixes $\lambda$.
At the poles $f_2$ and $f_6$ vanish (sec.~3.9.1 of \cite{DHoker:2017mds}), which with (\ref{eq:f2f6-alphabeta}) implies that $\alpha$ and $\beta$ both vanish. The condition (\ref{eq:susy-F1-0}) is thus formally satisfied at the poles.
The condition (\ref{eq:susy-F1}) can also be evaluated at the poles with the expressions derived in sec.~3.9 of \cite{DHoker:2017mds}, which shows that it is satisfied as well.

We now evaluate the on-shell action (\ref{eq:S-pq-gen}) more explicitly for $\frac{1}{2}$-BPS strings. For a fundamental string it reduces to
\begin{align}
 S_{(1,0)}&=-T \Vol_{AdS_2} e^\phi f_6^2~.
\end{align}
Noting that
\begin{align}
 e^{2\phi}&=\frac{\left|(T+1)\partial_{\bar w}\cG \left(\partial_w\cA_+-\partial_w\cA_-\right)+(T-1)\partial_w\cG\left(\partial_{\bar w}\bar \cA_+ -\partial_{\bar w}\bar \cA_-\right)\right|^2 }{4\kappa^2 |\partial_w\cG|^2 T}~,
\end{align}
and using that $T\rightarrow\infty$ at the boundary, the fundamental string action at a point on the boundary 
where (\ref{eq:BPS-F1-3}) is satisfied reduces to 
\begin{align}\label{eq:SF1-bps}
 S_{(1,0)}&=-T \Vol_{AdS_2} \left|\cA_+-\bar\cA_-+\bar\cA_+-\cA_-\right|~.
\end{align}

The BPS condition for generic $(p,q)$-strings can be obtained by an $SL(2,\RR)$ transformation. 
In (3.13) of \cite{DHoker:2017zwj} an $SL(2,\RR)$ matrix $Q$ was used to transform a D7-brane into a $[p,q]$ 7-brane, by transforming the monodromy matrix as follows,
\begin{align}\label{eq:Q}
 M_{[p,q]}&=Q M_{[1,0]} Q^{-1}~,
 &
 Q&=\begin{pmatrix}p & -q/(p^2+q^2)\\[1mm]q & p/(p^2+q^2)\end{pmatrix}~.
\end{align}
$Q$ transforms a fundamental string into a $(p,q)$ string.
The differentials $\partial_w\cA_\pm$ transform under $SU(1,1)$, with parameters related to the $SL(2,\RR)$ parameters $a,b,c,d$, as follows ((2.12) of \cite{DHoker:2017zwj}),
\begin{align}\label{eq:dApm-SU11}
 \begin{pmatrix}\partial_w\cA_+  \\ \partial_w\cA_- \end{pmatrix}
 &\rightarrow
 \begin{pmatrix}\partial_w\cA_+^\prime  \\ \partial_w\cA_-^\prime \end{pmatrix}=
 \begin{pmatrix} u & -v \\ -\bar v & \bar u\end{pmatrix}
 \begin{pmatrix}\partial_w\cA_+  \\ \partial_w\cA_- \end{pmatrix}~,
&&
\begin{matrix}
 u=\frac{1}{2}(+a+ib-ic+d)
 \\
 v=\frac{1}{2}(-a+ib+ic+d)
\end{matrix}~.
\end{align}
The BPS condition for generic $(p,q)$ strings can be obtained from the condition (\ref{eq:BPS-F1-3}) for the fundamental string
by transforming $(\partial_w\cA_+,\partial_w\cA_-)$ with the $SU(1,1)$ matrix corresponding to $Q$, which leads to
\begin{align}\label{eq:BPS-pq-string-app}
 \left(p+iq\right)\partial_w\cA_+&=\left(p-iq\right)\partial_w\cA_-~.
\end{align}
The on-shell action can similarly be obtained by $SU(1,1)$ transforming the result in (\ref{eq:SF1-bps}).
The $SU(1,1)$ action on the differentials $\partial_w\cA_\pm$ in general lifts to an $SU(1,1)\otimes \cC$ action on the functions $\cA_\pm$.
However, from (5.16) of \cite{DHoker:2016ujz} it follows that $\cA_+-\bar\cA_-$ is invariant under the shifts. Thus,
\begin{align}
 S_{(p,q)}&=-T\Vol_{AdS_2}\left|(p+iq)(\cA_+-\bar \cA_-)+(p-iq)(\bar\cA_+-\cA_-)\right|~.
\end{align}

\subsection{D3-brane BPS conditions}\label{sec:D3-susy}

We now discuss supersymmetry of D3-brane embeddings, using again the conventions of \cite{Karch:2015vra} for the form of $\Gamma_\kappa$
and the phase $e^{i\theta_\kappa}$ of \cite{Gutperle:2018vdd}, defined in (\ref{eq:theta-k}). This leads to
\begin{align}
 \Gamma_\kappa\epsilon &= \frac{-i}{\sqrt{\det(1+X)}}\left(
 \Gamma_{(0)}\epsilon 
 -\frac{1}{2}\gamma^{ij}X_{ij}\Gamma_{(0)}e^{i\theta_\kappa}C\epsilon^\star
 +\frac{1}{8}\gamma^{ijkl}X_{ij}X_{kl}\Gamma_{(0)}\epsilon\right)~,
\end{align}
where $X^{i}_{\hphantom{i}j}=g^{ik}\mathcal F_{kj}$ with $\mathcal F=F-B_2$.
For a D3-brane wrapping $AdS_2$ and $S^2$,
with flux along $AdS_2$ and $S^2$,
\begin{align}
 \Gamma_{(0)}&=\Gamma^{0167}~,
 &
 \frac{1}{2}\gamma^{ij}X_{ij}&=\Gamma^{01}X_{\underline{01}}+\Gamma^{67}X_{\underline{67}}~,
 &
 \frac{1}{8}\gamma^{ijkl}X_{ij}X_{kl}&=\Gamma^{0167}X_{\underline{01}}X_{\underline{67}}~.
\end{align}
This leads to
\begin{align}
 \Gamma_\kappa\epsilon &= \frac{-i}{\sqrt{\det(1+X)}}\left(
  \Gamma_{(0)}\epsilon  -X_{\underline{01}}X_{\underline{67}}\epsilon
  -e^{i\theta_\kappa}\left(\Gamma^{67}X_{\underline{01}}-\Gamma^{01}X_{\underline{67}}\right) C\epsilon^\star
 \right)~.
\end{align}
The $\kappa$-symmetry constraint becomes
\begin{align}
 \Gamma_{(0)}\epsilon-e^{i\theta_\kappa}\left(\Gamma^{67}X_{\underline{01}}-\Gamma^{01}X_{\underline{67}}\right) C\epsilon^\star
 &= h\epsilon~,
\end{align}
where
\begin{align}\label{eq:D3-h-def}
 h&\equiv i\sqrt{\det(1+X)}+X_{\underline{01}}X_{\underline{67}}~,
 &
 \det\left(1+X\right)&=\left(1-X_{\underline{01}}^2\right)\left(1+X_{\underline{67}}^2\right)~.
\end{align}
Similarly to the F1 discussion and notation we find
\begin{align}
 \Gamma^{01}&=\gamma_r\gamma_0\otimes \id_2\otimes\id_2~,
 &
 \Gamma^{67}&=i\,\id_8\otimes \sigma^3\otimes \id_2~,
 &
 \Gamma_{(0)}&=i\gamma_r\gamma_0\otimes \sigma^3\otimes \id_2~.
\end{align}
The D3 should preserve the same supersymmetries as the F1, singled out by the projector (\ref{eq:F1-proj}).
This leads to
\begin{align}
 \Gamma_{(0)}\epsilon&=i\lambda\sum_\eta\left(\chi^{-\eta,-}+i\nu\eta\chi^{-\eta,+}\right)\otimes \zeta_{\eta,+}~,
 \nonumber\\
 \Gamma^{67}C\epsilon^\star&=-\sum_\eta\left(\chi^{\eta,+}+i\nu\eta\chi^{\eta,-}\right)\otimes\sigma_2\zeta_{\eta,+}^\star~,
 \nonumber\\
 \Gamma^{01}C\epsilon^\star&=i\lambda \sum_\eta\left(\chi^{-\eta,-}+i\nu\eta\chi^{-\eta,+}\right)\otimes \sigma_2\zeta_{\eta,+}^\star~.
\end{align}
The $\kappa$-symmetry condition becomes
\begin{align}
 \sum_\eta\left(\chi^{\eta,+}+i\nu\eta\chi^{\eta,-}\right)\otimes
 \left[
 \nu\eta\lambda \zeta_{-\eta,+}
 +e^{i\theta_\kappa}X_{\underline{01}}\sigma_2\zeta_{\eta,+}^\star
 +e^{i\theta_\kappa}\nu\eta\lambda X_{\underline{67}}\sigma_2\zeta_{-\eta,+}^\star-h\zeta_{\eta,+}\right]&=0~.
\end{align}
The conditions for a $\tfrac{1}{2}$-BPS embedding thus are
\begin{align}
 \nu\eta\lambda \zeta_{-\eta,+}
 +e^{i\theta_\kappa}X_{\underline{01}}\sigma_2\zeta_{\eta,+}^\star
 +e^{i\theta_\kappa}\nu\eta\lambda X_{\underline{67}}\sigma_2\zeta_{-\eta,+}^\star-h\zeta_{\eta,+}
 &=0~.
\end{align}

These conditions will now be evaluated further, and solved for $X_{\underline{01}}$ and $X_{\underline{67}}$.
The conditions for $\eta=\pm$ are equivalent to each other. For $\eta=+$ the conditions become
\begin{align}
 0&=-(i\lambda+h) \bar\alpha -e^{i\theta_\kappa}\left(iX_{\underline{01}}+\lambda X_{\underline{67}}\right)\bar\beta ~,
 \nonumber\\
 0&=(i\lambda-h)\beta +e^{i\theta_\kappa}\left(iX_{\underline{01}}-\lambda X_{\underline{67}}\right)\alpha~.
 \label{eq:D3-kappa-3}
\end{align}
Multiplying the first equation by $-i\beta e^{-i\theta_\kappa}$ and the second by $i\bar\alpha e^{-i\theta_\kappa}$ and then taking the sum leads to
\begin{align}
 2\lambda e^{-i\theta_\kappa}\bar\alpha\beta&
 =-(\alpha\bar\alpha+\beta\bar\beta)X_{\underline{01}}
 +i\lambda(\beta\bar\beta-\alpha\bar\alpha)X_{\underline{67}}~.
\end{align}
The Killing spinor components can be related to $f_2$ and $f_6$ by eq.~(3.46) of \cite{DHoker:2016ujz},
\begin{align}\label{eq:f2f6-alphabeta}
 f_6&=\alpha\bar\alpha+\beta\bar\beta~,
 &
 3\nu f_2&=\beta\bar\beta-\alpha\bar\alpha~.
\end{align}
This leads to
\begin{align}
 2\lambda e^{-i\theta_\kappa}\bar\alpha\beta&
 =-f_6 X_{\underline{01}}+3i\nu\lambda f_2X_{\underline{67}}~.
\end{align}
Taking the real and imaginary parts of this equation leads to
\begin{align}\label{eq:D3-kappa-6}
 X_{\underline{01}}&=-\frac{2\lambda}{f_6}\Re\left(e^{-i\theta_\kappa}\bar\alpha\beta\right)~,
 &
 X_{\underline{67}}&=\frac{2\nu}{3f_2}\Im\left(e^{-i\theta_\kappa}\bar\alpha\beta\right)~.
\end{align}
This fixes the fluxes in terms of the position of the D3-brane on the Riemann surface.

The assignment (\ref{eq:D3-kappa-6}) solves one linear combination of the conditions in (\ref{eq:D3-kappa-3}).
It remains to check the remaining linear combination of the equations in (\ref{eq:D3-kappa-3}), which is
\begin{align}
 2ih\bar\alpha\beta e^{-i\theta_\kappa}-3\nu f_2 X_{\underline{01}}+i\lambda f_6 X_{\underline{67}}&=0~.
\end{align}
Using the expressions for $X_{\underline{01}}$ and $X_{\underline{67}}$ in (\ref{eq:D3-kappa-6}), it becomes
\begin{align}
 3if_2 f_6 h \bar \alpha \beta e^{-i\theta_\kappa}+9\nu\lambda f_2^2\Re\left(\bar\alpha \beta e^{-i\theta_\kappa}\right)+i\nu\lambda f_6^2\Im\left(\bar\alpha \beta e^{-i\theta_\kappa}\right)&=0~.
\end{align}
With (\ref{eq:f2f6-alphabeta}),
\begin{align}\label{eq:D3-kappa-h-cond}
 3if_2 f_6 h +\nu\lambda \left(\beta^2\bar\beta^2+\alpha^2\bar\alpha^2\right)-2\nu\lambda \alpha^2\bar\beta^2 e^{2i\theta_\kappa}&=0~.
\end{align}
To further evaluate this condition $h$ defined in (\ref{eq:D3-h-def}) has to be computed.
From (\ref{eq:D3-h-def}),
\begin{align}
 3f_2f_6h&=3if_2f_6\sqrt{\Big(1-X_{\underline{01}}^2\Big)\left(1+X_{\underline{67}}^2\right)}+3f_2f_6X_{\underline{01}}X_{\underline{67}}
 \nonumber\\
 &=
 i\sqrt{\left(f_6^2-4\Re\big(\bar\alpha \beta e^{-i\theta_\kappa}\big)^2\right)\left(9f_2^2+4\Im\big(\bar\alpha \beta e^{-i\theta_\kappa}\big)^2\right)}
 -4\lambda\nu\Re\big(\bar\alpha \beta e^{-i\theta_\kappa}\big)\Im\big(\bar\alpha \beta e^{-i\theta_\kappa}\big).
\end{align}
Using again the expressions for $f_2$ and $f_6$ in (\ref{eq:f2f6-alphabeta}) shows that the two factors in the square root are identical and uniformize the square root, leading to
\begin{align}
 3f_2f_6h&=
 i\left(\beta^2\bar\beta^2+\alpha^2\bar\alpha^2-\bar\alpha^2\beta^2 e^{-2i\theta_\kappa}-\alpha^2\bar\beta^2e^{2i\theta_\kappa}\right)
 +i\lambda\nu\left(\bar\alpha^2\beta^2 e^{-2i\theta_\kappa}-\alpha^2\bar\beta^2e^{2i\theta_\kappa}\right).
\end{align}
Using this expression for $h$ in the remaining condition (\ref{eq:D3-kappa-h-cond}) shows that it is satisfied if 
\begin{align}\label{eq:lambda-nu}
 \lambda&=\nu~.
\end{align}
Thus, the position of the D3-brane is not constrained by supersymmetry. But the fluxes have to be chosen as in (\ref{eq:D3-kappa-6}).

A relation to the BPS conditions for the probe D7-branes discussed in \cite{Gutperle:2018vdd} arises upon fixing the electric flux on the D3 and reading the resulting condition as a constraint on the position of the D3. Concretely, setting $X_{\underline{01}}=0$ leads, via (\ref{eq:D3-kappa-6}), to the constraint that $\bar\alpha \beta e^{-i\theta_\kappa}$ be imaginary. This is equivalent to the BPS condition for D7-branes in (3.27) of \cite{Gutperle:2018vdd}.

\subsection{D3-brane action and charge}\label{sec:D3-action-charge}

\let\addcontentsline\oldaddcontentsline

The on-shell action and charges for the $\frac{1}{2}$-BPS D3-brane are now evaluated, starting from the action in (\ref{eq:S-D3-gen}) and using the fluxes as given in (\ref{eq:D3-kappa-6}).
The notation of app.~\ref{sec:D3-susy} corresponds to
\begin{align}
 \mathfrak{f}_{\rm el}&=e^\phi f_6^2X_{\underline{01}}~, &
 \mathfrak{f}_{\rm m}-\Re(\cC)&=e^\phi f_2^2  X_{\underline{67}}~,
\end{align}
where, from (\ref{eq:D3-kappa-6}) with (\ref{eq:lambda-nu}), 
\begin{align}\label{eq:D3-X01-X67}
 X_{\underline{01}}&=-\frac{2\nu}{f_6}\Re\left(e^{-i\theta_\kappa}\bar\alpha\beta\right)~,
 &
 X_{\underline{67}}&=\frac{2\nu}{3f_2}\Im\left(e^{-i\theta_\kappa}\bar\alpha\beta\right)~.
\end{align}
This leads to the expressions in (\ref{eq:D3-fluxes}).
The expressions for the Killing spinor components $\alpha$, $\beta$  are in (\ref{eq:alphabeta-def}).
$\theta_\kappa$ is defined in terms of $B$ in (\ref{eq:theta-k}).

The quantization condition for the flux on $S^2$ and the conserved charge density associated with $\mathfrak{f}_{\rm el}$ integrated over $S^2$, as defined in (\ref{eq:ND1}) and (\ref{eq:NF1}), are given by
\begin{align}
 N_{\rm D1}&=2\left(\Re(\cC)+e^\phi f_2^2  X_{\underline{67}}\right)~,
 \\
  N_{\rm F1}&=
  -T_{\rm D3}V_{S^2}\left[\frac{1+X_{\underline{67}}^2}{L_{\rm DBI}}\,e^{-2\phi} \mathfrak{f}_{\rm el} f_2^4+\Im(\cC)+\chi e^{\phi}f_2^2 X_{\underline{67}}\right].
\end{align}
The DBI Lagrangian is given by
\begin{align}
 L_{\rm DBI}&=f_6^2 f_2^2\sqrt{\left(1-X_{\underline{01}}^2\right)\left(1+X_{\underline{67}}^2\right)}~.
\end{align}
As shown in app.~\ref{sec:D3-susy}, the two factors in the square root are related such that they form a square,
\begin{align}\label{eq:X01-X67-DBI}
 f_6^2\left(1-X_{\underline{01}}^2\right)&=9f_2^2\left(1+X_{\underline{67}}^2\right)
 ~.
\end{align}
This leads to
\begin{align}\label{eq:LDBI-sim}
 L_{\rm DBI}&=3f_6f_2^3\left(1+X_{\underline{67}}^2\right)~.
\end{align}
Consequently, 
\begin{align}\label{eq:NF1-app}
 N_{\rm F1}&=-T_{\rm D3}{\rm Vol}_{S^2}\left[\frac{1}{3}e^{-\phi} f_2 f_6X_{\underline{01}}+\chi e^\phi f_2^2 X_{\underline{67}}+\Im(\cC)\right].
\end{align}

The Legendre transform of $S_{\rm D3}$ is given by
\begin{align}
 S_{\rm D3} - \mathfrak{f}_{\rm el} \frac{\delta S_{\rm D3}}{\delta \mathfrak{f}_{\rm el}}
 &=
T_{\rm D3} {\rm Vol}_{AdS_2}{\rm Vol}_{S^2}\left[
-L_{\rm DBI}+\mathfrak{f}_{\rm el} \left(\Im(\cC)+\chi (\mathfrak{f}_{\rm m}-\Re(\cC))\right)
+\mathfrak{f}_{\rm el} N_{\rm F1}
\right].
\end{align}
The contribution of the WZ term cancels two of the terms in $N_{\rm F1}$.
The DBI Lagrangian is given by (\ref{eq:LDBI-sim}), leading to
\begin{align}
S_{\rm D3} - \mathfrak{f}_{\rm el} \frac{\delta S_{\rm D3}}{\delta \mathfrak{f}_{\rm el}}
&= T_{\rm D3} {\rm Vol}_{AdS_2}{\rm Vol}_{S^2}
f_2f_6\left[
-3f_2^2\left(1+X_{\underline{67}}^2\right)
-\frac{1}{3} f_6^2 X_{\underline{01}}^2
\right]
\nonumber\\
&=-\frac{2}{3}T_{\rm D3} {\rm Vol}_{AdS_2}{\rm Vol}_{S^2} \cG~.
\end{align}
For the second line (\ref{eq:X01-X67-DBI}) was used, and the expressions for $f_6^2$ and $f_2^2$ in (\ref{eqn:ansatz}).

It remains to evaluate $N_{\rm F1}$ and $N_{\rm D1}$. 
For $2\pi \alpha^\prime=1$ the overall factor in $N_{\rm F1}$ becomes $T_{\rm D3}\Vol_{S^2}=2$, such that
\begin{align}
 N_{\rm F1}+i N_{\rm D1}
 &=2i\cC-\frac{2}{3}e^{-\phi}f_2f_6 X_{\underline{01}}+2X_{\underline{67}}e^\phi f_2^2(i-\chi)
 \nonumber\\
 &=
 2i\cC+\frac{4}{3}\nu f_2 \left[e^{-\phi}\Re\big(e^{-i\theta_\kappa}\bar\alpha\beta\big)+e^\phi(i-\chi)\Im\big(e^{-i\theta_\kappa}\bar\alpha\beta\big)\right]~.
\end{align}
For the second line the expressions in (\ref{eq:D3-X01-X67}) were used.
This may be further simplified to 
\begin{align}\label{eq:NF-ND-app-0}
 N_{\rm F1}+i N_{\rm D1}
 &=
 2i\cC+\frac{4}{3}\nu f_2 e^\phi\left[-\Im\big(\bar \tau e^{-i\theta_\kappa}\bar\alpha\beta\big)+i\Im\big(e^{-i\theta_\kappa}\bar\alpha\beta\big)\right]~.
\end{align}
From (E.4) of \cite{DHoker:2016ujz},
\begin{align}
 \nu f_2&=\frac{1}{3\rho}\sqrt{\kappa^2\frac{1-R}{1+R}}=\frac{1}{3\rho}\sqrt{\frac{\kappa^2}{T}}~.
\end{align}
The axio-dilaton is given by
\begin{align}
 \tau&=i\,\frac{R\partial_w\cG \partial_{\bar w}(\bar\cA_++\bar\cA_-)-\partial_{\bar w}\cG\partial_w(\cA_++\cA_-)}{R\partial_w\cG\partial_{\bar w}(\bar\cA_+-\bar\cA_-)+\partial_{\bar w}\cG\partial_w(\cA_+-\cA_-)}~,
\end{align}
and with the explicit expressions for $\alpha$, $\beta$, $e^{2i\theta_\kappa}$ and $B$, one finds
\begin{align}
  \rho\bar\alpha\beta e^{-i\theta_\kappa}e^{\phi}&=\sqrt{R}\,\frac{ \partial_{\bar w}\cG\partial_w(\cA_+-\cA_-)+R\partial_w\cG \partial_{\bar w}(\bar\cA_+-\bar\cA_-)}{(R^2-1)|\partial_w\cG|}~.
\end{align}
With these expressions (\ref{eq:NF-ND-app-0}) simplifies to
\begin{align}\label{eq:NF1-ND1-cApm-app}
  N_{\rm F1}+i N_{\rm D1}&=\frac{4}{3}\left(\cA_+ +\bar \cA_-\right)~.
\end{align}

\section{General expression for \texorpdfstring{$\cG$}{cG}}\label{sec:cG-gen}

In this appendix a general expression for the function $\cG$, defined in (\ref{eq:kappa2-G}), for supergravity solutions corresponding to $(p,q)$ 5-brane junctions without 7-branes is derived.
The general form of the functions $\cA_\pm$ was derived in \cite{DHoker:2017mds}, they are given by
\begin{align}\label{eqn:cA}
 \cA_\pm &=\cA_\pm^0+\sum_{\ell=1}^L Z_\pm^\ell \ln(w-r_\ell)~,
 &
 \overline{\cA_\pm^0}&=-\cA_\mp^0~, & \overline{Z_\pm^\ell}&=-Z_\mp^\ell~.
\end{align}
The poles on the real line are at $r_\ell$, with residues given by $Z_\pm^\ell$.
Regular solutions require $\cG$ to be single-valued and positive in the interior of $\Sigma$ and vanishing on the boundary.
These requirements lead to the regularity conditions
\begin{align}
\label{eqn:constr}
 \cA_+^0 Z_-^k - \cA_-^0 Z_+^k 
+ \sum _{\stackrel{\ell=1}{\ell \neq k} }^LZ^{[\ell, k]} \ln |r_\ell - r_k| &=0~,
&k&=1,\cdots,L~,
\end{align}
where $Z^{[\ell, k]}\equiv Z_+^\ell Z_-^k-Z_+^k Z_-^\ell$.
If the sum of the residues vanishes, $\sum_{\ell=1}^L Z_\pm^\ell=0$, the functions $\cA_\pm$ are regular at infinity, and only $L-1$ of the conditions in (\ref{eqn:constr}) are independent.
If the sum of the residues does not vanish there is an additional pole at infinity;
an $SL(2,\RR)$ transformation of the upper half plane can then be used to map all poles to finite points.
The residue $Z_\pm^\ell$ encodes the charges of the $(p,q)$ 5-brane emerging at the pole $r_\ell$.
For a given choice of residues, subject to the constraint that they sum to zero, the conditions in (\ref{eqn:constr}) fix the remaining parameters.

Evaluating the function $\cG$ defined in (\ref{eq:kappa2-G}) for given $\cA_\pm$ involves an additional integration, to determine $\cB$.
We now show that, for generic $\cA_\pm$ as in (\ref{eqn:cA}), such that the conditions in (\ref{eqn:constr}) are satisfied,
$\cG$ is given by
\begin{align}\label{eq:cG-gen}
 \cG&=\sum_{\stackrel{\ell,k=1}{\ell\neq k}}^L 2iZ^{[\ell,k]} D\left(\frac{r_\ell-w}{r_\ell-r_k}\right)~,
\end{align}
where $Z^{[\ell, k]}\equiv Z_+^\ell Z_-^k-Z_+^k Z_-^\ell$ and 
$D$ is the Bloch-Wigner function
\begin{align}
 D(u)&=\Im\left(\Li_2(u)\right)+\arg(1-u)\ln |u|~.
\end{align}
This expression is manifestly real, since $Z^{[\ell,k]}$ is imaginary due to the conjugation relation in (\ref{eqn:cA}).
The sum in (\ref{eq:cG-gen}) may be restricted to $\ell<k$ at the expense of a factor two, noting that $Z^{[\ell,k]}$ is antisymmetric and $D(1-z)=-D(z)$.

\smallskip

To establish (\ref{eq:cG-gen}), we first note that, from the definition of $\cG$ and $\cB$ in (\ref{eq:kappa2-G}),
\begin{align}
 \partial_w\cG&=\left(\bar \cA_+-\cA_-\right)\partial_w\cA_+ + \left(\cA_+-\bar \cA_-\right)\partial_w\cA_-~.
\end{align}
Evaluating this expression for the functions $\cA_\pm$ in (\ref{eqn:cA}), and using the regularity conditions in
(\ref{eqn:constr}) to replace the combination $2\cA_+^0Z_-^k-2\cA_-^0Z_+^k$ leads to
\begin{align}
 \partial_w\cG
 &=\sum_{\stackrel{\ell,k=1}{\ell\neq k}}^L Z^{[\ell,k]} \frac{1}{w-r_k}\ln\left|\frac{w-r_\ell}{r_\ell-r_k}\right|^2
 \nonumber\\
 &=
 \sum_{\stackrel{\ell,k=1}{\ell\neq k}}^L Z^{[\ell,k]}\frac{1}{r_\ell-r_k} \left[ \frac{r_\ell-r_k}{w-r_k}\ln\left|\frac{w-r_\ell}{r_\ell-r_k}\right|- 
 \frac{r_\ell-r_k}{w-r_\ell}\ln\left|\frac{w-r_k}{r_\ell-r_k}\right|\right]~.
\end{align}
For the second line the antisymmetry of $Z^{[\ell,k]}$ under exchange of $\ell$ and $k$ was used.
The expression in the second line can be recognized as the derivative of $D$,
\begin{align}
 \partial_u D(u)&=\frac{i}{2}\left(\frac{1}{u}\ln|1-u|+\frac{1}{1-u}\ln|u|\right)~.
\end{align}
This leads to
\begin{align}\label{eq:dcG-gen}
 \partial_w\cG
 &=\partial_w\Bigg[\sum_{\stackrel{\ell,k=1}{\ell\neq k}}^L 2iZ^{[\ell,k]} D\left(\frac{r_\ell-w}{r_\ell-r_k}\right)\Bigg]~.
\end{align}
$\cG$ is real by construction, and the regularity conditions (\ref{eqn:constr}), which were assumed for the derivation, ensure that it is constant on the boundary of the upper half plane. 
Fixing the remaining integration constant so that $\cG$ vanishes on the real line thus leads to (\ref{eq:cG-gen}).

\bibliography{Wilson}
\end{document}